\documentclass[a4paper,11pt]{article}
\pdfoutput=1 

\usepackage{jheppub} 

\usepackage[T1]{fontenc} 

\usepackage{amsmath}
\usepackage{dsfont}
\usepackage{feynmp}
\usepackage{slashed}

\usepackage{longtable}
\usepackage{multirow}

\usepackage{hhline}
\usepackage{rotating}
\usepackage{array}
\usepackage{float}
\usepackage{afterpage}

\usepackage[dvipsnames]{xcolor}

  {}

\allowdisplaybreaks

\def\beq{\begin{equation}}
\def\eeq#1{\label{#1}\end{equation}}
\def\eeqn{\end{equation}}

\def\beqa{\begin{eqnarray}}
\def\eeqa#1{\label{#1}\end{eqnarray}}
\def\eeqan{\end{eqnarray}}
\def\CR{\nonumber \\}

\def\bseq{\begin{subequations}}
\def\eseq#1{\label{#1}\end{subequations}}
\def\eseqn{\end{subequations}}

\def\si#1{{\color{OliveGreen}{#1}}} 
\def\ci#1{{\color{BrickRed}{#1}}} 
\def\wi#1{{\color{Cerulean}{#1}}} 
\newcommand\qn[3]{\left(\,#1\,,\,#2\,,\,#3\,\right)} 
\def\sicolor{{\color{OliveGreen}{green}}} 
\def\cicolor{{\color{BrickRed}{brown}}} 
\def\wicolor{{\color{Cerulean}{blue}}} 

\def\L{{\cal L}}
\def\O{{\cal O}}
\def\I{{\cal I}}
\def\It{\tilde{\cal I}}
\def\Q{{\cal Q}}
\def\b{\text{b}}
\def\c{\text{c}}
\def\identity{\mathds{1}}
\def\Tr{\mathrm{Tr}}
\def\tr{\mathrm{tr}}
\def\Psl{\slashed{P}}

\def\lf{\frac{i}{16\pi^2}}

\def\lfr{\frac{1}{16\pi^2}}

\def\logm#1{\log\frac{#1}{Q^2}}
\def\dsq#1{\Delta_{#1}^2}

\def\mat#1{\boldsymbol{#1}}

\def\sw{s_\theta}
\def\cw{c_\theta}
\def\sb{s_\beta}
\def\sbp{s_{\beta'}}
\def\sbb{s_{2\beta}}
\def\cb{c_\beta}
\def\cbp{c_{\beta'}}
\def\cbb{c_{2\beta}}
\def\mHusq{m_{H_u}^2}
\def\mHdsq{m_{H_d}^2}

\def\ft{\tilde{f}}
\def\qt{\tilde{q}}
\def\ut{\tilde{u}}
\def\dt{\tilde{d}}
\def\lt{\tilde{l}}
\def\et{\tilde{e}}
\def\chit{\tilde{\chi}}
\def\Vt{\tilde{V}}
\def\gt{\tilde{g}}
\def\Wt{\tilde{W}}
\def\Bt{\tilde{B}}

\def\CasL{C_2^{SU(2)}}
\def\CasC{C_2^{SU(3)}}

\def\Qth{\Lambda}
\def\QGUT{Q_\text{GUT}}

\preprint{LCTP-17-06}

\title{Effective field theory approach to trans\,-TeV supersymmetry: covariant matching, Yukawa unification and Higgs couplings}

 \author{James~D.~Wells}
 \author{and Zhengkang~Zhang}
 \affiliation{Leinweber Center for Theoretical Physics (LCTP), University of Michigan, \\450 Church Street, Ann Arbor, MI 48109, USA}

\emailAdd{jwells@umich.edu}
\emailAdd{zzkevin@umich.edu}

\abstract{
Dismissing traditional naturalness concerns while embracing the Higgs boson mass measurement and unification motivates careful analysis of trans-TeV supersymmetric theories. 
We take an effective field theory (EFT) approach, matching the Minimal Supersymmetric Standard Model (MSSM) onto the Standard Model (SM) EFT by integrating out heavy superpartners, and evolving MSSM and SMEFT parameters according to renormalization group equations in each regime. 
Our matching calculation is facilitated by the recent covariant diagrams formulation of functional matching techniques, with the full one-loop SUSY threshold corrections encoded in just 30 diagrams. 
Requiring consistent matching onto the SMEFT with its parameters (those in the Higgs potential in particular) measured at low energies, and in addition requiring unification of bottom and tau Yukawa couplings at the scale of gauge coupling unification, we detail the solution space of superpartner masses from the TeV scale to well above. 
We also provide detailed views of parameter space where Higgs coupling measurements have probing capability at future colliders beyond the reach of direct superpartner searches at the LHC.
}

\begin{document} 
\maketitle
\flushbottom

\section{Introduction}
\label{sec:intro}

The lack of new physics discoveries at the LHC has led us to consider the possibility that beyond Standard Model (BSM) states responsible for solving the outstanding problems in particle physics, e.g.\ unification and dark matter, are much heavier than the weak scale. In this scenario, weak-scale phenomenology can be conveniently described by an effective field theory (EFT). With heavy new particles integrated out, their virtual effects are encoded in higher-dimensional effective operators involving the light Standard Model (SM) fields. Recent years have seen a growing EFT literature, many of which aim to carefully examine phenomenological impact of higher-dimensional operators; see e.g.~\cite{WillenbrockZhangEFTreview,FalkowskiEFTreview,CERNYR4} for reviews. Experimental data have put constraints on many of these operators, which can be translated into constraints on, e.g.\ masses and couplings of heavy new particles, once the BSM theory is specified.

On the other hand, the usefulness of EFT approaches to BSM physics extends beyond bottom-up studies. From a top-down perspective, we may be interested to ask whether some attractive speculative ideas -- supersymmetry (SUSY), unification, etc.\ -- can be realized in specific BSM setups, while being consistent with the SMEFT we have established at low energy. To address such questions requires careful matching between the full theory and EFT parameters across heavy particle thresholds. In particular, in addition to higher-dimensional operators being generated, changes in renormalizable operator coefficients across thresholds are often important to account for. These ``threshold corrections'' are invisible to low-energy experiment, but may be crucial for answering questions regarding high-scale physics, like the one on SUSY and unification posed above.

For example, in the context of the Minimal Supersymmetric Standard Model (MSSM), we would like to know what regions of parameter space can realize unification of not only the three gauge couplings, but also the bottom and tau Yukawa couplings, and meanwhile allow consistent matching onto the SMEFT with its parameters (those in the Higgs potential in particular) measured at low energy. Further, we would like to know what phenomenological implications, if any, such parameter choices may have.

These are questions we would like to investigate in this paper, taking a top-down EFT approach. We will compute the full one-loop contributions to the SM renormalizable operators when heavy BSM states in the MSSM are integrated out, from which SUSY threshold corrections to all SM parameters can be easily obtained. As we will see, threshold corrections to the bottom Yukawa and Higgs quartic couplings are of particular importance for achieving both $b$-$\tau$ Yukawa unification and consistent matching onto the SMEFT.

In this calculation, we find solutions for SUSY scales from TeV up to $10^{10}\,\text{GeV}$. However, only the lower edge of this broad trans-TeV window can be within experimental reach. Given the further motivation of a dark matter candidate, we will take a closer look at the $1$-$10\,\text{TeV}$ regime. In particular, we will extend our one-loop matching calculation to the dimension-six level, and obtain parametrically enhanced contributions to the operators affecting $hb\bar b$ and $h\tau^+\tau^-$ couplings, which can dominate over tree-level effects. We will show how precision Higgs measurements constitute a powerful indirect probe of TeV-scale SUSY with $b$-$\tau$ Yukawa unification that is complementary to direct superpartner searches.

We note that while the full one-loop SUSY threshold corrections (as well as sparticle mass corrections) in the MSSM have been known for some time~\cite{BMPZ}, growing interest in EFT formulations of the calculation is quite recent (see e.g.~\cite{DraperLeeWagner13,unnaturalSUSY,SUSYHD,LeeWagner15,FlexibleEFTHiggs,StaubPorod,Bagnaschi17}). This is of course largely due to higher-scale SUSY having been less attractive from the fine-tuning point of view. Here we shall adopt the perspective that the weak scale may indeed be fine-tuned to some degree, with new particles in the trans-TeV regime, justifying an EFT treatment. In this case, the MSSM still exhibits several attractive features, including gauge and Yukawa coupling unification as mentioned above, and may also provide a dark matter candidate. Our philosophy here is in line with earlier studies in~\cite{TobeWells,EHPR} on Yukawa unification with heavy superpartners. In particular, the capability of unification and dark matter requirements to severely constrain the SUSY parameter space has been recently emphasized in~\cite{EHPR}.

A new ingredient of our work is that, instead of computing low-energy observables such as SM particle masses and the Fermi constant, we obtain SUSY threshold corrections directly from the path integral by taking advantage of functional matching techniques, which have attracted much attention and undergone interesting developments recently~\cite{Gaillard85,Chan86,Cheyette87,HLM14,DEQY,HLM16,EQYZ16,FPR,CovDiag,EQYZ17} (see also~\cite{Huo:scalar,Huo:fermion,Huo:MSSM,DKS,BGP}). Such techniques can greatly simplify EFT matching calculations, thanks to preservation of gauge covariance and exhibition of a universal structure~\cite{DEQY,EQYZ16,EQYZ17}. In~\cite{CovDiag}, a concise diagrammatic formulation of functional matching at one-loop level was obtained: the low-energy effective Lagrangian directly derives from a sum of ``covariant diagrams,'' following a set of simple rules. This approach is general enough to overcome several limitations of previous formulations, and so will be used here.

From the technical point of view, our calculation also serves as a nontrivial test case for the covariant diagrams technique. It also further demonstrates the simplicity of the approach. In particular, with just 30 covariant diagrams, we are able to obtain full one-loop SUSY threshold corrections in agreement with existing results in the literature.

The paper is organized as follows. In Section~\ref{sec:overview}, we give a general overview of EFT approach to BSM physics. In particular, we aim to present a concise and cogent review of functional matching and covariant diagrams, for readers unfamiliar with such techniques to quickly grasp the essentials.

Starting from Section~\ref{sec:MSSMmatch}, we focus on the specific case of the MSSM. We first present an analytical calculation of matching the MSSM onto the SMEFT. Next, Sections~\ref{sec:YU} and~\ref{sec:Higgs} are dedicated to numerical analyses of implications of $b$-$\tau$ Yukawa unification on the SUSY spectrum and Higgs couplings. Finally, we conclude in Section~\ref{sec:concl}.

\section{Overview of EFT approach to BSM physics}
\label{sec:overview}

\subsection{EFT matching, threshold corrections and observables}

Consider a general BSM theory whose Lagrangian has the following form,
\beq
\L\,[\varphi_\text{BSM}, \varphi_\text{SM}] = \L_\text{SM}[\varphi_\text{SM}] + \L_\text{BSM}[\varphi_\text{BSM}, \varphi_\text{SM}] \,.
\eeqn
Here $\varphi_\text{SM}$, $\varphi_\text{BSM}$ collectively denote fields within and beyond the SM, respectively. The SM part of the Lagrangian reads
\beqa
\L_\text{SM} &=& |D_\mu \phi|^2 +\sum_{f=q,u,d,l,e} \bar f\, i\slashed{D} f -\frac{1}{4} G^A_{\mu\nu} G^{A\mu\nu} -\frac{1}{4} W^I_{\mu\nu} W^{I\mu\nu} -\frac{1}{4} B_{\mu\nu} B^{\mu\nu} \CR
&&  -m^2|\phi|^2 -\lambda|\phi|^4 -\bigl(\bar\psi_u \,\mat{y_u}\,\psi_q \cdot \epsilon \cdot \phi +\bar\psi_d \,\mat{y_d}\,\psi_q \cdot \phi^* +\bar\psi_e \,\mat{y_e}\,\psi_l \cdot \phi^* +\text{h.c.} \bigr) \,,\quad
\eeqa{LSM}
where $\epsilon=i\sigma^2$, and dots denote $SU(2)_L$ index contractions. $\psi_f$ are four-component spinors containing the SM chiral fermions $f$, e.g.\ $\psi_q = (q_a, 0)$, $\psi_u=(0, u^{\dagger\dot{a}})$, etc. 
Here and in the following, we use boldface, e.g.\ $\mat{y_u}$, $\mat{y_d}$, $\mat{y_e}$, for $3\times3$ matrices in generation space. 

The EFT approach applies when the BSM fields $\varphi_\text{BSM}$ are much heavier than the SM weak scale. In this case, integrating out $\varphi_\text{BSM}$ from the path integral results in a local effective Lagrangian for $\varphi_\text{SM}$,
\beqa
\int [D\varphi_\text{BSM}] [D\varphi_\text{SM}] \,e^{\,i\int d^dx \,(\L_\text{SM}+\L_\text{BSM})} 
&=& \int [D\varphi_\text{SM}] \,e^{\,i\int d^dx \,\bigl(\L_\text{SM} +\L_{(d\le4)} +\L_{(d=5)} +\L_{(d=6)} +\dots\bigr)} \CR
&\equiv& \int [D\varphi_\text{SM}] \,e^{\,i\int d^dx \,\L_\text{SMEFT}} \,.
\eeqa{pathintmatching}

As implied in the equation above, this procedure of matching $\L$ onto $\L_\text{SMEFT}$ generally produces extra renormalizable ($d\le4$) pieces in the EFT Lagrangian, in addition to $\L_\text{SM}$ that already exists in the full theory. However, they can be absorbed into $\L_\text{SM}$ via proper redefinitions of fields and couplings and thus do not have observable consequences at low energy. To be explicit, let us write
\beqa
\L_{(d\le4)} &=& \delta Z_\phi |D_\mu \phi|^2 +\sum_{f=q,u,d,l,e} \bar\psi_f\,\mat{\delta Z_f}\, i\slashed{D}\psi_f \CR
&& -\frac{1}{4} \delta Z_G G^A_{\mu\nu} G^{A\mu\nu} -\frac{1}{4}\delta Z_W W^I_{\mu\nu} W^{I\mu\nu} -\frac{1}{4}\delta Z_B B_{\mu\nu} B^{\mu\nu} \CR
&& +\delta m^2|\phi|^2 +\delta\lambda|\phi|^4 \CR
&& +\bigl(\bar\psi_u \,\mat{\delta y_u}\,\psi_q \cdot \epsilon \cdot \phi +\bar\psi_d \,\mat{\delta y_d}\,\psi_q \cdot \phi^* +\bar\psi_e \,\mat{\delta y_e}\,\psi_l \cdot \phi^* +\text{h.c.} \bigr) \,.
\eeqa{Ldle4}
Rescaling the SM fields to retain canonical normalization of their kinetic terms (up to terms of second order or higher in the $\delta Z$'s\footnote{Here $\delta Z_{\varphi_\text{SM}}$ is understood as the matrix $\mat{\delta Z_f}$ for $\varphi_\text{SM}=f$, which is symmetric in generation space as required by hermiticity of the Lagrangian.}),
\beq
\hat\varphi_\text{SM} = \Bigl(1+\frac{1}{2}\delta Z_{\varphi_\text{SM}} \Bigr) \varphi_\text{SM} \,,
\eeq{rescale}
and defining effective parameters as follows,
\beqa
&& g_3^\text{eff} = g_3\,\Bigl(1-\frac{1}{2}\, \delta Z_G\Bigr) \,, \qquad
g^\text{eff} = g\,\Bigl(1-\frac{1}{2}\, \delta Z_W\Bigr) \,, \qquad
{g'}^\text{eff} = g'\,\Bigl(1-\frac{1}{2} \, \delta Z_B\Bigr) \,, \CR
&& m_\text{eff}^2 = m^2\, (1-\delta Z_\phi) - \delta m^2 \,,\qquad
\lambda_\text{eff} = \lambda\,(1-2\,\delta Z_\phi) -\delta \lambda \,,\CR
&& \mat{y_u^\text{eff}} = \mat{y_u} -\mat{\delta y_u} -\frac{1}{2} (\mat{y_u}\, \mat{\delta Z_q} + \mat{\delta Z_u}\, \mat{y_u}  + \mat{y_u}\, \delta Z_\phi )\,,\CR
&& \mat{y_d^\text{eff}} = \mat{y_d} -\mat{\delta y_d} -\frac{1}{2} (\mat{y_d}\, \mat{\delta Z_q} + \mat{\delta Z_d}\, \mat{y_d}  + \mat{y_d}\, \delta Z_\phi )\,,\CR
&& \mat{y_e^\text{eff}} = \mat{y_e} -\mat{\delta y_e} -\frac{1}{2} (\mat{y_e}\, \mat{\delta Z_l} + \mat{\delta Z_e}\, \mat{y_e}  + \mat{y_e}\, \delta Z_\phi ) \,,
\eeqa{thrcor}
we obtain
\beqa
\L_\text{SMEFT} &=& |D_\mu \hat\phi|^2 +\sum_{f=q,u,d,l,e} \bar {\hat f}\, i\slashed{D} \hat f -\frac{1}{4} \hat G^A_{\mu\nu} \hat G^{A\mu\nu} -\frac{1}{4} \hat W^I_{\mu\nu} \hat W^{I\mu\nu} -\frac{1}{4} \hat B_{\mu\nu} \hat B^{\mu\nu} \CR
&&  -m_\text{eff}^2|\hat\phi|^2 -\lambda_\text{eff}|\hat\phi|^4 -\bigl(\bar{\hat\psi}_u \,\mat{y_u^\text{eff}}\, \hat\psi_q \cdot \epsilon \cdot \hat\phi +\bar{\hat\psi}_d \,\mat{y_d^\text{eff}}\, \hat\psi_q \cdot \hat\phi^* +\bar{\hat\psi}_e \,\mat{y_e^\text{eff}}\, \hat\psi_l \cdot \hat\phi^* +\text{h.c.} \bigr) \CR[4pt]
&& +\,\L_{(d=5)} +\,\L_{(d=6)} +\dots,
\eeqa{LSMEFTeff}
where
\beq
D_\mu = \partial_\mu -i\,g_3^\text{eff}\, t^A \hat G^A_\mu -i\,g^\text{eff}\, t^I \hat W^I_\mu -i\,{g'}^\text{eff}\, Y \hat B_\mu \,,
\eeqn
with $t^A$, $t^I$ being the $SU(3)_c$ and $SU(2)_L$ generators in the corresponding representation.

We see that while the renormalizable part of $\L_\text{SMEFT}$ contains the same operators as the $\L_\text{SM}$ part of the full theory Lagrangian, their coefficients, i.e.\ the parameters labeled by ``eff'' whose values we can extract from experiment, are generally different from their counterparts in the full theory. These differences are usually referred to as ``threshold corrections,'' and are important to take into account when studying higher-energy phenomena of the full theory, such as unification in the MSSM. It is clear from Eq.~\eqref{thrcor} that threshold corrections are directly related to operator coefficients in the $\L_{(d\le4)}$ piece generated from the matching procedure of Eq.~\eqref{pathintmatching}.

On the other hand, the non-renormalizable part of $\L_\text{SMEFT}$, i.e.\ $\L_{(d=5)} +\L_{(d=6)} +\dots$, can cause low-energy observations to deviate from expectations of the renormalizable SM: $\L_{(d=5)}$ contains just one operator which is responsible for non-zero neutrino masses, while $\L_{(d=6)}$ contains a large number of operators which contribute to e.g.\ electroweak, Higgs, and flavor observables. For example, consider the following dimension-six operators (neglecting differences between $\varphi_\text{SM}$ and $
\hat\varphi_\text{SM}$),
\beq
\L_{(d=6)} \supset |\phi|^2 \,(\bar\psi_q\,\mat{C_{d\phi}}\,\psi_d)\cdot\phi +|\phi|^2 \,(\bar\psi_l\,\mat{C_{e\phi}}\,\psi_e)\cdot\phi \,+\text{h.c.}
\eeqn
After RG evolved down to the weak scale, they affect couplings of the SM Higgs boson to down-type quarks and leptons, and hence observables like the Higgs boson partial widths. When fermion masses are used as inputs of the calculation, we have
\beq
\Gamma(h\to f\bar f)=(1+\delta\kappa_f)^2\,\Gamma(h\to f\bar f)_\text{SM} \,,
\eeqn
with
\beq
\delta\kappa_b = -\frac{C_{b\phi} v^2}{y_b^\text{eff}} \,,\qquad
\delta\kappa_\tau = -\frac{C_{\tau\phi} v^2}{y_\tau^\text{eff}} \,,
\eeq{dkf}
etc.\ at tree level, where $y_{b,\tau}^\text{eff}$, $C_{b\phi,\tau\phi}$ are $33$ elements of $\mat{y_{d,e}^\text{eff}}$, $\mat{C_{d\phi,e\phi}}$, respectively. 

Note that $C_{b\phi,\tau\phi}\sim\Lambda^{-2}$ with $\Lambda$ being the scale of new physics being integrated out, and therefore, the observable BSM effects $\delta\kappa_{b,\tau}$ decouple as $\frac{v^2}{\Lambda^2}$ as $\Lambda$ increases. This is in contrast to the (unobservable) threshold corrections discussed above, which originate from $d\le4$ operators and thus do not decouple. We will see in Section~\ref{sec:YU} that in the specific case of the MSSM with $b$-$\tau$ Yukawa unification, threshold corrections to $\lambda$ and $y_b$ are actually larger for higher SUSY scales.

Meanwhile, in addition to the $\Lambda$ power counting, the low-energy EFT is also organized by a loop counting. Take the calculation of $\Gamma(h\to b\bar b)$ for example. Higher order corrections come from both EFT matching for $d>4$ operators ($C_{b\phi}=C_{b\phi}^\text{tree}+C_{b\phi}^\text{1-loop}+\dots$ in the present case) and loop level Feynman diagrams in the EFT. Generally speaking, when $\Lambda$ is much higher than the weak scale, the only such corrections that are essential to take into account are the non-decoupling ones from the renormalizable SM loops, namely corrections to $\Gamma(h\to b\bar b)_\text{SM}$ (see~\cite{BCKhbb,BFKhbb,DSWhbb} for state-of-the-art calculations). An exception is when $\O(\frac{1}{16\pi^2\Lambda^2})$ corrections are parametrically enhanced, e.g.\ by $\tan\beta\gg1$ in the case of the MSSM. We will see in Section~\ref{sec:Higgs} that such enhanced contributions to $C_{b\phi}^\text{1-loop}$ can dominate over $C_{b\phi}^\text{tree}$ in some regions of the MSSM parameter space, and are therefore also essential to take into account when making predictions for $\Gamma(h\to b\bar b)$ in the EFT. The results can of course be further refined by computing non-enhanced contributions, and even higher order terms in both the $\Lambda^{-1}$ and loop expansions (see e.g.~\cite{NothSpira,MihailaReisser,CrivellinGreub,GauldPecjakScott}). See also~\cite{BCKpowercounting,BCKsilh,GJMMpowercounting,BCCKcomment} for related discussions on EFT power counting.

\subsection{Functional matching and covariant diagrams}
\label{sec:CDrev}

We now review the procedure of carrying out the matching calculation of Eq.~\eqref{pathintmatching}. The idea is to derive the EFT Lagrangian as a sum of {\it gauge-invariant operators} directly from the path integral, {\it without computing correlation functions}. In particular, we will sketch the procedure leading to the systematic formulation of covariant diagrams, the details of which can be found in Ref.~\cite{CovDiag}. This can be viewed as the analog of the procedure of deriving Feynman rules for correlation functions. However, it should be clear that here we are taking a different route from the path integral than deriving Feynman rules, which is in fact a most economic route where we gather just enough information for the purpose of obtaining the EFT Lagrangian. 
We shall use a more general notation $\varphi_\text{BSM}\to\varphi_H$ (``H'' for ``heavy''), $\varphi_\text{SM}\to\varphi_L$ (``L'' for ``light''), since the applicability of the approach is not restricted to the specific case of matching decoupled BSM theories onto the SMEFT. 

Our goal is to integrate out the heavy fields $\varphi_H$ in the UV theory path integral, as in Eq.~\eqref{pathintmatching}. At tree (classical) level, the result is given by the stationary point approximation,
\beq
\L_\text{EFT}^\text{tree}[\varphi_L] = \L_\text{UV}\bigl[\varphi_{H,\c}[\varphi_L], \varphi_L\bigr] \,,
\eeq{LEFTtree}
where $\varphi_{H,\c}[\varphi_L]$ is a gauge-invariant local operator expansion that solves the classical equations of motion for $\varphi_H$,
\beq
\left.\frac{\delta\L_\text{UV}}{\delta\varphi_H} \right|_{\varphi_H = \varphi_{H,\c}} = 0 \,.
\eeq{eom}

Moving on to one-loop level, we take into account the leading quantum corrections using the background field method,
\beq
\varphi_H = \varphi_{H,\c}[\varphi_{L,\b}] +\varphi_H' \,,\quad \varphi_L = \varphi_{L,\b} +\varphi_L' \,,
\eeq{bkgfld}
\beqa
\Rightarrow \L_\text{UV}[\varphi_H,\varphi_L] +J_L\varphi_L &=& \L_\text{UV}\bigl[\varphi_{H,\c}[\varphi_{L,\b}],\varphi_{L,\b}\bigr] +J_L\varphi_{L,\b} \CR
&& -\frac{1}{2} \bigl(\varphi_H^{\prime\, T} \;\; \varphi_L^{\prime\, T} \bigr) \,\Q_\text{UV} \bigl[\varphi_{H,\c}[\varphi_{L,\b}],\varphi_{L,\b}\bigr] \left(
\begin{matrix}
\varphi_H' \\ \varphi_L'
\end{matrix}
\right) +\O(\varphi'^3) \,.\qquad
\eeqa{LUVexpand}
Up to this order in the quantum fluctuation fields $\varphi_{H,L}'$, the path integral is Gaussian, and evaluates to the functional determinant of the quadratic operator $\Q_\text{UV}$. This is in fact the familiar procedure of computing one-particle-irreducible (1PI) effective actions. Here, since we have set the background heavy fields to $\varphi_{H,\c}[\varphi_{L,\b}]$, which corresponds to setting heavy fields' currents  $J_H = -\left.\frac{\delta\L_\text{UV}}{\delta\varphi_H}\right|_{\varphi=\varphi_\b}$ to zero in the Legendre transform, what we obtain is the one-light-particle-irreducible (1LPI) effective action,
\beqa
\Gamma_\text{L,UV}^\text{1-loop} [\varphi_{L,\b}] &=& i\,c_s\, \log\det\Q_\text{UV} \bigl[\varphi_{H,\c}[\varphi_{L,\b}], \varphi_{L,\b}\bigr]  \CR
&=& i\,c_s\, \Tr\,\log\Q_\text{UV}
= i\,c_s\int d^dx \int\frac{d^dq}{(2\pi)^d}\, \tr\,\log\left.\Q_\text{UV}\right|_{P_\mu\to P_\mu-q_\mu} ,
\eeqa{GammaL}
where $c_s$ is a spin factor, e.g.\ $c_s=\frac{1}{2}$ ($-\frac{1}{2}$) for real scalars and vectors (Weyl fermions). The last equation follows from evaluating the spacetime part of the functional trace, and the remaining ``tr'' is over internal degrees of freedom only. We have introduced the notation $P_\mu\equiv iD_\mu$, understood to be acting on everything to the right (so that $(D_\mu\varphi) = -i\,[P_\mu,\varphi]$, etc.). A convenient feature of $P_\mu$ is that it is a hermitian operator,
\beq
(\dots A P_\mu B\dots)^\dagger =\, \dots B^\dagger (-i\overleftarrow{D}_\mu) A^\dagger \dots \,=\, \dots B^\dagger P_\mu A^\dagger\dots
\eeqn

We would like to derive $\L_\text{EFT}^\text{1-loop}$ from $\Gamma_\text{L,UV}^\text{1-loop}$. The fact that $\int d^dx\, \L_\text{EFT}^\text{1-loop}[\varphi_L] \ne \Gamma_\text{L,UV}^\text{1-loop} [\varphi_L]$ has caused some confusion in the previous literature on the capability of functional methods to compute $\L_\text{EFT}^\text{1-loop}$. However, as argued in~\cite{FPR} and proved in~\cite{CovDiag}, there is actually a simple relation between the two quantities,
\beq
\int d^dx\, \L_\text{EFT}^\text{1-loop}[\varphi_L] = \left.\Gamma_\text{L,UV}^\text{1-loop} [\varphi_L]\right|_\text{hard} ,
\eeq{hard}
where ``hard'' means taking the hard region contribution to the loop integral. Technically, using dimensional regularization, one simply expands the integrand for $|q^2|\sim m_{\varphi_H}^2 \gg m_{\varphi_L}^2$ (with $q$ being the loop momentum) before integrating over the full momentum space~\footnote{Correspondingly, the soft region contribution is obtained by expanding the integrand for $|q^2|\sim m_{\varphi_L}^2 \ll m_{\varphi_H}^2$ before integrating over the full momentum space. The sum of hard and soft region contributions is equal to the original integral; see e.g.~\cite{EBR-BenekeS,EBR-Smirnov,EBR-Jantzen}.}.

The physical intuition of Eq.~\eqref{hard} is the following. With an integral over the full momentum space $\int\frac{d^dq}{(2\pi)^d}$ as in Eq.~\eqref{GammaL}, the 1LPI effective action $\Gamma_\text{L,UV}^\text{1-loop}$ encodes quantum fluctuations at {\it all scales} in the UV theory (when used at classical level). It is non-local in general due to long-distance contributions. By taking the hard region contribution as in Eq.~\eqref{hard}, we essentially extract {\it short-distance} information from $\Gamma_\text{L,UV}^\text{1-loop}$, manifest as {\it local} effective operators in $\L_\text{EFT}^\text{1-loop}$.

Combining Eqs.~\eqref{GammaL} and~\eqref{hard}, we have
\beq
\L_\text{EFT}^\text{1-loop} [\varphi_{L}] 
= i\,c_s \,\mathrm{tr}\int\frac{d^dq}{(2\pi)^d}\, \left[\,\log\left.\Q_\text{UV}\right|_{P_\mu\to P_\mu-q_\mu}\right]_{\text{expand for }|q^2|\sim m_{\varphi_H}^2 \gg |m_{\varphi_L}^2|} .
\eeq{LEFTloop}
We would like to perform the expansion in a gauge-covariant way, without separating the $P_\mu$'s contained in $\Q_\text{UV}$ into partial derivatives and gauge fields. In other words, we would like to perform a covariant derivative expansion (CDE), which automatically ensures gauge invariance of the final result. To begin with, we assume the following general (gauge-covariant) form for the quadratic operator,
\beq
\Q_\text{UV} = {\bf K}[P_\mu; m_{\varphi_H}, m_{\varphi_L}] + {\bf X} [\varphi, P_\mu] \,,
\eeq{QUV}
where ${\bf K}$ is the diagonal kinetic operator with elements $-P^2+m^2$ for bosonic fields and $-\Psl+m$ for fermionic fields\footnote{\label{conjfootnote}Note that this requires $\big(\varphi_H^{\prime\, T} \; \varphi_L^{\prime\, T} \bigr)$ on the left side of $\Q_\text{UV}$ in Eq.~\eqref{LUVexpand} should be replaced by a conjugate field multiplet that contains $\phi^\dagger$ and $\bar\psi$ for complex scalars and fermions, respectively. Also, to have the right prefactor $-\frac{1}{2}$ for the quadratic terms, we can represent each complex scalar $\phi$ by a multiplet $(\phi,\phi^*)^T$, and each Dirac fermion $\psi$ by a multiplet $(\psi, \psi^c)^T$.}, and
\beq
{\bf X}[\varphi, P_\mu] = \left(
\begin{matrix}
X_H &\,\, X_{HL} \\
X_{LH} &\,\, X_L
\end{matrix}
\right) = {\bf U}[\varphi] +P_\mu\, {\bf Z}^\mu[\varphi] +{\bf Z}^{\dagger\mu}[\varphi]\, P_\mu +\O(P^2)
\eeq{Xmat}
contains interactions among the fields. The expansion and momentum integration in Eq.~\eqref{LEFTloop} can then be carried out in a general way, in terms of $P_\mu$ and components of $\bf U[\varphi]$, $\bf Z[\varphi]$ matrices. 

The CDE procedure outline above, albeit conceptually simple, can be quite tedious technically due to the large number of terms produced by the expansion. The idea of covariant diagrams is to significantly reduce the technical complexity by introducing a systematic bookkeeping device that collects identical terms. As it turns out, the rules for using covariant diagrams, as derived in~\cite{CovDiag}, are very simple. We will review some of these rules in Section~\ref{sec:CDrev-rules}.

It is worth noting that the philosophy here is quite similar to using Feynman diagrams to keep track of the (non-gauge-covariant) expansion of correlation functions. However, unlike Feynman diagrams, covariant diagrams represent expressions made of covariant derivatives $P_\mu$ and the light fields $\varphi_L$ contained in the $\bf U[\varphi]$, $\bf Z[\varphi]$ matrices (after setting $\varphi_H=\varphi_{H,\c}[\varphi_L]$), which combine into gauge-invariant operators. Therefore, by enumerating covariant diagrams, we are able to directly obtain all operators generated by one-loop matching, avoiding the detour of computing correlation functions.

\subsubsection{Rules for covariant diagrams}
\label{sec:CDrev-rules}

Covariant diagrams consist of propagators and vertex insertions of the form shown in Table~\ref{tab:cdrules}.\footnote{We are using a slightly different notation than~\cite{CovDiag}: here we prefer to make the distinction between bosonic and fermionic propagators more transparent by using different types of lines (dashed vs.\ solid). In~\cite{CovDiag}, on the other hand, more emphasis is put on the treatment of heavy vs.\ light fields, and solid (dashed) propagators are used for heavy (light) propagators regardless of spin.} Some of them carry Lorentz indices, which we connect in pairs with dotted lines to indicate Lorentz contraction. There are different types of vertex insertions, represented by different symbols. Only two of them, $P$ and $U$, are shown in the table, which are all we need for the calculation in this paper\footnote{From the CDE procedure summarized above it should be clear that additional types of vertex insertions include $Z$, $Z^\dagger$ and $m_{\varphi_L}$ insertions. There are only a few nonzero entries of the MSSM $\bf Z$ matrix because of minimal coupling, none of which contributes to the operators we will calculate. Also, $m_{\varphi_L}$ insertions are not considered because they lead to terms suppressed by powers of $\frac{m^2}{\Lambda^2}$ compared to those retained in our results.}.

\begin{table}[t]
\begin{center}
\begin{tabular}{|c|l|l|}
\hline
Building blocks & \multicolumn{1}{c|}{Bosonic} & \multicolumn{1}{c|}{Fermionic} \\
\hhline{|===|}
\multirow{2}{*}{\parbox[c][72pt][t]{90pt}{\centering \vspace{24pt}
Propagators}} & 
\multirow{2}{*}{
\qquad\parbox[c][60pt][c]{42pt}{\centering 
\begin{fmffile}{cd-prop-b}
\begin{fmfgraph*}(40,20)
\fmfleft{i1}
\fmfright{o1}
\fmf{dashes,label=$i$,l.s=left,l.d=3pt}{i1,o1}
\end{fmfgraph*}
\end{fmffile}}
$\;\; =\; 1$
} & 
\quad\parbox[c][36pt][c]{42pt}{\centering
\begin{fmffile}{cd-prop-f0}
\begin{fmfgraph*}(40,20)
\fmfleft{i1}
\fmfright{o1}
\fmf{plain,label=$i$,l.s=left,l.d=3pt}{i1,o1}
\end{fmfgraph*}
\end{fmffile}}
$\;\; =\;\biggl\{\begin{matrix} & M_i && \text{(heavy)}\\ & 0 && \text{(light)} \end{matrix}$
\\
& & \quad\parbox[c][36pt][c]{42pt}{\centering
\begin{fmffile}{cd-prop-f1}
\begin{fmfgraph*}(40,20)
\fmfleft{i1,i2,i3}
\fmfright{o1,o2,o3}
\fmf{plain}{i2,v}
\fmf{plain}{v,o2}
\fmf{phantom}{i1,b,o1}
\fmf{dots,width=thick,tension=0}{b,v}
\fmfv{label=$i$,l.a=90,l.d=3pt}{v}
\end{fmfgraph*}
\end{fmffile}}
$\;\; =\; -\gamma_\mu $ 
\\
\hhline{|===|}
$P$ insertions & 
\qquad \parbox[c][36pt][c]{42pt}{\centering
\begin{fmffile}{cd-Pins-b}
\begin{fmfgraph*}(40,25)
\fmfleft{i1,i2,i3}
\fmfright{o1,o2,o3}
\fmf{dashes,label=$i$,l.s=left,l.d=3pt}{i2,vP}
\fmf{dashes,label=$i$,l.s=right,l.d=3pt}{o2,vP}
\fmf{phantom}{i1,b,o1}
\fmf{dots,width=thick,tension=0}{b,vP}
\fmfv{decor.shape=circle,decor.filled=full,decor.size=3thick}{vP}
\end{fmfgraph*}
\end{fmffile}}
$\;\; =\; 2P_\mu \qquad$ & 
\quad\parbox[c][36pt][c]{42pt}{\centering
\begin{fmffile}{cd-Pins-f}
\begin{fmfgraph*}(40,25)
\fmfleft{i1}
\fmfright{o1}
\fmf{plain,label=$i$,l.s=left,l.d=3pt}{i1,vP}
\fmf{plain,label=$i$,l.s=right,l.d=3pt}{o1,vP}
\fmfv{decor.shape=circle,decor.filled=full,decor.size=3thick}{vP}
\end{fmfgraph*}
\end{fmffile}}
$\;\; = \; -\Psl $ 
\\
\hline
$U$ insertions & \multicolumn{2}{l|}{
\quad\parbox[c][36pt][c]{42pt}{\centering
\begin{fmffile}{cd-Uins-bb}
\begin{fmfgraph*}(40,25)
\fmfleft{i1}
\fmfright{o1}
\fmf{dashes,label=$i$,l.s=left,l.d=3pt}{i1,vU}
\fmf{dashes,label=$j$,l.s=right,l.d=2pt}{o1,vU}
\fmfv{decor.shape=circle,decor.filled=empty,decor.size=3thick}{vU}
\end{fmfgraph*}
\end{fmffile}}
\;,\;
\parbox[c][36pt][c]{42pt}{\centering
\begin{fmffile}{cd-Uins-bf}
\begin{fmfgraph*}(40,25)
\fmfleft{i1}
\fmfright{o1}
\fmf{dashes,label=$i$,l.s=left,l.d=3pt}{i1,vU}
\fmf{plain,label=$j$,l.s=right,l.d=2pt}{o1,vU}
\fmfv{decor.shape=circle,decor.filled=empty,decor.size=3thick}{vU}
\end{fmfgraph*}
\end{fmffile}}
\;,\;
\parbox[c][36pt][c]{42pt}{\centering
\begin{fmffile}{cd-Uins-fb}
\begin{fmfgraph*}(40,25)
\fmfleft{i1}
\fmfright{o1}
\fmf{plain,label=$i$,l.s=left,l.d=3pt}{i1,vU}
\fmf{dashes,label=$j$,l.s=right,l.d=2pt}{o1,vU}
\fmfv{decor.shape=circle,decor.filled=empty,decor.size=3thick}{vU}
\end{fmfgraph*}
\end{fmffile}}
\;,\;
\parbox[c][36pt][c]{42pt}{\centering
\begin{fmffile}{cd-Uins-ff}
\begin{fmfgraph*}(40,25)
\fmfleft{i1}
\fmfright{o1}
\fmf{plain,label=$i$,l.s=left,l.d=3pt}{i1,vU}
\fmf{plain,label=$j$,l.s=right,l.d=2pt}{o1,vU}
\fmfv{decor.shape=circle,decor.filled=empty,decor.size=3thick}{vU}
\end{fmfgraph*}
\end{fmffile}}
$\;\; = \;U_{ij} [\varphi]$\;\;
} \\
\hhline{|===|}
Contractions & \multicolumn{2}{c|}{\parbox[c][36pt][c]{42pt}{\centering
\begin{fmffile}{cd-contr}
\begin{fmfgraph*}(40,25)
\fmfleft{i1}
\fmfright{o1}
\fmf{dots,width=thick,label=$\mu\quad$,l.s=left,l.d=2pt}{i1,vP}
\fmf{dots,width=thick,label=$\quad\;\nu$,l.s=left,l.d=3pt}{vP,o1}
\fmfv{}{vP}
\end{fmfgraph*}
\end{fmffile}}
$\;\;=\; g^{\mu\nu}$} \\
\hline
\end{tabular}
\caption{\label{tab:cdrules}
Building blocks of covariant diagrams needed in this paper. $i,j$ represent fields that can be either heavy or light unless specified otherwise. A covariant diagram must contain at least one heavy propagator so that its prefactor \eqref{prefactor} is nonzero. See~\cite{CovDiag} for a detailed derivation of the rules for covariant diagrams from evaluating the CDE series of Eq.~\eqref{LEFTloop}.}
\end{center}
\end{table}

Each covariant diagram drawn this way represents a collection of terms in the CDE of Eq.~\eqref{LEFTloop} that reads ``prefactor$\,\cdot\,\tr\,\O[P_\mu, \varphi]$.'' The operator structure $\O[P_\mu, \varphi]$ is obtained simply by sequentially reading off each building block according to Table~\ref{tab:cdrules}. The prefactor takes care of combinatorics (which can be quite tedious to work out if one were to proceed algebraically as in previous CDE literature). For a covariant diagram with $n_i$ ($n_j, \dots$) heavy propagators of tree-level mass $M_i$ ($M_j, \dots$), $n_L$ light propagators, and $n_c$ dotted lines (Lorentz contractions), we have
\beq
\text{prefactor} \;=\,
-i\,c_s\,\frac{1}{S}\,\I[q^{2n_c}]_{ij\dots 0}^{n_i n_j \dots n_L} 
=\frac{c_s}{16\pi^2}\,\frac{1}{S}\,\It[q^{2n_c}]_{ij\dots 0}^{n_i n_j \dots n_L} \,,
\eeq{prefactor}
where the spin factor $c_s$ is determined by the spin of the propagator from which one starts reading the diagram. The symmetry factor $\frac{1}{S}$ is present if the diagram has a $\mathbb{Z}_S$ symmetry under rotation, and the {\it master integrals} $\,\I[q^{2n_c}]_{ij\dots 0}^{n_i n_j \dots n_L} = \frac{i}{16\pi^2}\,\It[q^{2n_c}]_{ij\dots 0}^{n_i n_j \dots n_L}$ are defined by
\beq
\int\frac{d^dq}{(2\pi)^d} \frac{q^{\mu_1}\cdots q^{\mu_{2n_c}}}{(q^2-M_i^2)^{n_i}(q^2-M_j^2)^{n_j}\cdots (q^2)^{n_L}}
\,\equiv\, g^{\mu_1\dots\mu_{2n_c}} \,\I[q^{2n_c}]_{ij\dots 0}^{n_i n_j\dots n_L} ,
\eeqn
with $g^{\mu_1\dots\mu_{2n_c}}$ being the completely symmetric tensor, e.g.\ $g^{\mu\nu\rho\sigma}=g^{\mu\nu}g^{\rho\sigma} +g^{\mu\rho}g^{\nu\sigma} +g^{\mu\sigma}g^{\nu\rho}$. Note that in dimensional regularization, $\I[q^{2n_c}]_0^{n_L}=0$ because they are scaleless integrals, which means a covariant diagram must contain at least one heavy propagator to be nonzero. We derive a general decomposition formula that allows us to easily evaluate arbitrary master integrals in Appendix~\ref{app:MI}, where we also give explicit expressions of the master integrals appearing in this paper.

By connecting vertex insertions with propagators and contracting Lorentz indices in all possible ways, we can derive all independent operators in $\L_\text{EFT}^\text{1-loop}$. In practice, however, one is often interested in obtaining just a few specific operators. To decide what covariant diagrams should be computed, we note that since all fields must come from $\bf U[\varphi]$ or $\bf Z[\varphi]$, we can simply look for elements of these matrices involving the same fields as contained in the operators of interest, and enumerate combinations of them that can be connected by propagators to form loops.

As a technical note, when enumerating covariant diagrams, we can omit those with operator structure $\tr(\dots P^\mu P_\mu \dots)=\tr(\dots P^2 \dots)$. This is because the remaining diagrams, giving rise to $\tr(\dots P^\mu\dots P_\mu\dots)$ with no ``adjacent $P_\mu$ contractions,'' already contain sufficient information for determining all the independent EFT operator coefficients.

\section{Matching the MSSM onto the SMEFT}
\label{sec:MSSMmatch}

\subsection{MSSM fields and interactions}
\label{sec:MSSMmatch-fi}

The techniques reviewed in Section~\ref{sec:CDrev} are generally applicable to matching any perturbative Lorentz-invariant UV theory onto a Lorentz-invariant EFT. We now focus on the specific case of matching the MSSM onto the SMEFT. To begin with, we need to extract the field content (including gauge quantum numbers of each field which determine the form of $P_\mu$) and the interaction matrix $\bf U[\varphi]$ of the MSSM. 

The complete MSSM field multiplet $(\varphi_H ,\varphi_L)^T$ is given in Tables~\ref{tab:HeavyFields} and~\ref{tab:LightFields}. We have explicitly written out the internal indices carried by each field, in various colors for clarity. In particular, we use $\ci{i}$, $\ci{A}$, $\wi{\alpha}$, $\wi{I}$, $\si{a}$ and $\si{\dot{a}}$ for $\ci{SU(3)_c}$ fundamental and adjoint, $\wi{SU(2)_L}$ fundamental and adjoint, and \si{spinor} indices on the conjugate fields on the left side of the quadratic operator $\Q_\text{UV}$, and $\ci{j}$, $\ci{B}$, $\wi{\beta}$, $\wi{J}$, $\si{b}$ and $\si{\dot{b}}$ for those on the fields on the right side.

\begin{table}[p]
\begin{center}
\begin{tabular}{|c|c|c|}
\hline
MSSM field & $SU(3)_c \times SU(2)_L \times U(1)_Y$ & Conjugate field \\
\hhline{|===|}
\multicolumn{3}{|c|}{Heavy spin-0 \,($\,c_s=\frac{1}{2}\,$)} \\
\hline
$\Phi_\wi{\beta}$ & $\qn{1}{2}{\frac{1}{2}}$ & $\Phi^{*\wi{\alpha}}$ \\
$\Phi^{*\wi{\beta}}$ & $\qn{1}{\bar{2}}{-\frac{1}{2}}$ & $\Phi_\wi{\alpha}$ \\
\hline
$\tilde q_{\ci{j}\wi{\beta}} = \wi{\bigl(}\,u_{L\ci{j}} \,,\, d_{L\ci{j}}\,\wi{\bigr)}$ & $\qn{3}{2}{\frac{1}{6}}$ & $\tilde q^{*\ci{i}\wi{\alpha}}$ \\
$\tilde q^{*\ci{j}\wi{\beta}} = \wi{\bigl(}\,u_L^{*\ci{j}} \,,\, d_L^{*\ci{j}}\,\wi{\bigr)}$ & $\qn{\bar{3}}{\bar{2}}{-\frac{1}{6}}$ & $\tilde q_{\ci{i}\wi{\alpha}}$ \\
\hline
$\tilde u_\ci{j} = \tilde u_{R\ci{j}}$ & $\qn{3}{1}{\frac{2}{3}}$ & $\tilde u^{*\ci{i}}$ \\
$\tilde u^{*\ci{j}} = \tilde u_R^{*\ci{j}}$ & $\qn{\bar 3}{1}{-\frac{2}{3}}$ & $\tilde u_\ci{i}$ \\ 
\hline
$\tilde d_\ci{j} = \tilde d_{R\ci{j}}$ & $\qn{3}{1}{-\frac{1}{3}}$ & $\tilde d^{*\ci{i}}$ \\
$\tilde d^{*\ci{j}} = \tilde d_R^{*\ci{j}}$ & $\qn{\bar 3}{1}{\frac{1}{3}}$ & $\tilde d_\ci{i}$ \\ 
\hline
$\tilde l_\wi{\beta} = \wi{\bigl(}\,\nu_L \,,\, e_L\,\wi{\bigr)}$ & $\qn{1}{2}{-\frac{1}{2}}$ & $\tilde l^{*\wi{\alpha}}$ \\
$\tilde l^{*\wi{\beta}} = \wi{\bigl(}\,\nu_L^* \,,\, e_L^*\,\wi{\bigr)}$ & $\qn{1}{\bar{2}}{\frac{1}{2}}$ & $\tilde l_\wi{\alpha}$ \\
\hline
$\tilde e = \tilde e_R$ & $\qn{1}{1}{-1}$ & $\tilde e^*$ \\
$\tilde e^* = \tilde e_R^*$ & $\qn{1}{1}{1}$ & $\tilde e$ \\ 
\hhline{|===|}
\multicolumn{3}{|c|}{Heavy spin-1/2 \,($\,c_s=-\frac{1}{2}\,$)} \\
\hline
$\tilde\chi_\wi{\beta}=\si{\Biggl(}\begin{matrix} \tilde\chi_{u\si{b}\wi{\beta}} \\ \epsilon_\wi{\beta\delta} \tilde\chi_d^{\dagger\si{\dot{b}}\wi{\delta}} \end{matrix}\si{\Biggr)}$ 
& 
$\qn{1}{2}{\frac{1}{2}}$ 
& 
$\overline{\tilde\chi}^\wi{\alpha}=\si{\bigl(}\,\epsilon^\wi{\alpha\gamma}\tilde\chi_{d\wi{\gamma}}^{\si{a}} \,,\, \chi_{u\si{\dot{a}}}^{\dagger\wi{\alpha}} \, \si{\bigr)}$ \\
$\tilde\chi^{\c\wi{\beta}}=\si{\Biggl(}\begin{matrix}\epsilon^\wi{\beta\delta} \tilde\chi_{d\si{b}\wi{\delta}} \\  \tilde\chi_u^{\dagger\si{\dot{b}}\wi{\beta}} \end{matrix}\si{\Biggr)}$ 
& 
$\qn{1}{\bar{2}}{-\frac{1}{2}}$ 
& 
$\overline{\tilde\chi}^\c_\wi{\alpha}=
\si{\bigl(}\,\tilde\chi_{u\wi{\alpha}}^{\si{a}} \,,\, \epsilon_\wi{\alpha\gamma}\chi_{d\si{\dot{a}}}^{\dagger\wi{\gamma}} \, \si{\bigr)}$ \\
\hline
$\tilde g^\ci{B} = \si{\Biggl(}\begin{matrix} \lambda_{g\si{b}}^{\ci{B}} \\ \lambda_g^{\dagger\si{\dot{b}}\ci{B}} \end{matrix}\si{\Biggr)}$ 
& 
$\qn{8}{1}{0}$ 
& 
$\overline{\tilde g}^\ci{A} = \si{\bigl(}\, \lambda_g^{\si{a}\ci{A}} \,,\, \lambda_{g\si{\dot{a}}}^{\dagger\ci{A}} \,\si{\bigr)}$ \\
\hline
$\tilde W^\wi{J} = \si{\Biggl(}\begin{matrix} \lambda_{W\si{b}}^{\wi{J}} \\ \lambda_W^{\dagger\si{\dot{b}}\wi{J}} \end{matrix}\si{\Biggr)}$ 
& 
$\qn{1}{3}{0}$
& 
$\overline{\tilde W}^\wi{I} = \si{\bigl(}\, \lambda_W^{\si{a}\wi{I}} \,,\, \lambda_{W\si{\dot{a}}}^{\dagger\wi{I}} \,\si{\bigr)}$ \\
\hline
$\tilde B = \si{\Biggl(}\begin{matrix} \lambda_{B\si{b}} \\ \lambda_B^{\dagger\si{\dot{b}}} \end{matrix}\si{\Biggr)}$ 
& 
$\qn{1}{1}{0}$
& 
$\overline{\tilde B} = \si{\bigl(}\, \lambda_B^{\si{a}} \,,\, \lambda_{B\si{\dot{a}}}^\dagger \,\si{\bigr)}$ \\
\hline
\end{tabular}
\caption{\label{tab:HeavyFields}
Heavy fields $\varphi_H$ in the MSSM, their gauge quantum numbers, and conjugate fields (which appear on the left side of $\Q_\text{UV}$, see footnote~\ref{conjfootnote}). For clarity, \ci{$SU(3)_c$}, \wi{$SU(2)_L$} and \si{spinor} indices are in \cicolor, \wicolor~and \sicolor, respectively.}
\end{center}
\end{table}

\begin{table}[p]
\begin{center}
\begin{tabular}{|c|c|c|}
\hline
MSSM field & $SU(3)_c \times SU(2)_L \times U(1)_Y$ & Conjugate field \\
\hhline{|===|}
\multicolumn{3}{|c|}{Light spin-0 \,($\,c_s=\frac{1}{2}\,$)} \\
\hline
$\phi_\wi{\beta}$ & $\qn{1}{2}{\frac{1}{2}}$ & $\phi^{*\wi{\alpha}}$ \\
$\phi^{*\wi{\beta}}$ & $\qn{1}{\bar{2}}{-\frac{1}{2}}$ & $\phi_\wi{\alpha}$ \\
\hhline{|===|}
\multicolumn{3}{|c|}{Light spin-1/2 \,($\,c_s=-\frac{1}{2}\,$)} \\
\hline
$\psi_{q\ci{j}\wi{\beta}} = \si{\Biggl(}\begin{matrix} q_{\si{b}\ci{j}\wi{\beta}} \\ q'^{\dagger\si{\dot{b}}}_{\ci{j}\wi{\beta}} \end{matrix}\si{\Biggr)}$ 
& 
$\qn{3}{2}{\frac{1}{6}}$ 
& 
$\overline{\psi}_q^{\ci{i}\wi{\alpha}} = \si{\bigl(}\, q'^{\si{a}\ci{i}\wi{\alpha}} \,,\, q^{\dagger\ci{i}\wi{\alpha}}_\si{\dot{a}} \,\si{\bigr)}$ \\
$\psi_q^{\c\ci{j}\wi{\beta}} = \si{\Biggl(}\begin{matrix} q'^{\ci{j}\wi{\beta}}_{\si{b}} \\ q^{\dagger\si{\dot{b}}\ci{j}\wi{\beta}} \end{matrix}\si{\Biggr)}$ 
& 
$\qn{\bar{3}}{\bar{2}}{-\frac{1}{6}}$ 
& 
$\overline{\psi}_{q\ci{i}\wi{\alpha}}^\c = \si{\bigl(}\, q^{\si{a}}_{\ci{i}\wi{\alpha}} \,,\, q'^\dagger_{\si{\dot{a}}\ci{i}\wi{\alpha}} \,\si{\bigr)}$ \\
\hline
$\psi_{u\ci{j}} = \si{\Biggl(}\begin{matrix} u'_{\si{b}\ci{j}} \\ u^{\dagger\si{\dot{b}}}_{\ci{j}} \end{matrix}\si{\Biggr)}$ 
& 
$\qn{3}{1}{\frac{2}{3}}$ 
& 
$\overline{\psi}_u^{\ci{i}} = \si{\bigl(}\, u^{\si{a}\ci{i}} \,,\, u'^{\dagger\ci{i}}_\si{\dot{a}} \,\si{\bigr)}$ \\
$\psi_u^{\c\ci{j}} = \si{\Biggl(}\begin{matrix} u^{\ci{j}}_{\si{b}} \\ u'^{\dagger\si{\dot{b}}\ci{j}} \end{matrix}\si{\Biggr)}$ 
& 
$\qn{\bar 3}{1}{-\frac{2}{3}}$ 
& 
$\overline{\psi}_{u\ci{i}}^\c = \si{\bigl(}\, u'^{\si{a}}_{\ci{i}} \,,\, u^\dagger_{\si{\dot{a}}\ci{i}} \,\si{\bigr)}$ \\
\hline
$\psi_{d\ci{j}} = \si{\Biggl(}\begin{matrix} d'_{\si{b}\ci{j}} \\ d^{\dagger\si{\dot{b}}}_{\ci{j}} \end{matrix}\si{\Biggr)}$ 
& 
$\qn{3}{1}{-\frac{1}{3}}$ 
& 
$\overline{\psi}_d^{\ci{i}} = \si{\bigl(}\, d^{\si{a}\ci{i}} \,,\, d'^{\dagger\ci{i}}_\si{\dot{a}} \,\si{\bigr)}$ \\
$\psi_d^{\c\ci{j}} = \si{\Biggl(}\begin{matrix} d^{\ci{j}}_{\si{b}} \\ d'^{\dagger\si{\dot{b}}\ci{j}} \end{matrix}\si{\Biggr)}$ 
& 
$\qn{\bar 3}{1}{\frac{1}{3}}$ 
& 
$\overline{\psi}_{d\ci{i}}^\c = \si{\bigl(}\, d'^{\si{a}}_{\ci{i}} \,,\, d^\dagger_{\si{\dot{a}}\ci{i}} \,\si{\bigr)}$ \\
\hline
$\psi_{l\wi{\beta}} = \si{\Biggl(}\begin{matrix} l_{\si{b}\wi{\beta}} \\ l'^{\dagger\si{\dot{b}}}_{\wi{\beta}} \end{matrix}\si{\Biggr)}$ 
& 
$\qn{1}{2}{-\frac{1}{2}}$ 
& 
$\overline{\psi}_l^{\wi{\alpha}} = \si{\bigl(}\, l'^{\si{a}\wi{\alpha}} \,,\, l^{\dagger\wi{\alpha}}_\si{\dot{a}} \,\si{\bigr)}$ \\
$\psi_l^{\c\wi{\beta}} = \si{\Biggl(}\begin{matrix} l'^{\wi{\beta}}_{\si{b}} \\ l^{\dagger\si{\dot{b}}\wi{\beta}} \end{matrix}\si{\Biggr)}$ 
& 
$\qn{1}{\bar{2}}{\frac{1}{2}}$ 
& 
$\overline{\psi}_{l\wi{\alpha}}^\c = \si{\bigl(}\, l^{\si{a}}_{\wi{\alpha}} \,,\, l'^\dagger_{\si{\dot{a}}\wi{\alpha}} \,\si{\bigr)}$ \\
\hline
$\psi_e = \si{\Biggl(}\begin{matrix} e'_\si{b} \\ e^{\dagger\si{\dot{b}}} \end{matrix}\si{\Biggr)}$ 
& 
$\qn{1}{1}{-1}$ 
& 
$\overline{\psi}_e = \si{\bigl(}\, e^\si{a} \,,\, e'^\dagger_\si{\dot{a}} \,\si{\bigr)}$ \\
$\psi_e^\c = \si{\Biggl(}\begin{matrix} e_{\si{b}} \\ e'^{\dagger\si{\dot{b}}} \end{matrix}\si{\Biggr)}$ 
& 
$\qn{1}{1}{1}$ 
& 
$\overline{\psi}_e^\c = \si{\bigl(}\, e'^{\si{a}} \,,\, e^\dagger_\si{\dot{a}} \,\si{\bigr)}$ \\
\hhline{|===|}
\multicolumn{3}{|c|}{Light spin-1 \,($\,c_s=\frac{1}{2}\,$)} \\
\hline
$G_\nu^{\ci{B}}$ & $\qn{8}{1}{0}$ & $G_\mu^{\ci{A}}$ \\
$W_\nu^{\wi{J}}$ & $\qn{1}{3}{0}$ & $W_\mu^{\wi{I}}$ \\
$B_\nu$ & $\qn{1}{1}{0}$ & $B_\mu$ \\
\hline
\end{tabular}
\caption{\label{tab:LightFields}
Light fields $\varphi_L$ in the MSSM, their gauge quantum numbers, and conjugate fields (which appear on the left side of $\Q_\text{UV}$, see footnote~\ref{conjfootnote}). For clarity, \ci{$SU(3)_c$}, \wi{$SU(2)_L$} and \si{spinor} indices are in \cicolor, \wicolor~and \sicolor, respectively. Primed Weyl fermion fields are unphysical auxiliary fields, to be set to zero at the end of the calculation.
}
\end{center}
\end{table}

\begin{sidewaystable}
\begin{center}
\begin{tabular}{|c||cccccc|cccc||c|ccccc|ccc|}
\hline
 & $\Phi$ & $\tilde q$ & $\tilde u$ & $\tilde d$ & $\tilde l$ & $\tilde e$ & $\tilde\chi$ & $\tilde g$ & $\tilde W$ & $\tilde B$ & $\phi$ & $q$ & $u$ & $d$ & $l$ & $e$ & $G$ & $W$ & $B$ \\
\hhline{|=#===================|}
$\Phi$ & $\varphi^2$ & & & & & & & & & & $\varphi^2$ & $u,d$ & $q$ & $q$ & $e$ & $l$ & & $D\Phi$ & $D\Phi$ \\
$\tilde q$ & & $\varphi^2$ & $\varphi$ & $\varphi$ & & & $u,d$ & $q$ & $q$ & $q$ & & & & & & & & & \\
$\tilde u$ & & $\varphi$ & $\varphi^2$ & $\Phi\phi$ & & & $q$ & $u$ & & $u$ & & & & & & & & & \\
$\tilde d$ & & $\varphi$ & $\Phi\phi$ & $\varphi^2$ & & & $q$ & $d$ & & $d$ & & & & & & & & & \\
$\tilde l$ & & & & & $\varphi^2$ & $\varphi$ & $e$ & & $l$ & $l$ & & & & & & & & & \\
$\tilde e$ & & & & & $\varphi$ & $\varphi^2$ & $l$ & & & $e$ & & & & & & & & & \\
\hhline{|-||-------------------|}
$\tilde\chi$ & & $u,d$ & $q$ & $q$ & $e$ & $l$ & & & $\varphi$ & $\varphi$ & & & & & & & & & \\
$\tilde g$ & & $q$ & $u$ & $d$ & & & & & & & & & & & & & & & \\
$\tilde W$ & & $q$ & & & $l$ & & $\varphi$ & & & & & & & & & & & & \\
$\tilde B$ & & $q$ & $u$ & $d$ & $l$ & $e$ & $\varphi$ & & & & & & & & & & & & \\
\hhline{|=||======|====#=|=====|===|}
$\phi$ & $\varphi^2$ & & & & & & & & & & $\varphi^2$ & $u,d$ & $q$ & $q$ & $e$ & $l$ & & $D\phi$ & $D\phi$ \\
\hhline{|-||-------------------|}
$q$ & $u,d$ & & & & & & & & & & $u,d$ & & $\varphi$ & $\varphi$ & & & $q$ & $q$ & $q$ \\
$u$ & $q$ & & & & & & & & & & $q$ & $\varphi$ & & & & & $u$ & & $u$ \\
$d$ & $q$ & & & & & & & & & & $q$ & $\varphi$ & & & & & $d$ & & $d$ \\
$l$ & $e$ &  & & & & & & & & & $e$ & & & & & $\varphi$ & & $l$ & $l$ \\
$e$ & $l$ & & & & & & & & & & $l$ & & & & $\varphi$ & & & & $e$ \\
\hhline{|-||-------------------|}
$G$ & & & & & & & & & & & & $q$ & $u$ & $d$ & & & $G_{\mu\nu}$ & & \\
$W$ & $D\Phi$ & & & & & & & & & & $D\phi$ & $q$ & & & $l$ & & & $W_{\mu\nu},\phi^2,\Phi^2$ & \\
$B$ & $D\Phi$ & & & & & & & & & & $D\phi$ & $q$ & $u$ & $d$ & $l$ & $e$ & & & $\phi^2,\Phi^2$ \\
\hline
\end{tabular}
\caption{\label{tab:Umatrix}
Fields contained in the nonzero entries of the MSSM $\bf U$ matrix, with the heavy fields $\varphi_H$ to be set to $\varphi_{H,c}$ (which is nonvanishing only for the heavy Higgs $\Phi$). In this table, $\varphi$ collectively denotes the heavy and light Higgs fields $\Phi$ and $\phi$, e.g.\ $\varphi^2$ represents $\Phi^2,\Phi\phi,\phi^2$. Also, $\psi_f$, the Dirac spinors containing SM fermions, are written simply as $f$ for clarity. Detailed expressions of the $\bf U$ matrix can be found in Appendix~\ref{app:Umatrix}.}
\end{center}
\end{sidewaystable}
\afterpage{\clearpage}

The scalar sector of the MSSM consists of sfermions and two Higgs doublets. For the latter, we choose a basis $(\Phi,\phi)$ where the mass matrix in the electroweak-symmetric phase is diagonal,
\beqa
\L_\text{MSSM} &\supset& -(\mu^2+\mHusq)|H_u|^2 -(\mu^2+\mHdsq)|H_d|^2 -b\,(H_u\cdot\epsilon\cdot H_d+\text{h.c.}) \CR
&=& -m^2|\phi|^2 -M_\Phi^2|\Phi|^2 \,.
\eeqan
The $(\Phi,\phi)$ and $(H_u,H_d)$ bases are related by
\beq
\Phi = \cbp \,H_u +\sbp \, \epsilon\cdot H_d^*\,,\quad \phi = \sbp \,H_u -\cbp \, \epsilon\cdot H_d^* \,.
\eeq{Phiphidef}
where we have abbreviated $\sin\beta'\equiv\sbp$, $\cos\beta'\equiv\cbp$, with $\beta'$ defined by
\beq
\tan 2\beta' = \frac{2b}{\mHusq-\mHdsq} \,.
\eeq{betaprimedef}
Note that $\beta'$ is different from what is usually referred to as $\beta$ as in $\tan\beta=\frac{v_u}{v_d}$, the ratio of vacuum expectation values (vevs) of $H_u$ and $H_d$. In fact, at one-loop level, minimizing the effective potential, we can see that the two are related by\footnote{Here we are defining $\beta$ in a tadpole-free scheme, where $v_u,v_d$ denote the location of the minimum of the loop-corrected effective potential (and are gauge dependent). An alternative scheme that is also commonly used defines $v_u,v_d$ by the location of the minimum of the tree-level Higgs potential, independent of gauge choice; in that scheme, $\tan\beta$ differs from $\tan\beta'$ only by $\O(\Lambda^{-2})$ terms.}
\beq
\beta = \beta'+\frac{1}{M_\Phi^2}\sb\cb\Bigl(\frac{t_u}{v_u}-\frac{t_d}{v_d}\Bigr)_{\O\bigl(\frac{\Lambda^2}{16\pi^2}\bigr)} +\O(\Lambda^{-2}) \,,
\eeq{betadiff}
in the decoupling limit $|m^2|\ll M_{\Phi,\ft,\chit,\Vt}^2 \sim \O(\Lambda^2)$ that we are interested in. Here $t_u, t_d$ are one-loop tadpoles, whose analytical expressions can be found in e.g.~\cite{BMPZ}. 

As for fermions, we choose to work with four-component spinor fields. In particular, we write the Higgsinos as a Dirac spinor, and the gauginos as Majorana spinors. The SM chiral fermions $f$ are embedded into Dirac spinors $\psi_f$, in which we also retain unphysical wrong-chirality Weyl fermions $f'$, and set them to zero only at the end of the calculation\footnote{Generally, such embedding would require additionally writing projection operators in interaction terms to pick up the physical fermion fields. However, this is not necessary in the special case of $R$-parity-conserving MSSM considered here.}. 

Interactions among the MSSM fields in Tables~\ref{tab:HeavyFields} and~\ref{tab:LightFields} are encoded in the covariant derivative $P_\mu$ and interaction matrices $\bf U[\varphi]$, $\bf Z[\varphi]$ in our functional matching formalism. They are extracted from the terms in the MSSM Lagrangian that are quadratic in quantum fluctuation fields, following Eqs.~\eqref{bkgfld}, \eqref{LUVexpand}, \eqref{QUV} and~\eqref{Xmat}. It turns out that the $\bf Z$ matrix does not contribute to the operators computed in this paper (up to $\Lambda^{-1}$ suppressed corrections), and so will not be considered further.

To write down the $\bf U$ matrix, we follow the conventions in~\cite{MartinPrimer} (Ref.~\cite{BMPZ} uses an opposite sign for the tree-level Higgsino mass parameter $\mu$), assuming $R$-parity is conserved. We assume a trivial flavor structure for the soft SUSY breaking parameters for simplicity,
\beqa
\L_\text{MSSM} &\supset& -M_{\tilde q}^2\, \tilde q^* \,\identity\, \tilde q -M_{\tilde u}^2\, \tilde u^* \,\identity\, \tilde u -M_{\tilde d}^2\, \tilde d^* \,\identity\, \tilde d -M_{\tilde l}^2\, \tilde l^* \,\identity\, \tilde l -M_{\tilde e}^2\, \tilde e^* \,\identity\, \tilde e \CR
&& -A_u \,\tilde u^* \,\mat{\lambda_u}\, \tilde q \cdot \epsilon \cdot H_u +A_d \,\tilde d^* \,\mat{\lambda_d}\, \tilde q \cdot \epsilon \cdot H_d +A_e \,\tilde e^* \,\mat{\lambda_e}\, \tilde l \cdot \epsilon \cdot H_d \,,
\eeqa{soft}
where $\mat{\lambda_u},\mat{\lambda_d},\mat{\lambda_e}$ are Yukawa matrices in the MSSM. Our results can be easily extended to include flavor mixing, at the cost of making the analytical expressions more complicated. Furthermore, we assume $\mu v, A_f v \ll M_{\ft}^2$, so that matching in the electroweak-symmetric phase without sfermion mass mixing is justified.

We summarize the fields contained in each nonzero entry of the MSSM $\bf U$ matrix in Table~\ref{tab:Umatrix}, relegating detailed expressions to Appendix~\ref{app:Umatrix}. This $\bf U$ matrix exhibits a block-diagonal structure because of the assumed $R$-parity: if $i$ and $j$ have opposite $R$-parity, $U_{ij}$ would be proportional to a heavy $R$-parity-odd field, which should be set to $\varphi_{H,\c}=0$ (so that $\frac{\delta\L_\text{MSSM}}{\delta\varphi_H}\bigr|_{\varphi_H=\varphi_{H,\c}}\propto\varphi_{H,\c}=0$). We will demonstrate in Sections~\ref{sec:MSSMmatch-loop} and~\ref{sec:MSSMmatch-dsix} how to use Table~\ref{tab:Umatrix} to quickly pick out the $U$ insertions containing the right fields to make up a desired operator in our one-loop matching calculation.

\subsection{Tree-level matching}
\label{sec:MSSMmatch-tree}

The tree-level effective Lagrangian is obtained by solving the equations of motion of the heavy fields; see Eq.~\eqref{LEFTtree}. As mentioned in the previous subsection, in the absence of $R$-parity violation, $\frac{\delta\L_\text{MSSM}}{\delta\varphi_H}=0$ is trivially solved by $\varphi_{H,\c}=0$ for all the heavy fields in the MSSM except the $R$-parity-even heavy Higgs doublet $\Phi$, for which
\beqa
\frac{\delta\L_\text{MSSM}}{\delta\Phi^{*\wi{\alpha}}} &=& \bigl[ (P^2)\wi{_\alpha^\beta} -M_\Phi^2\,\delta\wi{_\alpha^\beta} \bigr] \Phi_\wi{\alpha} \CR
&& +\frac{1}{8}(g^2+g'^2) s_{4\beta'} |\phi|^2\phi_\wi{\beta} +\cbp\,\epsilon_\wi{\alpha\beta} \bar\psi_q^{\wi{\beta}} \,\mat{\lambda_u^\dagger}\, \psi_u 
+\sbp\, \bar\psi_d \,\mat{\lambda_d}\, \psi_{q\wi{\alpha}} 
+\sbp\, \bar\psi_e \,\mat{\lambda_e}\, \psi_{l\wi{\alpha}} \CR
&& +\Bigl[\frac{1}{4}(g^2+g'^2)c_{2\beta'}^2-\frac{1}{2}g^2\Bigr]|\phi|^2\Phi_{\wi{\alpha}}
 -\Bigl[\frac{1}{4}(g^2+g'^2)s_{2\beta'}^2-\frac{1}{2}g^2\Bigr](\phi^*\Phi)\,\phi_{\wi{\alpha}} \CR
&& -\frac{1}{4}(g^2+g'^2)s_{2\beta'}^2(\Phi^*\phi)\,\phi_{\wi{\alpha}} +\O(\Phi^2\phi, \Phi^3) \,. 
\eeqa{PhiEoM}
in the $\overline{\text{DR}}$ scheme. In the $\overline{\text{MS}}$ scheme, on the other hand, the scalar quartic couplings (and hence the scalar cubic terms in Eq.~\eqref{PhiEoM}) receive $\O(\frac{g^4}{16\pi^2},\frac{g^2g'^2}{16\pi^2},\frac{g'^4}{16\pi^2})$ corrections~\cite{MartinVaughnDRbar}.

We can solve the equation of motion $\frac{\delta\L_\text{MSSM}}{\delta\Phi^*}=0$ for $\Phi_\c$ perturbatively, as a power series in $M_\Phi^{-1}$,
\beq
\Phi_\c = \Phi_\c^{(1)} +\Phi_\c^{(2)} +\dots \qquad\quad \text{where}\quad \Phi_\c^{(n)} \sim \O(M_\Phi^{-2n})\,.
\eeqn
The first and second order solutions read
\beqa
\Phi_{\c\wi{\alpha}}^{(1)} &=& \frac{1}{M_\Phi^2} \biggl[\frac{1}{8}(g^2+g'^2) s_{4\beta'} |\phi|^2 \phi_\wi{\alpha} +\cbp\,\epsilon_\wi{\alpha\beta} \bar\psi_q^{\wi{\beta}} \,\mat{\lambda_u^\dagger}\, \psi_u 
+\sbp\, \bar\psi_d \,\mat{\lambda_d}\, \psi_{q\wi{\alpha}} 
+\sbp\, \bar\psi_e \,\mat{\lambda_e}\, \psi_{l\wi{\alpha}} 
\biggr] \,,\CR\\
\Phi_{\c\wi{\alpha}}^{(2)} &=& \frac{1}{M_\Phi^2} \Bigl\{ -\bigl(D^2\Phi_\c^{(1)}\bigr)_\wi{\alpha} +\Bigl[\frac{1}{4}(g^2+g'^2)c_{2\beta'}^2-\frac{1}{2}g^2\Bigr]|\phi|^2\Phi_{\c\wi{\alpha}}^{(1)} \CR
&&\qquad\;\; -\Bigl[\frac{1}{4}(g^2+g'^2)s_{2\beta'}^2-\frac{1}{2}g^2\Bigr](\phi^*\Phi_\c^{(1)})\,\phi_{\wi{\alpha}} -\frac{1}{4}(g^2+g'^2)s_{2\beta'}^2(\Phi_\c^{(1)*}\phi)\,\phi_{\wi{\alpha}} \Bigr\} \,.
\eeqan
Only $\Phi_\c^{(1)}$ is needed in tree-level matching up to dimension six. We have
\beqa
\L_\text{SMEFT}^\text{tree} &=& \left.\L_\text{MSSM} \right|_{\varphi_H\to\varphi_{H,\c}} = \L_\text{SM} +M_\Phi^2 \bigl|\Phi_\c^{(1)}\bigr|^2 +\O(\Lambda^{-4}) \CR
&=& \L_\text{SM} +\sum_i C_i^\text{tree} \O_i^{(d=6)} +\O(\Lambda^{-4}) \,,
\eeqa{LSMEFTtree}
where the dimension-six operators $\O_i^{(d=6)}$ generated and their coefficients $C_i^\text{tree}$ are listed in Table~\ref{tab:tlm}. We have used the basis of~\cite{WarsawBasis}, known as the Warsaw basis, for dimension-six operators. Fierz identities have been used to transform some of the four-fermion operators into this basis. Note that with the tree-level matching of Eq.~\eqref{LSMEFTtree} alone, each appearance of $\beta$ in Table~\ref{tab:tlm} should really read $\beta'$. However, as we will see shortly, part of one-loop matching result can be absorbed into a redefinition of $\beta'\to\beta$ in the tree-level operator coefficients.

\begin{table}[t]
\begin{center}
\begin{tabular}{|c|c|}
\hline
Coefficient & Operator \\
\hhline{|==|}
$C_\phi^\text{tree} = \frac{1}{64M_\Phi^2}s_{4\beta}^2 \, (g^2+g'^2)^2$ & $\O_\phi = |\phi|^6$ \\
\hline
$\bigl[\mat{C_{u\phi}^\text{tree}}\bigr]_{pr} = -\frac{1}{8M_\Phi^2}s_{4\beta}\cb\,(g^2+g'^2) \bigl[\mat{\lambda_u^\dagger}\bigr]_{pr}$ & $\bigl[\O_{u\phi}\bigr]^{pr} = |\phi|^2 \,(\bar\psi_q^{\;p}\psi_u^{\;r})\cdot\epsilon\phi^*$ \\
$\bigl[\mat{C_{d\phi}^\text{tree}}\bigr]_{pr} = \frac{1}{8M_\Phi^2}s_{4\beta}\sb\,(g^2+g'^2) \bigl[\mat{\lambda_d^\dagger}\bigr]_{pr}$ & $\bigl[\O_{d\phi}\bigr]^{pr} = |\phi|^2 \,(\bar\psi_q^{\;p}\psi_d^{\;r})\cdot\phi$ \\
$\bigl[\mat{C_{e\phi}^\text{tree}}\bigr]_{pr} = \frac{1}{8M_\Phi^2}s_{4\beta}\sb\,(g^2+g'^2) \bigl[\mat{\lambda_e^\dagger}\bigr]_{pr}$ & $\bigl[\O_{e\phi}\bigr]^{pr} = |\phi|^2 \,(\bar\psi_l^{\;p}\psi_e^{\;r})\cdot\phi$ \\
\hline
$\bigl[\mat{C_{qu}^{(1)\text{tree}}}\bigr]_{prst} = -\frac{1}{6M_\Phi^2}\cb^2\,\bigl[\mat{\lambda_u^\dagger}\bigr]_{pt}\bigl[\mat{\lambda_u}\bigr]_{sr}$ & $\bigl[\O_{qu}^{(1)}\bigr]^{prst} = (\bar\psi_q^{\;p}\gamma_\mu\psi_q^{\;r}) (\bar\psi_u^{\;s}\gamma^\mu\psi_u^{\;t})$ \\
$\bigl[\mat{C_{qu}^{(8)\text{tree}}}\bigr]_{prst} = -\frac{1}{M_\Phi^2}\cb^2\,\bigl[\mat{\lambda_u^\dagger}\bigr]_{pt}\bigl[\mat{\lambda_u}\bigr]_{sr}$ & $\bigl[\O_{qu}^{(8)}\bigr]^{prst} = (\bar\psi_q^{\;p}\gamma_\mu T^A\psi_q^{\;r}) (\bar\psi_u^{\;s}\gamma^\mu T^A\psi_u^{\;t})$ \\
$\bigl[\mat{C_{qd}^{(1)\text{tree}}}\bigr]_{prst} = -\frac{1}{6M_\Phi^2}\sb^2\,\bigl[\mat{\lambda_d^\dagger}\bigr]_{pt}\bigl[\mat{\lambda_d}\bigr]_{sr}$ & $\bigl[\O_{qd}^{(1)}\bigr]^{prst} = (\bar\psi_q^{\;p}\gamma_\mu\psi_q^{\;r}) (\bar\psi_d^{\;s}\gamma^\mu\psi_d^{\;t})$ \\
$\bigl[\mat{C_{qd}^{(8)\text{tree}}}\bigr]_{prst} = -\frac{1}{M_\Phi^2}\sb^2\,\bigl[\mat{\lambda_d^\dagger}\bigr]_{pt}\bigl[\mat{\lambda_d}\bigr]_{sr}$ & $\bigl[\O_{qd}^{(8)}\bigr]^{prst} = (\bar\psi_q^{\;p}\gamma_\mu T^A\psi_q^{\;r}) (\bar\psi_d^{\;s}\gamma^\mu T^A\psi_d^{\;t})$ \\
$\bigl[\mat{C_{le}^{\text{tree}}}\bigr]_{prst} = -\frac{1}{2M_\Phi^2}\sb^2\,\bigl[\mat{\lambda_e^\dagger}\bigr]_{pt}\bigl[\mat{\lambda_e}\bigr]_{sr}$ & $\bigl[\O_{le}\bigr]^{prst} = (\bar\psi_l^{\;p}\gamma_\mu\psi_l^{\;r}) (\bar\psi_e^{\;s}\gamma^\mu\psi_e^{\;t})$ \\
$\bigl[\mat{C_{quqd}^{(1)\text{tree}}}\bigr]_{prst} = -\frac{1}{M_\Phi^2}\sb\cb\,\bigl[\mat{\lambda_u^\dagger}\bigr]_{pr}\bigl[\mat{\lambda_d^\dagger}\bigr]_{st}$ & $\bigl[\O_{quqd}^{(1)}\bigr]^{prst} = (\bar\psi_q^{\;p}\psi_u^{\;r})\cdot\epsilon\cdot(\bar\psi_q^{\;s}\psi_d^{\;t})$ \\
$\bigl[\mat{C_{lequ}^{(1)\text{tree}}}\bigr]_{prst} = \frac{1}{M_\Phi^2}\sb\cb\,\bigl[\mat{\lambda_e^\dagger}\bigr]_{pr}\bigl[\mat{\lambda_u^\dagger}\bigr]_{st}$ & $\bigl[\O_{lequ}^{(1)}\bigr]^{prst} = (\bar\psi_l^{\;p}\psi_e^{\;r})\cdot\epsilon\cdot(\bar\psi_q^{\;s}\psi_u^{\;t})$ \\
$\bigl[\mat{C_{ledq}^{\text{tree}}}\bigr]_{prst} = \frac{1}{M_\Phi^2}\sb^2\,\bigl[\mat{\lambda_e^\dagger}\bigr]_{pr}\bigl[\mat{\lambda_d}\bigr]_{st}$ & $\bigl[\O_{ledq}\bigr]^{prst} = (\bar\psi_l^{\;p}\psi_e^{\;r})(\bar\psi_d^{\;s}\psi_q^{\;t})$ \\
\hline
\end{tabular}
\caption{\label{tab:tlm}
Dimension-six operators generated at tree level when matching the MSSM onto the SMEFT. $p,r,s,t$ are generation indices. Tree-level matching alone produces the operator coefficients listed here, but with $\beta'$ in place of $\beta$. As explained in Section~\ref{sec:MSSMmatch-tree-beta}, adding the one-loop-generated piece $c_{\Phi\phi}(\Phi_\c^*\phi+\phi^*\Phi_\c)$ to $\L_\text{SMEFT}^\text{tree}$ amounts to replacing $\beta'$ by $\beta$ in all $C_i^\text{tree}$.}
\end{center}
\end{table}

\subsubsection{$\beta$ redefinition}
\label{sec:MSSMmatch-tree-beta}

An interesting observation can be made on the tree-level effective Lagrangian computed above. Differentiating $\L_\text{SMEFT}^\text{tree}$ with respect to $\beta'$, we find
\beqa
\frac{\partial}{\partial\beta'}\L_\text{SMEFT}^{\text{tree}\,(d=4)} &=& \frac{\partial}{\partial\beta'}\biggl[ -\frac{1}{8}\bigl(g^2+g'^2\bigr)c_{2\beta'}^2|\phi|^4 \CR
&&\qquad\; -\bigl(\sbp\bar\psi_u\,\mat{\lambda_u}\,\psi_q\cdot\epsilon\cdot\phi +\cbp\bar\psi_d\,\mat{\lambda_d}\,\psi_q\cdot\phi^* +\cbp\bar\psi_e\,\mat{\lambda_e}\,\psi_l\cdot\phi^* +\text{h.c.}\bigr)\biggr] \CR
&=& \frac{1}{4}\bigl(g^2+g'^2\bigr) s_{4\beta'} |\phi|^4 \CR
&&\;\; + \Bigl[\phi^*\bigl(\cbp\epsilon\cdot\bar\psi_q\,\mat{\lambda_u^\dagger}\,\psi_u +\sbp\bar\psi_d\,\mat{\lambda_d}\,\psi_q +\sbp\bar\psi_e\,\mat{\lambda_e}\,\psi_l \bigr) +\text{h.c.}\Bigr] \CR
&=& M_\Phi^2 \bigl(\Phi_\c^{(1)*}\phi+\phi^*\Phi_\c^{(1)}\bigr) \,,\\
\frac{\partial}{\partial\beta'}\L_\text{SMEFT}^{\text{tree}\,(d=6)} &=& M_\Phi^2 \biggl(\Phi_\c^{(1)*}\cdot\frac{\partial\Phi_\c^{(1)}}{\partial\beta'} +\text{h.c.}\biggr) \CR
&=& \Phi_\c^{(1)*}\cdot \biggl[\frac{1}{2}\bigl(g^2+g'^2\bigr) c_{4\beta'} |\phi|^2\phi \CR
&&\qquad\quad\;\; -\sbp\,\epsilon\cdot\bar\psi_q \,\mat{\lambda_u^\dagger}\, \psi_u 
+\cbp\, \bar\psi_d \,\mat{\lambda_d}\, \psi_q
+\cbp\, \bar\psi_e \,\mat{\lambda_e}\, \psi_l \biggr] +\text{h.c.} \CR
&\overset{\text{EoM}}{=}& \Phi_\c^{(1)*}\cdot \biggl[-(D^2\phi) +\frac{1}{4}\bigl(g^2+g'^2\bigr) \bigl(c_{2\beta'}^2-2s_{2\beta'}^2\bigr) |\phi|^2\phi \biggr] +\text{h.c.} \CR
&\overset{\text{IBP}}{=}& M_\Phi^2 \bigl(\Phi_\c^{(2)*}\phi+\phi^*\Phi_\c^{(2)}\bigr) \,.
\eeqan
In the equations above, we have used the fact that both $m^2=\mu^2+\mHusq\sbp^2+\mHdsq\cbp^2-bs_{2\beta'}$ and $M_\Phi^2=\mu^2+\mHusq\cbp^2+\mHdsq\sbp^2+bs_{2\beta'}$ have vanishing first derivative with respect to $\beta'$, when Eq.~\eqref{betaprimedef} is satisfied. ``$\overset{\text{EoM}}{=}$'' and ``$\overset{\text{IBP}}{=}$'' mean equivalence with the use of the renormalizable SM equations of motion and integration by parts, respectively --- operations that are allowed when we are dealing with dimension-six operators in the SMEFT Lagrangian. We have neglected the $m^2$ piece and one-loop threshold corrections to the relation $\lambda=\frac{1}{8}(g^2+g'^2)c_{2\beta'}^2$ when applying the equation of motion for $\phi$, because they lead to $\frac{m^2}{\Lambda^2}$ suppressed and loop-suppressed terms compared to those retained in our results.

Meanwhile, as we will see explicitly in the next subsection, matching the MSSM onto the SMEFT at one-loop level generates
\beq
\L_\text{SMEFT}^\text{1-loop} \supset c_{\Phi\phi}(\Phi_\c^*\phi+\phi^*\Phi_\c) \,,
\eeq{LPhiphi}
with $c_{\Phi\phi}\sim\O(\frac{\Lambda^2}{16\pi^2})$ given in Eq.~\eqref{cPhiphi}. The observation we made above, namely
\beq
\frac{\partial}{\partial\beta'}\L_\text{SMEFT}^{\text{tree}} = M_\Phi^2 \bigl(\Phi_\c^*\phi+\phi^*\Phi_\c\bigr)
\eeqn
suggests that we can absorb the part of $\L_\text{SMEFT}^\text{1-loop}$ shown in Eq.~\eqref{LPhiphi} into $\L_\text{SMEFT}^\text{tree}$ via a redefinition of $\beta'$,
\beq
\L_\text{SMEFT}^\text{tree}(\beta') +c_{\Phi\phi}(\Phi_\c^*\phi+\phi^*\Phi_\c) = \L_\text{SMEFT}^\text{tree}\Bigl(\beta'+\frac{c_{\Phi\phi}}{M_\Phi^2}\Bigr) \,,
\eeqn
up to two-loop corrections. Comparing $c_{\Phi\phi}$ presented below in Eq.~\eqref{cPhiphi} and analytical expressions of one-loop tadpoles in~\cite{BMPZ}, we can actually show that
\beq
c_{\Phi\phi} = \sb\cb\Bigl(\frac{t_u}{v_u}-\frac{t_d}{v_d}\Bigr)_{\O\bigl(\frac{\Lambda^2}{16\pi^2}\bigr)}\,.
\eeqn
Therefore,
\beq
\L_\text{SMEFT}^\text{tree}(\beta') +c_{\Phi\phi}(\Phi_\c^*\phi+\phi^*\Phi_\c) = \L_\text{SMEFT}^\text{tree}(\beta) \,,
\eeq{betaredef}
with $\beta$ defined by the minimum of the (loop-corrected) 1PI effective potential, i.e.\ $\tan\beta=\frac{v_u}{v_d}$ in the tadpole-free scheme; see Eq.~\eqref{betadiff}. We see that adding the one-loop-generated piece $c_{\Phi\phi}(\Phi_\c^*\phi+\phi^*\Phi_\c)$ to $\L_\text{SMEFT}^\text{tree}$ amounts to simply replacing $\beta'$ by $\beta$ in all tree-level operator coefficients.

There is a simple power-counting argument for the relation Eq.~\eqref{betaredef}. If instead of Eq.~\eqref{Phiphidef}, we define $\Phi,\phi$ to be related to $H_u,H_d$ by an angle $\beta$ (as opposed to $\beta'$) rotation, we would have $\langle\phi\rangle=\frac{v}{\sqrt{2}}\simeq174\,\text{GeV}$, while $\langle\Phi\rangle=0$. In this basis (usually referred to as the Higgs basis), integrating out heavy superpartners must not produce $(\Phi_\c^*\phi+\phi^*\Phi_\c)$ with $\O(\frac{\Lambda^2}{16\pi^2})$ coefficient, because otherwise, the same contribution would be present if we compute the 1PI effective potential of the MSSM --- this would lead to an $\O(\frac{v}{16\pi^2})$ contribution to $\langle\Phi\rangle$ which, in fact, is the only possible contribution at this order, thus contradicting $\langle\Phi\rangle=0$. Technically, what happens is a cancellation of $\O(\frac{\Lambda^2}{16\pi^2})\cdot(\Phi_\c^*\phi+\phi^*\Phi_\c)$ pieces between $\L_\text{SMEFT}^\text{tree}$ and $\L_\text{SMEFT}^\text{1-loop}$: the same $\L_\text{SMEFT}^\text{1-loop}\supset c_{\Phi\phi}(\Phi_\c^*\phi+\phi^*\Phi_\c)$ is generated by one-loop matching, while $\L_\text{SMEFT}^\text{tree}\supset -c_{\Phi\phi}(\Phi_\c^*\phi+\phi^*\Phi_\c)$ because the UV theory Lagrangian now contains a mass mixing term,
\beq
\L_\text{MSSM} \supset \biggl[ b\,\cbb -\frac{1}{2}(\mHusq-\mHdsq)\sbb \biggr] (\Phi^*\phi+\phi^*\Phi)
= -\sb\cb\Bigl(\frac{t_u}{v_u}-\frac{t_d}{v_d}\Bigr)_{\O\bigl(\frac{\Lambda^2}{16\pi^2}\bigr)} (\Phi^*\phi+\phi^*\Phi) \,,
\eeqn
up to $\frac{m^2}{\Lambda^2}$ suppressed and higher-loop corrections. Note that the presence of mass mixing in this basis does not invalidate our functional matching formalism (which assumes a diagonal mass matrix), if we treat it as a small constant term in the $\bf U$ matrix. However, the Higgs basis is not a convenient choice for tree-level matching, because $\Phi_\c$ has to be solved as a double series in $\Lambda^{-1}$ and $\lfr$.

\subsection{One-loop matching: $d\le4$ operators and SUSY threshold corrections}
\label{sec:MSSMmatch-loop}

\subsubsection{Enumerating covariant diagrams}

To match the MSSM onto the SMEFT at one-loop level, we draw covariant diagrams contributing to each SMEFT operator of interest, starting from the $d\le4$ ones which encode SUSY threshold corrections. Enumerating covariant diagrams is straightforward by looking for desired fields from the MSSM $\bf U$ matrix.

Let us demonstrate the procedure with an example operator $\bar\psi_d \,\mat{\delta y_d}\, \psi_q \cdot \phi^*+\text{h.c}$. Obviously we should look for a $d$, a $q$ and a $\phi$ in Table~\ref{tab:Umatrix}. To begin with, there are several options to get a $d$, such as from $U_{\qt\chit}$, or from $U_{\dt\gt}$. Let us pick $U_{\qt\chit}$ first. This $U_{\qt\chit}$ insertion should be followed by a $\chit$ propagator, and then another $U$ insertion containing either $q$ or $\phi$. For this second $U$ insertion, we need to enumerate all viable choices, one of them being $U_{\chit\ut}\sim q$. With this particular choice, we can then close the loop with a $\ut$ propagator, followed by a $U_{\ut\qt}\sim\phi$ insertion, and then a $\qt$ propagator connecting back to our starting point $U_{\qt\chit}$. We thus end up with the following covariant diagram,
\beq
\begin{gathered}
\begin{fmffile}{cd-d4-sfho}
\begin{fmfgraph*}(40,40)
\fmfsurround{vUqu,vUgq,vUug}
\fmf{dashes,left=0.6,label=$\ut$,l.d=3pt}{vUgq,vUqu}
\fmf{dashes,left=0.6,label=$\qt$,l.d=3pt}{vUqu,vUug}
\fmf{plain,left=0.6,label=$\chit$,l.d=3pt}{vUug,vUgq}
\fmfv{decor.shape=circle,decor.filled=empty,decor.size=3thick}{vUqu,vUgq,vUug}
\end{fmfgraph*}
\end{fmffile}
\end{gathered}
\quad = -\frac{i}{2}\mu \,\I_{\qt\ut\chit}^{111} \,\tr\bigl(U_{\qt\chit} U_{\chit\ut} U_{\ut\qt} \bigr)
\eeq{d4-sfho}
Plugging in explicit expressions of $U_{\qt\chit}$, $U_{\chit\ut}$, $U_{\ut\qt}$ from Eqs.~\eqref{Uqtchit}, \eqref{Uutchit}, \eqref{Uutqt}, we obtain
\beqa
\tr\bigl(U_{\qt\chit} U_{\chit\ut} U_{\ut\qt} \bigr) &\supset& \sb\,(A_u-\mu\cot\beta)\bigl(
\bar\psi_d \,\mat{\lambda_d \lambda_u^\dagger \lambda_u}\,\psi_q \cdot\phi^*
+\bar\psi_d^c \,\mat{\lambda_d^* \lambda_u^T \lambda_u^*} \,\psi_q^c \cdot \phi
\bigr) \CR
&=& (A_u\tan\beta-\mu)\bigl(
\bar\psi_d \,\mat{y_d \lambda_u^\dagger \lambda_u}\,\psi_q \cdot\phi^* +\text{h.c.}\bigr) \,,
\eeqan
where we have dropped similar terms involving $\Phi$ which, after setting $\Phi$ to $\Phi_\c$, contribute to $\bar\psi_d \,\mat{\delta y_d}\,\psi_q \cdot \phi^*+\text{h.c.}$ only at higher order in $\frac{1}{\Lambda^2}$. Noting that there is an identical contribution from the mirror reflection of the diagram of Eq.~\eqref{d4-sfho}, we can write the squark-Higgsino loop contribution to $\mat{\delta y_d}$ as
\beq
\mat{\delta y_d} \supset \mat{y_d}\,\mat{\bar\delta y_d^{(\qt\ut\chit)}} \,,
\eeqn
where
\beq
16\pi^2\, \mat{\bar\delta y_d^{(\qt\ut\chit)}} = \mat{\lambda_u^\dagger \lambda_u}\, \mu(A_u\tan\beta-\mu)\,\It_{\qt\ut\chit}^{111} \,.
\eeqn
The master integral involved here has the following explicit expression,
\beq
\It_{ijk}^{111} \equiv \,\I_{ijk}^{111}/\frac{i}{16\pi^2} = \frac{M_j^2}{(M_i^2-M_j^2)(M_j^2-M_k^2)}\log\frac{M_j^2}{M_i^2}
+ \frac{M_k^2}{(M_j^2-M_k^2)(M_k^2-M_i^2)}\log\frac{M_k^2}{M_i^2} \,,
\eeqn
see Appendix~\ref{app:MI}.

An alternative route we can take to obtain $\bar\psi_d \,\mat{\delta y_d}\,\psi_q \cdot \phi^*+\text{h.c.}$ is to start from $U_{\dt\gt}$, and form a $\dt$-$\gt$-$\qt$ loop,
\beq
\begin{gathered}
\begin{fmffile}{cd-d4-sfgo}
\begin{fmfgraph*}(40,40)
\fmfsurround{vUqu,vUgq,vUug}
\fmf{dashes,left=0.6,label=$\qt$,l.d=3pt}{vUgq,vUqu}
\fmf{dashes,left=0.6,label=$\dt$,l.d=3pt}{vUqu,vUug}
\fmf{plain,left=0.6,label=$\gt$,l.d=3pt}{vUug,vUgq}
\fmfv{decor.shape=circle,decor.filled=empty,decor.size=3thick}{vUqu,vUgq,vUug}
\end{fmfgraph*}
\end{fmffile}
\end{gathered}
\quad = -\frac{i}{2}M_3 \,\I_{\qt\dt\gt}^{111} \,\tr\bigl(U_{\dt\gt} U_{\gt\qt} U_{\qt\dt} \bigr)\,.
\eeq{d4-sfgo}
Evaluating the trace and adding the mirror diagram, we obtain the squark-gluino loop contribution to $\mat{\delta y_d}$,
\beq
\mat{\delta y_d} \supset 
\mat{y_d}\,\bar\delta y_d^{(\qt\dt\gt)} \,,
\eeqn
where
\beq
16\pi^2\, \bar\delta y_d^{(\qt\dt\gt)} = -2\, (A_d-\mu\tan\beta)\, g_3^2\,\CasC M_3 \,\It_{\qt\dt\gt}^{111} \,,
\eeqn
with $\CasC=\frac{4}{3}$ being the quadratic Casimir of the fundamental representation of $SU(3)_c$. It is worth noting that both the squark-Higgsino loop and the squark-gluino loop computed above can be viewed as part of a single covariant diagram,
\beq
\begin{gathered}
\begin{fmffile}{cd-U3-bbf}
\begin{fmfgraph*}(40,40)
\fmfsurround{vU2,vU1,vU3}
\fmf{dashes,left=0.6,label=$i$,l.s=left,l.d=2pt}{vU1,vU2}
\fmf{dashes,left=0.6,label=$j$,l.s=left,l.d=2pt}{vU2,vU3}
\fmf{plain,left=0.6,label=$k$,l.s=left,l.d=3pt}{vU3,vU1}
\fmfv{decor.shape=circle,decor.filled=empty,decor.size=3thick}{vU1,vU2,vU3}
\end{fmfgraph*}
\end{fmffile}
\end{gathered}
\quad = -\frac{i}{2}M_k \,\I_{ijk}^{111} \,\tr\bigl(U_{ij} U_{jk} U_{ki} \bigr)\,,
\eeqn
with summation over $i,j,k$ understood.

\begin{table}[t]
\begin{center}
\begin{tabular}{|c|c|c|}
\hline
\multicolumn{3}{|c|}{\large {\bf 14}~$U$-only covariant diagrams contributing to $\delta m^2$, $\delta\lambda$, $\mat{\delta y_f}$} \\
\hhline{|===|}
&
\parbox[c][60pt][c]{60pt}{\centering
\begin{fmffile}{cd-U1}
\begin{fmfgraph*}(40,40)
\fmfsurround{v0,vU}
\fmf{dashes,left=1}{vU,v0,vU}
\fmfv{decor.shape=circle,decor.filled=empty,decor.size=3thick}{vU}
\fmfv{label=$i$,l.s=left,l.d=3pt}{v0}
\end{fmfgraph*}
\end{fmffile}}
 & $i = \Phi;\, \ft$ \\
\cline{2-3}
$\delta m^2$ & 
\parbox[c][66pt][c]{60pt}{\centering
\begin{fmffile}{cd-U2-f0}
\begin{fmfgraph*}(40,40)
\fmfsurround{vU2,vU1}
\fmf{plain,left=1,label=$i$,l.s=left,l.d=2pt}{vU1,vU2}
\fmf{plain,left=1,label=$j$,l.s=left,l.d=2pt}{vU2,vU1}
\fmfv{decor.shape=circle,decor.filled=empty,decor.size=3thick}{vU1,vU2}
\end{fmfgraph*}
\end{fmffile}}
\parbox[c][66pt][c]{60pt}{\centering
\begin{fmffile}{cd-U2-f1}
\begin{fmfgraph}(40,40)
\fmfsurround{vU1,vP1,vU2,vP2}
\fmf{plain,left=0.4}{vU2,vP1,vU1}
\fmf{plain,left=0.4}{vU1,vP2,vU2}
\fmfv{decor.shape=circle,decor.filled=empty,decor.size=3thick}{vU1,vU2}
\fmf{dots,width=thick}{vP1,vP2}
\end{fmfgraph}
\end{fmffile}}
& $ij = \chit\Wt,\, \chit\Bt$ \\
\cline{2-3}
& 
\multirow{2}{*}{
\parbox[c][60pt][c]{60pt}{\centering
\begin{fmffile}{cd-U2-b}
\begin{fmfgraph*}(40,40)
\fmfsurround{vU2,vU1}
\fmf{dashes,left=1,label=$i$,l.s=left,l.d=2pt}{vU1,vU2}
\fmf{dashes,left=1,label=$j$,l.s=left,l.d=2pt}{vU2,vU1}
\fmfv{decor.shape=circle,decor.filled=empty,decor.size=3thick}{vU1,vU2}
\end{fmfgraph*}
\end{fmffile}}
}
& \parbox[c][36pt][c]{150pt}{\centering $ij = \qt\ut,\, \qt\dt,\, \lt\et$} \\
\hhline{|=~=|}
& & \parbox[c][36pt][c]{150pt}{\centering $ij = \Phi\Phi,\, \Phi\phi;\, \ft\ft$} \\
\cline{2-3}
& 
\parbox[c][60pt][c]{60pt}{\centering
\begin{fmffile}{cd-U3-b}
\begin{fmfgraph*}(40,40)
\fmfsurround{vU2,vU1,vU3}
\fmf{dashes,left=0.6,label=$i$,l.s=left,l.d=3pt}{vU3,vU1}
\fmf{dashes,left=0.6,label=$j$,l.s=left,l.d=2pt}{vU1,vU2}
\fmf{dashes,left=0.6,label=$k$,l.s=left,l.d=2pt}{vU2,vU3}
\fmfv{decor.shape=circle,decor.filled=empty,decor.size=3thick}{vU1,vU2,vU3}
\end{fmfgraph*}
\end{fmffile}}
& $ijk = \qt\qt\ut,\, \ut\ut\qt,\, \qt\qt\dt,\, \dt\dt\qt,\, \lt\lt\et,\, \et\et\lt$ \\
\cline{2-3}
$\delta\lambda$ & 
\parbox[c][60pt][c]{60pt}{\centering
\begin{fmffile}{cd-U4-b}
\begin{fmfgraph*}(40,40)
\fmfsurround{vU3,vU2,vU1,vU4}
\fmf{dashes,left=0.4,label=$i$,l.s=left,l.d=2pt}{vU4,vU1}
\fmf{dashes,left=0.4,label=$j$,l.s=left,l.d=2pt}{vU1,vU2}
\fmf{dashes,left=0.4,label=$k$,l.s=left,l.d=2pt}{vU2,vU3}
\fmf{dashes,left=0.4,label=$l$,l.s=left,l.d=2pt}{vU3,vU4}
\fmfv{decor.shape=circle,decor.filled=empty,decor.size=3thick}{vU1,vU2,vU3,vU4}
\end{fmfgraph*}
\end{fmffile}}
& $ijkl = \qt\ut\qt\ut,\, \qt\dt\qt\dt,\, \lt\et\lt\et$ \\
\cline{2-3}
& 
\parbox[c][60pt][c]{60pt}{\centering
\begin{fmffile}{cd-U4-f0}
\begin{fmfgraph*}(40,40)
\fmfsurround{vU3,vU2,vU1,vU4}
\fmf{plain,left=0.4,label=$i$,l.s=left,l.d=2pt}{vU4,vU1}
\fmf{plain,left=0.4,label=$j$,l.s=left,l.d=2pt}{vU1,vU2}
\fmf{plain,left=0.4,label=$k$,l.s=left,l.d=2pt}{vU2,vU3}
\fmf{plain,left=0.4,label=$l$,l.s=left,l.d=2pt}{vU3,vU4}
\fmfv{decor.shape=circle,decor.filled=empty,decor.size=3thick}{vU1,vU2,vU3,vU4}
\end{fmfgraph*}
\end{fmffile}}
\parbox[c][60pt][c]{60pt}{\centering
\begin{fmffile}{cd-U4-f1a}
\begin{fmfgraph}(40,40)
\fmfsurround{v3,v23,v2,v12,v1,v41,v4,v34}
\fmf{plain,left=0.25}{v1,v12,v2,v23,v3,v34,v4,v41,v1}
\fmfv{decor.shape=circle,decor.filled=empty,decor.size=3thick}{v1,v2,v3,v4}
\fmf{dots,width=thick,right=0.6}{v12,v23}
\end{fmfgraph}
\end{fmffile}}
\parbox[c][60pt][c]{60pt}{\centering
\begin{fmffile}{cd-U4-f1b}
\begin{fmfgraph}(40,40)
\fmfsurround{v3,v23,v2,v12,v1,v41,v4,v34}
\fmf{plain,left=0.25}{v1,v12,v2,v23,v3,v34,v4,v41,v1}
\fmfv{decor.shape=circle,decor.filled=empty,decor.size=3thick}{v1,v2,v3,v4}
\fmf{dots,width=thick}{v12,v34}
\end{fmfgraph}
\end{fmffile}}
& \multirow{2}{*}{
\parbox[c][120pt][t]{150pt}{\centering \vspace{36pt}
$ijkl = \chit\Wt\chit\Wt,\, \chit\Wt\chit\Bt,\, \chit\Bt\chit\Bt$}} \\
& 
\parbox[c][60pt][c]{60pt}{\centering
\begin{fmffile}{cd-U4-f2a}
\begin{fmfgraph}(40,40)
\fmfsurround{v3,v23,v2,v12,v1,v41,v4,v34}
\fmf{plain,left=0.25}{v1,v12,v2,v23,v3,v34,v4,v41,v1}
\fmfv{decor.shape=circle,decor.filled=empty,decor.size=3thick}{v1,v2,v3,v4}
\fmf{dots,width=thick,right=0.6}{v12,v23}
\fmf{dots,width=thick,right=0.6}{v34,v41}
\end{fmfgraph}
\end{fmffile}}
\parbox[c][60pt][c]{60pt}{\centering
\begin{fmffile}{cd-U4-f2b}
\begin{fmfgraph}(40,40)
\fmfsurround{v3,v23,v2,v12,v1,v41,v4,v34}
\fmf{plain,left=0.25}{v1,v12,v2,v23,v3,v34,v4,v41,v1}
\fmfv{decor.shape=circle,decor.filled=empty,decor.size=3thick}{v1,v2,v3,v4}
\fmf{dots,width=thick}{v12,v34}
\fmf{dots,width=thick}{v23,v41}
\end{fmfgraph}
\end{fmffile}}
& \\
\hhline{|===|}
\multirow{2}{*}{\parbox[c][120pt][t]{24pt}{\centering \vspace{36pt}
$\mat{\delta y_f}$}}
& 
\parbox[c][60pt][c]{60pt}{\centering
\begin{fmffile}{cd-U3-bff0}
\begin{fmfgraph*}(40,40)
\fmfsurround{vU2,vU1,vU3}
\fmf{plain,left=0.6,label=$j$,l.s=left,l.d=2pt}{vU1,vU2}
\fmf{plain,left=0.6,label=$k$,l.s=left,l.d=2pt}{vU2,vU3}
\fmf{dashes,left=0.6,label=$i$,l.s=left,l.d=3pt}{vU3,vU1}
\fmfv{decor.shape=circle,decor.filled=empty,decor.size=3thick}{vU1,vU2,vU3}
\end{fmfgraph*}
\end{fmffile}}
\parbox[c][60pt][c]{60pt}{\centering
\begin{fmffile}{cd-U3-bff1}
\begin{fmfgraph}(40,40)
\fmfsurround{vU1,vP1,vU4,vU3,vU2,vP2}
\fmf{plain,left=0.25}{vU4,vP1,vU1,vP2,vU2}
\fmf{dashes,left=0.25}{vU2,vU3,vU4}
\fmfv{decor.shape=circle,decor.filled=empty,decor.size=3thick}{vU1,vU2,vU4}
\fmf{dots,width=thick,right=0.4}{vP1,vP2}
\end{fmfgraph}
\end{fmffile}}
& $ijk = \Phi qu,\, \Phi qd;\, \ft\chit\Vt$ \\
\cline{2-3}
& 
\parbox[c][60pt][c]{60pt}{\centering
\begin{fmffile}{cd-U3-bbf}
\begin{fmfgraph*}(40,40)
\fmfsurround{vU2,vU1,vU3}
\fmf{dashes,left=0.6,label=$i$,l.s=left,l.d=2pt}{vU1,vU2}
\fmf{dashes,left=0.6,label=$j$,l.s=left,l.d=2pt}{vU2,vU3}
\fmf{plain,left=0.6,label=$k$,l.s=left,l.d=3pt}{vU3,vU1}
\fmfv{decor.shape=circle,decor.filled=empty,decor.size=3thick}{vU1,vU2,vU3}
\end{fmfgraph*}
\end{fmffile}}
& $ijk=\qt\ut\chit,\,\qt\dt\chit;\,\qt\ut\Vt,\,\qt\dt\Vt,\,\lt\et\Vt$ \\
\hline
\end{tabular}
\caption{\label{tab:cd-U}
Covariant diagrams contributing to Higgs potential and Yukawa interactions.
}
\end{center}
\end{table}

\begin{table}[t]
\begin{center}
\begin{tabular}{|c|c|c|}
\hline
\multicolumn{3}{|c|}{\large {\bf 16}~$P$-dependent covariant diagrams contributing to $\delta Z$} \\
\hhline{|===|}
&
\parbox[c][60pt][c]{60pt}{\centering
\begin{fmffile}{cd-P2U2-b}
\begin{fmfgraph*}(40,40)
\fmfsurround{vP2,vU1,vP1,vU2}
\fmf{dashes,left=0.4,label=$i$,l.s=left,l.d=2pt}{vU2,vP1,vU1}
\fmf{dashes,left=0.4,label=$j$,l.s=left,l.d=2pt}{vU1,vP2,vU2}
\fmfv{decor.shape=circle,decor.filled=full,decor.size=3thick}{vP1,vP2}
\fmfv{decor.shape=circle,decor.filled=empty,decor.size=3thick}{vU1,vU2}
\fmf{dots,width=thick}{vP1,vP2}
\end{fmfgraph*}
\end{fmffile}}
& $ij=\qt\ut,\, \qt\dt,\, \lt\et$ \\[10pt]
\cline{2-3}
$\delta Z_\phi$ 
&
\parbox[c][60pt][c]{60pt}{\centering
\begin{fmffile}{cd-P2U2-f0}
\begin{fmfgraph*}(40,40)
\fmfsurround{vP2,vU1,vP1,vU2}
\fmf{plain,left=0.4,label=$i$,l.s=left,l.d=2pt}{vU2,vP1,vU1}
\fmf{plain,left=0.4,label=$j$,l.s=left,l.d=2pt}{vU1,vP2,vU2}
\fmfv{decor.shape=circle,decor.filled=full,decor.size=3thick}{vP1,vP2}
\fmfv{decor.shape=circle,decor.filled=empty,decor.size=3thick}{vU1,vU2}
\end{fmfgraph*}
\end{fmffile}}
\parbox[c][60pt][c]{60pt}{\centering
\begin{fmffile}{cd-P2U2-f1a}
\begin{fmfgraph}(40,40)
\fmfsurround{v3,v23,v2,v12,v1,v41,v4,v34}
\fmf{plain,left=0.25}{v1,v12,v2,v23,v3,v34,v4,v41,v1}
\fmfv{decor.shape=circle,decor.filled=full,decor.size=3thick}{v1,v3}
\fmfv{decor.shape=circle,decor.filled=empty,decor.size=3thick}{v2,v4}
\fmf{dots,width=thick,right=0.6}{v41,v12}
\end{fmfgraph}
\end{fmffile}}
\parbox[c][60pt][c]{60pt}{\centering
\begin{fmffile}{cd-P2U2-f1b}
\begin{fmfgraph}(40,40)
\fmfsurround{v3,v23,v2,v12,v1,v41,v4,v34}
\fmf{plain,left=0.25}{v1,v12,v2,v23,v3,v34,v4,v41,v1}
\fmfv{decor.shape=circle,decor.filled=full,decor.size=3thick}{v1,v3}
\fmfv{decor.shape=circle,decor.filled=empty,decor.size=3thick}{v2,v4}
\fmf{dots,width=thick,right=0.6}{v12,v23}
\end{fmfgraph}
\end{fmffile}}
\parbox[c][60pt][c]{60pt}{\centering
\begin{fmffile}{cd-P2U2-f1c}
\begin{fmfgraph}(40,40)
\fmfsurround{v3,v23,v2,v12,v1,v41,v4,v34}
\fmf{plain,left=0.25}{v1,v12,v2,v23,v3,v34,v4,v41,v1}
\fmfv{decor.shape=circle,decor.filled=full,decor.size=3thick}{v1,v3}
\fmfv{decor.shape=circle,decor.filled=empty,decor.size=3thick}{v2,v4}
\fmf{dots,width=thick}{v41,v23}
\end{fmfgraph}
\end{fmffile}}
 & \multirow{2}{*}{
\parbox[c][120pt][t]{72pt}{\centering \vspace{36pt}
$ij=\chit\Wt,\, \chit\Bt$}
} \\
& 
\parbox[c][60pt][c]{60pt}{\centering
\begin{fmffile}{cd-P2U2-f2a}
\begin{fmfgraph}(40,40)
\fmfsurround{v3,v23,v2,v12,v1,v41,v4,v34}
\fmf{plain,left=0.25}{v1,v12,v2,v23,v3,v34,v4,v41,v1}
\fmfv{decor.shape=circle,decor.filled=full,decor.size=3thick}{v1,v3}
\fmfv{decor.shape=circle,decor.filled=empty,decor.size=3thick}{v2,v4}
\fmf{dots,width=thick,right=0.6}{v41,v12}
\fmf{dots,width=thick,right=0.6}{v23,v34}
\end{fmfgraph}
\end{fmffile}}
\parbox[c][60pt][c]{60pt}{\centering
\begin{fmffile}{cd-P2U2-f2b}
\begin{fmfgraph}(40,40)
\fmfsurround{v3,v23,v2,v12,v1,v41,v4,v34}
\fmf{plain,left=0.25}{v1,v12,v2,v23,v3,v34,v4,v41,v1}
\fmfv{decor.shape=circle,decor.filled=full,decor.size=3thick}{v1,v3}
\fmfv{decor.shape=circle,decor.filled=empty,decor.size=3thick}{v2,v4}
\fmf{dots,width=thick,right=0.6}{v12,v23}
\fmf{dots,width=thick,right=0.6}{v34,v41}
\end{fmfgraph}
\end{fmffile}}
\parbox[c][60pt][c]{60pt}{\centering
\begin{fmffile}{cd-P2U2-f2c}
\begin{fmfgraph}(40,40)
\fmfsurround{v3,v23,v2,v12,v1,v41,v4,v34}
\fmf{plain,left=0.25}{v1,v12,v2,v23,v3,v34,v4,v41,v1}
\fmfv{decor.shape=circle,decor.filled=full,decor.size=3thick}{v1,v3}
\fmfv{decor.shape=circle,decor.filled=empty,decor.size=3thick}{v2,v4}
\fmf{dots,width=thick}{v41,v23}
\fmf{dots,width=thick}{v12,v34}
\end{fmfgraph}
\end{fmffile}}
& \\
\hhline{|===|}

$\mat{\delta Z_f}$
&
\parbox[c][60pt][c]{60pt}{\centering
\begin{fmffile}{cd-P1U2-bff0}
\begin{fmfgraph*}(40,40)
\fmfsurround{vU2,vU1,vU3}
\fmf{plain,left=0.6,label=$j$,l.s=left,l.d=1pt}{vU1,vU2,vU3}
\fmf{dashes,left=0.6,label=$i$,l.s=left,l.d=3pt}{vU3,vU1}
\fmfv{decor.shape=circle,decor.filled=empty,decor.size=3thick}{vU1,vU3}
\fmfv{decor.shape=circle,decor.filled=full,decor.size=3thick}{vU2}
\end{fmfgraph*}
\end{fmffile}}
\parbox[c][60pt][c]{60pt}{\centering
\begin{fmffile}{cd-P1U2-bff1}
\begin{fmfgraph}(40,40)
\fmfsurround{vU1,vP1,vU4,vU3,vU2,vP2}
\fmf{plain,left=0.25}{vU4,vP1,vU1,vP2,vU2}
\fmf{dashes,left=0.25}{vU2,vU3,vU4}
\fmfv{decor.shape=circle,decor.filled=full,decor.size=3thick}{vU1}
\fmfv{decor.shape=circle,decor.filled=empty,decor.size=3thick}{vU2,vU4}
\fmf{dots,width=thick,right=0.4}{vP1,vP2}
\fmfv{label=$i$,l.s=left,l.d=3pt}{vU3}
\fmfv{label=$j$,l.s=left,l.d=2pt}{vP1,vP2}
\end{fmfgraph}
\end{fmffile}}
\parbox[c][60pt][c]{60pt}{\centering
\begin{fmffile}{cd-P1U2-bbf}
\begin{fmfgraph*}(40,40)
\fmfsurround{vP2,vU2,vU1,vP1,vU4,vU3}
\fmf{dashes,left=0.25}{vU3,vU4,vP1,vU1,vU2}
\fmf{plain,left=0.25}{vU2,vP2,vU3}
\fmfv{decor.shape=circle,decor.filled=full,decor.size=3thick}{vP1}
\fmfv{decor.shape=circle,decor.filled=empty,decor.size=3thick}{vU2,vU3}
\fmfv{label=$i$,l.s=left,l.d=2pt}{vU1,vU4}
\fmfv{label=$j$,l.s=left,l.d=3pt}{vP2}
\fmf{dots,width=thick}{vP1,vP2}
\end{fmfgraph*}
\end{fmffile}}
& $ij = \Phi f;\, \ft\chit,\, \ft\Vt$ \\
\hhline{|===|}
\multirow{2}{*}{\parbox[c][120pt][t]{42pt}{\centering \vspace{36pt}
$\delta Z_{G,W,B}$}}
&
\parbox[c][60pt][c]{60pt}{\centering
\begin{fmffile}{cd-P4-b}
\begin{fmfgraph*}(40,40)
\fmfsurround{vP3,vP2,vP1,vP4}
\fmf{dashes,left=0.4,label=$i$,l.s=left,l.d=2pt}{vP1,vP2,vP3,vP4,vP1}
\fmfv{decor.shape=circle,decor.filled=full,decor.size=3thick}{vP1,vP2,vP3,vP4}
\fmf{dots,width=thick}{vP1,vP3}
\fmf{dots,width=thick}{vP2,vP4}
\end{fmfgraph*}
\end{fmffile}}
& $i = \Phi,\, \ft$ \\
\cline{2-3}
&
\parbox[c][60pt][c]{60pt}{\centering
\begin{fmffile}{cd-P4-f0}
\begin{fmfgraph*}(40,40)
\fmfsurround{v3,v2,v1,v4}
\fmf{plain,left=0.4,label=$i$,l.s=left,l.d=2pt}{v1,v2,v3,v4,v1}
\fmfv{decor.shape=circle,decor.filled=full,decor.size=3thick}{v1,v2,v3,v4}
\end{fmfgraph*}
\end{fmffile}}
\parbox[c][60pt][c]{60pt}{\centering
\begin{fmffile}{cd-P4-f1}
\begin{fmfgraph}(40,40)
\fmfsurround{v3,v23,v2,v12,v1,v41,v4,v34}
\fmf{plain,left=0.25}{v1,v12,v2,v23,v3,v34,v4,v41,v1}
\fmfv{decor.shape=circle,decor.filled=full,decor.size=3thick}{v1,v2,v3,v4}
\fmf{dots,width=thick,right=0.6}{v12,v23}
\end{fmfgraph}
\end{fmffile}}
\parbox[c][60pt][c]{60pt}{\centering
\begin{fmffile}{cd-P4-f2a}
\begin{fmfgraph}(40,40)
\fmfsurround{v3,v23,v2,v12,v1,v41,v4,v34}
\fmf{plain,left=0.25}{v1,v12,v2,v23,v3,v34,v4,v41,v1}
\fmfv{decor.shape=circle,decor.filled=full,decor.size=3thick}{v1,v2,v3,v4}
\fmf{dots,width=thick,right=0.6}{v12,v23}
\fmf{dots,width=thick,right=0.6}{v34,v41}
\end{fmfgraph}
\end{fmffile}}
\parbox[c][60pt][c]{60pt}{\centering
\begin{fmffile}{cd-P4-f2b}
\begin{fmfgraph}(40,40)
\fmfsurround{v3,v23,v2,v12,v1,v41,v4,v34}
\fmf{plain,left=0.25}{v1,v12,v2,v23,v3,v34,v4,v41,v1}
\fmfv{decor.shape=circle,decor.filled=full,decor.size=3thick}{v1,v2,v3,v4}
\fmf{dots,width=thick}{v12,v34}
\fmf{dots,width=thick}{v23,v41}
\end{fmfgraph}
\end{fmffile}}
& $i = \chit,\, \gt,\, \Wt$ \\
\hline
\end{tabular}
\caption{\label{tab:cd-P}
Covariant diagrams contributing to kinetic terms.}
\end{center}
\end{table}

There are several other routes that can take us to the operator $\bar\psi_d \,\mat{\delta y_d}\,\psi_q \cdot \phi^*+\text{h.c.}$, and many others that can take us to other SMEFT operators. Following the procedure demonstrated with the examples above, we enumerate covariant diagrams contributing to each $d\le4$ operator in Tables~\ref{tab:cd-U} and~\ref{tab:cd-P}. In particular, Table~\ref{tab:cd-U} contains covariant diagrams contributing to the Higgs potential and Yukawa interactions, which involve $U$ insertions only and no $P$ insertions. The kinetic terms (wavefunction renormalization factors $\delta Z$), on the other hand, come from covariant diagrams that involve $P$ insertions, as shown in Table~\ref{tab:cd-P}. The $c_{\Phi\phi}(\Phi_\c^{*}\phi +\phi^*\Phi_\c)$ piece, which we choose to absorb into $\L_\text{SMEFT}^\text{tree}$ via a redefinition of $\beta$ as explained in Section~\ref{sec:MSSMmatch-tree-beta}, is computed from the same covariant diagrams contributing to $\delta m^2|\phi|^2$. In diagrams where permutations of propagator labels produce inequivalent diagrams, such permutations are implicitly assumed to be included. We refrain from elaborating on how to compute each of the tabulated covariant diagrams, as the general procedure should already be clear from the examples given above.

From Tables~\ref{tab:cd-U} and~\ref{tab:cd-P}, we can see an advantage of our approach is that despite the large number of terms in the final results of one-loop SUSY threshold corrections (which we will present below), {\it they all derive from just 30 covariant diagrams}. The small number of covariant diagrams can be understood on dimensional grounds. Generally, we have
\beq
\text{dim}(P_\mu) = 1 \,,\quad \text{dim}(U_{ij}[\varphi]) \ge 1 \,,
\eeqn
where ``dim'' means operator dimension. $d\le4$ operators can therefore only come from covariant diagrams with at most 4 vertex insertions, as enumerated in the tables for the case of the MSSM\footnote{Similarly, dimension-six operators can be obtained from covariant diagrams with at most 6 vertex insertions. This is true regardless of the UV theory, as long as it is Lorentz-invariant and satisfies the general form of Eq.~\eqref{QUV}. This simple observation of finite combinatorics underlies the idea of deriving universal formulas for one-loop effective Lagrangians~\cite{DEQY,EQYZ16,EQYZ17}.}.

\subsubsection{Results}

Now we present the results of one-loop-level coefficients of all $d\le4$ SMEFT operators, i.e.\ $\delta Z_{\phi,f,V}$, $\delta m^2$, $\delta\lambda,$ $\mat{\delta y_f}$ defined in Eq.~\eqref{Ldle4}, which are calculated from the 30 covariant diagrams in Tables~\ref{tab:cd-U} and~\ref{tab:cd-P}. These coefficients, together with the tree-level relations,
\beqa
&& m^2 \equiv \mu^2 +\mHusq\sb^2 +\mHdsq\cb^2 -b\sbb \,,\quad \lambda \equiv \frac{1}{8} (g^2 +g'^2)\, \cbb^2 \,, \CR
&& \mat{y_u} = \mat{\lambda_u}\, \sb \,,\quad \mat{y_d} = \mat{\lambda_d}\, \cb \,,\quad \mat{y_e} = \mat{\lambda_e}\, \cb \,,
\eeqa{SMparam}
%
can be readily plugged into Eq.~\eqref{thrcor} to obtain one-loop SUSY threshold corrections (there is a one-loop correction to the equation for $\lambda$ if we work with the $\overline{\text{MS}}$ scheme~\cite{MartinVaughnDRbar}). We will use parenthesized subscripts or superscripts to indicate the covariant diagram each term comes from, and mention the reduction formulas used on the master integrals so that all results can be easily reproduced.

We have cross-checked our results against conventional Feynman diagram calculations reported in~\cite{BMPZ} and found complete agreement; see Appendix~\ref{app:BMPZ}. Note in particular that at one-loop level, MSSM threshold corrections are the same in both $\overline{\text{MS}}$ and $\overline{\text{DR}}$ schemes, as is clear from the absence of $\epsilon$-scalar loops in our matching calculation. 

Our notation is the following. $N_c=3$ is the number of colors. 
$\CasC=\frac{4}{3}$ and $\CasL=\frac{3}{4}$ are quadratic Casimirs of fundamental representations of $SU(3)_c$ and $SU(2)_L$, respectively. The $U(1)_Y$ hypercharges are
\beq
\bigl\{\, Y_\phi \,,\; Y_q \,,\; Y_u \,,\; Y_d \,,\; Y_l \,,\; Y_e \, \bigr\} = \Bigl\{\, \frac{1}{2} \,,\; \frac{1}{6} \,,\; \frac{2}{3} \,,\; -\frac{1}{3} \,,\; -\frac{1}{2} \,,\; -1 \, \Bigr\} \,.
\eeqn
The master integrals $\,\It\equiv\,\I/\frac{i}{16\pi^2}$ are functions of tree-level masses of the heavy particles. Their analytical expressions in terms of tree-level heavy particle masses can be found in Appendix~\ref{app:MI}.

\paragraph{Higgs potential.} 

The one-loop coefficient of the $d=2$ operator $|\phi|^2$ reads
\beq
\delta m^2 = \delta m^2_{(\Phi)} +\delta m^2_{(\ft)} +\delta m^2_{(\ft\ft)} +\delta m^2_{(\chit\Vt)} \,,
\eeqn
where
\beqa
16\pi^2\, \delta m^2_{(\Phi)} &=& \Bigl[\frac{3}{4}\, g^2\sbb^2 +g'^2\, Y_\phi^2\, (\sbb^2-2\cbb^2)\Bigr]\,\It_{\Phi}^1 \,,\\[4pt]
16\pi^2\, \delta m^2_{(\ft)} &=& N_c\,\tr(\mat{\lambda_u^\dagger \lambda_u})\,\sb^2\, \bigl(\,\It_{\qt}^1 +\,\It_{\ut}^1\bigr) \CR
&& +N_c\,\tr(\mat{\lambda_d^\dagger \lambda_d})\,\cb^2\, \bigl(\,\It_{\qt}^1 +\,\It_{\dt}^1\bigr) +\tr(\mat{\lambda_e^\dagger \lambda_e}) \,\cb^2\, \bigl(\,\It_{\lt}^1 +\,\It_{\et}^1\bigr) \CR
&& -g'^2\,Y_\phi\cbb \bigl(2N_c Y_q \,\It_{\qt}^1 -N_c Y_u \,\It_{\ut}^1 -N_c Y_d \,\It_{\dt}^1 +2 Y_l \,\It_{\lt}^1 - Y_e \,\It_{\et}^1\bigr) \,,\\[4pt]
16\pi^2\, \delta m^2_{(\ft\ft)} &=& N_c\,\tr(\mat{\lambda_u^\dagger \lambda_u}) \,\sb^2\, (A_u-\mu\cot\beta)^2 \,\It_{\qt\ut}^{11} \CR
&& +N_c\,\tr(\mat{\lambda_d^\dagger \lambda_d}) \,\sb^2\,(A_d\cot\beta-\mu)^2 \,\It_{\qt\dt}^{11} \CR
&& +\,\tr(\mat{\lambda_e^\dagger \lambda_e})\,\sb^2 (A_e\cot\beta-\mu)^2 \,\It_{\lt\et}^{11} \,,\\[4pt]
16\pi^2\, \delta m^2_{(\chit\Vt)} &=& -4\,g^2\,\CasL \Bigl[\frac{M_2(M_2+\sbb\mu)}{M_2^2-\mu^2} \,\It_{\Wt}^1 -\frac{\mu(\sbb M_2+\mu)}{M_2^2-\mu^2}\,\It_{\chit}^1 \Bigr] \CR
&& -4\,g'^2\, Y_\phi^2\Bigl[\frac{M_1(M_1+\sbb\mu)}{M_1^2-\mu^2} \,\It_{\Bt}^1 -\frac{\mu(\sbb M_1+\mu)}{M_1^2-\mu^2}\,\It_{\chit}^1 \Bigr] \,.\label{dm-chitVt}
\eeqan
Note that terms proportional to $\gamma^\mu\gamma_\mu=4-\epsilon$ are generally encountered when computing loops involving two fermionic fields. To arrive at Eq.~\eqref{dm-chitVt}, we have used Eq.~\eqref{MIred-nc1} to reduce $(4-\epsilon)\,\It[q^2]_{\chit\Vt}^{11}$ to $\,\It_{\chit}^{1}$, $\,\It_{\Vt}^{1}$ and $\,\It_{\chit\Vt}^{11}$, and further used Eq.~\eqref{MIred-h} to reduce $\,\It_{\chit\Vt}^{11}$ to $\,\It_{\chit}^{1}$ and $\,\It_{\Vt}^{1}$.

From the expressions of master integrals in Eqs.~\eqref{Ii1} and~\eqref{Iij11}, it is clear that each term in the equations above is $\O(\frac{\Lambda^2}{16\pi^2})$. Quite generally, the $|\phi|^2$ operator receives threshold corrections that are quadratically sensitive to the EFT cutoff scale $\Lambda$ when a high-energy BSM theory is matched onto the SMEFT, as a manifestation of a potential hierarchy problem.

As noted before, the same covariant diagrams contributing to $\delta m^2|\phi|^2$ can also be used to compute the $c_{\Phi\phi}(\Phi_\c^{*}\phi +\phi^*\Phi_\c)$ piece, for which we obtain
\beq
c_{\Phi\phi} = c_{\Phi\phi}^{(\Phi)} +c_{\Phi\phi}^{(\ft)} +c_{\Phi\phi}^{(\ft\ft)} +c_{\Phi\phi}^{(\chit\Vt)} \,,
\eeq{cPhiphi}
where
\beqa
16\pi^2\, c_{\Phi\phi}^{(\Phi)} &=& \frac{3}{8}(g^2+g'^2) \,s_{4\beta} \,\It_\Phi^1 \,,\\[4pt]
16\pi^2\, c_{\Phi\phi}^{(\ft)} &=& N_c\,\tr(\mat{\lambda_u^\dagger \lambda_u})\,\sb\cb \bigl(\,\It_{\qt}^1 +\,\It_{\ut}^1\bigr) \CR
&& -N_c\,\tr(\mat{\lambda_d^\dagger \lambda_d}) \,\sb\cb \bigl(\,\It_{\qt}^1 +\,\It_{\dt}^1\bigr) -\tr(\mat{\lambda_e^\dagger \lambda_e}) \,\sb\cb \bigl(\,\It_{\lt}^1 +\,\It_{\et}^1\bigr) \CR
&& +2g'^2\,Y_\phi\sb\cb \bigl(2N_c Y_q \,\It_{\qt}^1 -N_c Y_u \,\It_{\ut}^1 -N_c Y_d \,\It_{\dt}^1 +2 Y_l \,\It_{\lt}^1 - Y_e \,\It_{\et}^1\bigr) \,,\\[4pt]
16\pi^2\, c_{\Phi\phi}^{(\ft\ft)} &=& N_c\,\tr(\mat{\lambda_u^\dagger \lambda_u})\,\sb^2\, (A_u-\mu\cot\beta)(A_u\cot\beta+\mu) \,\It_{\qt\ut}^{11} \CR
&& -N_c\,\tr(\mat{\lambda_d^\dagger \lambda_d})\,\sb^2\, (A_d\cot\beta-\mu)(A_d+\mu\cot\beta) \,\It_{\qt\dt}^{11} \CR
&& -\,\tr(\mat{\lambda_e^\dagger \lambda_e}) \,\sb^2\,(A_e\cot\beta-\mu)(A_e+\mu\cot\beta) \,\It_{\lt\et}^{11} \,,\\[4pt]
16\pi^2\, c_{\Phi\phi}^{(\chit\Vt)} &=& -4\,g^2\,\CasL \cbb M_2\mu\,\It_{\chit\Wt}^{11} -4\,g'^2\,Y_\phi^2 \cbb M_1\mu\,\It_{\chit\Bt}^{11}\,.
\eeqan
As discussed in Section~\ref{sec:MSSMmatch-tree-beta}, we absorb this piece into the tree-level effective Lagrangian via a redefinition of $\beta$, and thus do not consider it as contributing to threshold corrections.

The $d=4$ operator $|\phi|^4$ has the following one-loop coefficient,
\beq
\delta\lambda = \delta\lambda_{(\Phi\varphi)} +\delta\lambda_{(\ft\ft)} +\delta\lambda_{(\ft\ft\ft)} +\delta\lambda_{(\ft\ft\ft\ft)} +\delta\lambda_{(\chit\Vt\chit\Vt)} \,,
\eeqn
where
\beqa
16\pi^2\, \delta\lambda_{(\Phi\varphi)} &=& 
\frac{1}{16}\Bigl[(g^2+g'^2)^2\bigl(\sbb^4-\sbb^2\cbb^2+\cbb^4\bigr)-2\,g^2(g^2+g'^2)\cbb^2 +2\,g^4\Bigr]\,\It_\Phi^2 \CR
&& +\frac{3}{8}(g^2+g'^2)^2 \sbb^2\cbb^2 \,\It_{\Phi0}^{11} \,,\label{dlambdaPhi}\\[4pt]
16\pi^2\, \delta\lambda_{(\ft\ft)} &=& 
\frac{1}{2}\,\tr\, \biggl\{ \,N_c\biggl[\Bigl(\mat{\lambda_u^\dagger \lambda_u}\sb^2 +\frac{1}{4}g^2\cbb -g'^2\, Y_\phi Y_q \cbb \Bigr)^2 \CR
&&\qquad\qquad\; +\Bigl(\mat{\lambda_d^\dagger \lambda_d}\cb^2  -\frac{1}{4}g^2\cbb -g'^2\, Y_\phi Y_q \cbb \Bigr)^2 \biggr] \,\It_{\qt}^2 \CR
&&\qquad +N_c\, \bigl(\mat{\lambda_u \lambda_u^\dagger}\sb^2 +g'^2\,Y_\phi Y_u\cbb\bigr)^2 \,\It_{\ut}^2
+N_c\, \bigl(\mat{\lambda_d \lambda_d^\dagger}\cb^2 +g'^2\,Y_\phi Y_d\cbb\bigr)^2 \,\It_{\dt}^2 \CR
&&\qquad +\biggl[\Bigl(\mat{\lambda_e^\dagger \lambda_e}\cb^2  -\frac{1}{4}g^2\cbb -g'^2\, Y_\phi Y_l \cbb \Bigr)^2 + \Bigl(\frac{1}{4}g^2\cbb -g'^2\, Y_\phi Y_l \cbb \Bigr)^2\biggr] \,\It_{\lt}^2 \CR
&&\qquad + \bigl(\mat{\lambda_e \lambda_e^\dagger}\cb^2 +g'^2\,Y_\phi Y_e\cbb\bigr)^2 \,\It_{\et}^2 \biggr\} \,,\\[4pt]
16\pi^2\, \delta\lambda_{(\ft\ft\ft)} &=& \tr\, \biggl\{ N_c \,(A_u-\mu\cot\beta)^2 \,\mat{\lambda_u^\dagger \lambda_u}\sb^2\biggl[ \Bigl(\mat{\lambda_u^\dagger \lambda_u}\sb^2 +\frac{1}{4} g^2 \cbb - g'^2\, Y_\phi Y_q \cbb \Bigr)\,\It_{\qt\ut}^{21} \CR
&&\qquad\qquad\qquad\qquad\qquad\qquad\qquad +\bigl(\mat{\lambda_u^\dagger \lambda_u}\sb^2 +g'^2\,Y_\phi Y_u \cbb\bigr)\,\It_{\qt\ut}^{12}\biggr] \CR
&&\quad +N_c\,(A_d\cot\beta-\mu)^2 \,\mat{\lambda_d^\dagger \lambda_d}\sb^2 \biggl[ \Bigl(\mat{\lambda_d^\dagger \lambda_d}\cb^2 -\frac{1}{4} g^2 \cbb - g'^2\, Y_\phi Y_q \cbb \Bigr)\,\It_{\qt\dt}^{21} \CR
&&\qquad\qquad\qquad\qquad\qquad\qquad\qquad +\bigl(\mat{\lambda_d^\dagger \lambda_d}\cb^2 +g'^2\, Y_\phi Y_d \cbb\bigr)\,\It_{\qt\dt}^{12}\biggr] \CR
&&\quad +(A_e\cot\beta-\mu)^2 \,\mat{\lambda_e^\dagger \lambda_e}\sb^2 \biggl[ \Bigl(\mat{\lambda_e^\dagger \lambda_e}\cb^2 -\frac{1}{4} g^2 \cbb - g'^2\, Y_\phi Y_l \cbb \Bigr)\,\It_{\lt\et}^{21} \CR
&&\qquad\qquad\qquad\qquad\qquad\qquad\qquad +\bigl(\mat{\lambda_e^\dagger \lambda_e}\cb^2 +g'^2\, Y_\phi Y_e \cbb\bigr)\,\It_{\lt\et}^{12}\biggr] \biggr\} \,,\label{dlambda-ft3}\\[4pt]
16\pi^2\, \delta\lambda_{(\ft\ft\ft\ft)} &=& \frac{1}{2}\, N_c\,\tr\bigl(\mat{\lambda_u^\dagger \lambda_u \lambda_u^\dagger \lambda_u} \bigr) \,\sb^4\, (A_u-\mu\cot\beta)^4\,\It_{\qt\ut}^{22} \CR
&& +\frac{1}{2}\, N_c\,\tr\bigl(\mat{\lambda_d^\dagger \lambda_d \lambda_d^\dagger \lambda_d} \bigr)\,\sb^4\, (A_d\cot\beta-\mu)^4\,\It_{\qt\dt}^{22} \CR
&& +\frac{1}{2}\,\tr\bigl(\mat{\lambda_e^\dagger \lambda_e \lambda_e^\dagger \lambda_e} \bigr)\,\sb^4\, (A_e\cot\beta-\mu)^4\,\It_{\lt\et}^{22} \,,\label{dlambda-ft4} \\[4pt]
16\pi^2\, \delta\lambda_{(\chit\Vt\chit\Vt)} &=& -\frac{3}{4}\, g^4 \Bigl\{ \,\It_{\chit\Wt}^{11} -\cbb^2 M_2^2 \mu^2 \,\It_{\chit\Wt}^{22} \CR
&&\qquad\quad +4\bigl[2M_2^2 +4\sbb M_2 \mu +(1+\sbb^2)\mu^2 \bigr]\,\It[q^2]_{\chit\Wt}^{22} \Bigr\} \CR
&& -\frac{1}{2}\,g^4\cbb^2 \bigl(\,\It_{\chit\Wt}^{11} -M_2^2\mu^2\,\It_{\chit\Wt}^{22} +4\mu^2\,\It[q^2]_{\chit\Wt}^{22} \bigr) \\
&& - g^2 g'^2\, Y_\phi^2 \Bigl\{ \,\It_{\chit\Wt}^{11} +\,\It_{\chit\Bt}^{11} - \bigl(M_2^2+M_1^2-2\sbb^2 M_2 M_1 \bigr) \mu^2 \,\It_{\chit\Wt\Bt}^{211} \CR
&&\qquad\quad +4\bigl[(M_2+M_1)^2 +4\sbb (M_2+M_1) \mu +2(1+\sbb^2)\mu^2 \bigr]\,\It[q^2]_{\chit\Wt\Bt}^{211} \Bigr\} \CR
&& -4\,g'^4\, Y_\phi^4 \Bigl\{ \,\It_{\chit\Bt}^{11} -\cbb^2 M_1^2 \mu^2 \,\It_{\chit\Bt}^{22} \CR
&&\qquad\quad +4\bigl[2M_1^2 +4\sbb M_1 \mu +(1+\sbb^2)\mu^2 \bigr]\,\It[q^2]_{\chit\Bt}^{22} \Bigr\} \,.\label{dlam-chitV}
\eeqan
For the $\chit\Vt\chit\Vt$ loops, we have used Eq.~\eqref{MIred-nc} to eliminate the $\,\It[q^4]$ master integrals (coming from covariant diagrams with two Lorentz contractions) from Eq.~\eqref{dlam-chitV}. 

\paragraph{Yukawa interactions.}

The $d=4$ Yukawa interaction operators $\bar\psi_u \,\mat{\delta y_u}\,\psi_q \cdot \epsilon \cdot \phi +\bar\psi_d \,\mat{\delta y_d}\,\psi_q \cdot \phi^* +\bar\psi_e \,\mat{\delta y_e}\,\psi_l \cdot \phi^* +\text{h.c.}$ are obtained with the following one-loop coefficients,
\bseq
\beqa
\mat{\delta y_u} &=& \mat{y_u\,\bar\delta y_u} = \mat{y_u} \bigl( \mat{\bar\delta y_u^{(\Phi qd)}} +\mat{\bar\delta y_u^{(\qt\dt\chit)}} +\bar\delta y_u^{(\qt\ut\Vt)} +\bar\delta y_u^{(\ft\chit\Vt)}
\bigr) ,\\
\mat{\delta y_d} &=& \mat{y_d\,\bar\delta y_d} = \mat{y_d} \bigl( \mat{\bar\delta y_d^{(\Phi qu)}} +\mat{\bar\delta y_d^{(\qt\ut\chit)}} +\bar\delta y_d^{(\qt\dt\Vt)} +\bar\delta y_d^{(\ft\chit\Vt)}
\bigr) ,\\
\mat{\delta y_e} &=& \mat{y_e\,\bar\delta y_e} = \mat{y_e} \bigl( \bar\delta y_e^{(\lt\et\Vt)} +\bar\delta y_e^{(\ft\chit\Vt)}
\bigr) ,
\eeqan
\eseqn
where
\bseq
\beqa
16\pi^2\, \mat{\bar\delta y_u^{(\Phi qd)}} &=& \mat{\lambda_d^\dagger \lambda_d}\,\cb^2 \,\It_{\Phi0}^{11} \,,\\
16\pi^2\, \mat{\bar\delta y_d^{(\Phi qu)}} &=& \mat{\lambda_u^\dagger \lambda_u}\,\sb^2 \,\It_{\Phi0}^{11} \,,
\eeqan
\eseq{dy-Phiff}
\vspace{-20pt}
\bseq
\beqa
16\pi^2\, \mat{\bar\delta y_u^{(\qt\dt\chit)}} &=& \mat{\lambda_d^\dagger \lambda_d}\, \mu\,(A_d\cot\beta-\mu)\,\It_{\qt\dt\chit}^{111} \,,\\
16\pi^2\, \mat{\bar\delta y_d^{(\qt\ut\chit)}} &=& \mat{\lambda_u^\dagger \lambda_u} \,\mu\,(A_u\tan\beta-\mu)\,\It_{\qt\ut\chit}^{111} \,,\label{dyb-ho}
\eeqan
\eseq{dy-ftftchit}
\vspace{-20pt}
\bseq
\beqa
16\pi^2\, \bar\delta y_u^{(\qt\ut\Vt)} &=& -2\, (A_u-\mu\cot\beta)\bigl(g_3^2\,\CasC M_3 \,\I_{\qt\ut\gt}^{111} +g'^2\,Y_q Y_u M_1 \,\I_{\qt\ut\Bt}^{111}\bigr) \,,\\
16\pi^2\, \bar\delta y_d^{(\qt\dt\Vt)} &=& -2\, (A_d-\mu\tan\beta)\bigl(g_3^2\,\CasC M_3 \,\I_{\qt\dt\gt}^{111} +g'^2\,Y_q Y_d M_1 \,\I_{\qt\dt\Bt}^{111}\bigr) \,,\label{dyb-go}\\
16\pi^2\, \bar\delta y_e^{(\lt\et\Vt)} &=& -2\, (A_e-\mu\tan\beta) \,g'^2\,Y_l Y_e M_1 \,\I_{\lt\et\Bt}^{111} \,,
\eeqan
\eseq{dy-ftftVt}
\vspace{-20pt}
\bseq
\beqa
16\pi^2\, \bar\delta y_u^{(\ft\chit\Vt)} &=& -2\,g^2\, \CasL \Bigl[ \frac{M_2(M_2+\mu\cot\beta)}{M_2^2-\mu^2} \,\It_{\qt\Wt}^{11} -\frac{\mu(\mu+M_2\cot\beta)}{M_2^2-\mu^2} \,\It_{\qt\chit}^{11}\, \Bigr] \CR
&& +2\,g'^2\, Y_\phi \Bigl[ \frac{M_1(M_1+\mu\cot\beta)}{M_1^2-\mu^2} \bigl(Y_q\,\It_{\qt\Bt}^{11}-Y_u\,\It_{\ut\Bt}^{11}\bigr) \CR
&&\qquad\qquad\, -\frac{\mu(\mu+M_1\cot\beta)}{M_1^2-\mu^2} \bigl(Y_q\,\It_{\qt\chit}^{11} -Y_u\,\It_{\ut\chit}^{11}\bigr) \Bigr] \,,\\[4pt]
16\pi^2\, \bar\delta y_d^{(\ft\chit\Vt)} &=& -2\,g^2\, \CasL \Bigl[ \frac{M_2(M_2+\mu\tan\beta)}{M_2^2-\mu^2} \,\It_{\qt\Wt}^{11} -\frac{\mu(\mu+M_2\tan\beta)}{M_2^2-\mu^2} \,\It_{\qt\chit}^{11}\, \Bigr] \CR
&& -2\,g'^2\, Y_\phi \Bigl[ \frac{M_1(M_1+\mu\tan\beta)}{M_1^2-\mu^2} \bigl(Y_q\,\It_{\qt\Bt}^{11}-Y_d\,\It_{\dt\Bt}^{11}\bigr) \CR
&&\qquad\qquad\, -\frac{\mu(\mu+M_1\tan\beta)}{M_1^2-\mu^2} \bigl(Y_q\,\It_{\qt\chit}^{11} -Y_d\,\It_{\dt\chit}^{11}\bigr) \Bigr] \,,\\[4pt]
16\pi^2\, \bar\delta y_e^{(\ft\chit\Vt)} &=& -2\,g^2\, \CasL \Bigl[ \frac{M_2(M_2+\mu\tan\beta)}{M_2^2-\mu^2} \,\It_{\lt\Wt}^{11} -\frac{\mu(\mu+M_2\tan\beta)}{M_2^2-\mu^2} \,\It_{\lt\chit}^{11}\, \Bigr] \CR
&& -2\,g'^2\, Y_\phi \Bigl[ \frac{M_1(M_1+\mu\tan\beta)}{M_1^2-\mu^2} \bigl(Y_l\,\It_{\lt\Bt}^{11}-Y_e\,\It_{\et\Bt}^{11}\bigr) \CR
&&\qquad\qquad\, -\frac{\mu(\mu+M_1\tan\beta)}{M_1^2-\mu^2} \bigl(Y_l\,\It_{\lt\chit}^{11} -Y_e\,\It_{\et\chit}^{11}\bigr) \Bigr] \,.
\eeqan
\eseq{dy-ftchitVt}
We have used Eq.~\eqref{MIred-nc1-0} to reduce $(4-\epsilon)\,\It[q^2]_{\Phi0}^{12}$ to $\,\It_{\Phi0}^{11}$ in Eq.~\eqref{dy-Phiff}, and Eqs.~\eqref{MIred-h} and~\eqref{MIred-nc1} to reduce $\,\It_{\ft\Vt\chit}^{111}$ and $(4-\epsilon)\,\It[q^2]_{\ft\chit\Vt}^{111}$ to $\,\It_{\ft\chit}^{11}$ and $\,\It_{\ft\Vt}^{11}$ in Eq.~\eqref{dy-ftchitVt}. 

In these results, of particular interest is the appearance of terms proportional to $\tan\beta$, originating from $\mat{\lambda_{d,e}}=\mat{y_{d,e}}\cb^{-1}=\mat{y_{d,e}}\sb^{-1}\tan\beta$. Since matching calculations are done with UV theory parameters, it is expected here that $\mat{\delta y_{d,e}}$ contain terms of order $\lfr\,\mat{\lambda_{d,e}}\propto\frac{\tan\beta}{16\pi^2}\,\mat{y_{d,e}}$. A large $\tan\beta$ can partially overcome the loop suppression, giving rise to sizable SUSY threshold corrections, which in turn is important for achieving $b$-$\tau$ Yukawa unification. More on this in Section~\ref{sec:YU}.

\paragraph{Higgs kinetic term.}

The one-loop coefficient of the $d=4$ Higgs kinetic term $|D_\mu\phi|^2$ is
\beq
\delta Z_\phi = \delta Z_\phi^{(\ft\ft)} +\delta Z_\phi^{(\chit\Vt)} \,,
\eeqn
where
\beqa
16\pi^2\,\delta Z_\phi^{(\ft\ft)} &=& 
-2\,N_c\,\tr(\mat{\lambda_u^\dagger \lambda_u})\,\sb^2\, (A_u-\mu\cot\beta)^2 \,\It[q^2]_{\qt\ut}^{22} \CR
&& -2\,N_c\,\tr(\mat{\lambda_d^\dagger \lambda_d})\,\sb^2\, (A_d\cot\beta-\mu)^2 \,\It[q^2]_{\qt\dt}^{22} 
\CR && 
-2\,\tr(\mat{\lambda_e^\dagger \lambda_e})\,\sb^2\, (A_e\cot\beta-\mu)^2 \,\It[q^2]_{\lt\et}^{22} \,,\\[4pt]
16\pi^2\,\delta Z_\phi^{(\chit\Vt)} &=& 
2\,g^2\, \CasL \Bigl[\,\It_{\chit\Wt}^{11} +2(M_2^2+\mu^2+2M_2\mu\sbb)\,\It[q^2]_{\chit\Wt}^{22} \Bigr] 
\CR && 
+2\,g'^2\, Y_\phi^2\Bigl[\,\It_{\chit\Bt}^{11} +2(M_1^2+\mu^2+2M_1\mu\sbb)\,\It[q^2]_{\chit\Bt}^{22} \Bigr]\,.\label{dZphi-chitVt}
\eeqan
Again, we have used Eq.~\eqref{MIred-nc} to eliminate $\It[q^4]$ in order to arrive at Eq.~\eqref{dZphi-chitVt}. Note that $D_\mu\phi_\wi{\alpha}$ is written as $-i\,[P_\mu,\phi_\wi{\alpha}]$ in our approach (recall $P_\mu$ acts on everything to its right). The covariant diagrams listed in Table~\ref{tab:cd-P} give us the $\tr(P^\mu\phi^*P_\mu\phi)$ piece of $|D_\mu\phi|^2=-\tr([P^\mu,\phi^*][P_\mu,\phi])=\tr(P^2\phi^*\phi)+\tr(P^2\phi\phi^*)-2\,\tr(P^\mu\phi^*P_\mu\phi)$, which is sufficient to fix the coefficient of $|D_\mu\phi|^2$.

Note that unlike $\mat{\delta y_{d,e}}\sim\lfr\mat{\lambda_{d,e}}\sim\frac{\tan\beta}{16\pi^2}\mat{y_{d,e}}$, contributions to the threshold corrections $\mat{y_{d,e}} -\mat{y_{d,e}^\text{eff}}$ from $\delta Z_\phi$ (and also $\mat{\delta Z_f}$ below) are only $\sim \lfr\mat{y_{d,e}}$ (see Eq.~\eqref{thrcor}), and are thus subleading in the large $\tan\beta$ limit.

\paragraph{Fermion kinetic terms.}

The $d=4$ fermion kinetic terms $\sum_{f} \bar\psi_f\,\mat{\delta Z_f}\, i\slashed{D}\psi_f$ are obtained with the following one-loop coefficients,
\bseq
\beqa
\mat{\delta Z_q} &=& \mat{\delta Z_q^{(\Phi f)}} +\mat{\delta Z_q^{(\ft\chit)}} +\delta Z_q^{(\qt\Vt)} \,\identity \,,\\
\mat{\delta Z_u} &=& \mat{\delta Z_u^{(\Phi q)}} +\mat{\delta Z_u^{(\qt\chit)}} +\delta Z_u^{(\ut\Vt)} \,\identity \,,\\
\mat{\delta Z_d} &=& \mat{\delta Z_d^{(\Phi q)}} +\mat{\delta Z_d^{(\qt\chit)}} +\delta Z_d^{(\dt\Vt)} \,\identity \,,\\
\mat{\delta Z_l} &=& \mat{\delta Z_l^{(\Phi e)}} +\mat{\delta Z_l^{(\et\chit)}} +\delta Z_l^{(\lt\Vt)} \,\identity \,,\\
\mat{\delta Z_e} &=& \mat{\delta Z_e^{(\Phi l)}} +\mat{\delta Z_e^{(\lt\chit)}} +\delta Z_e^{(\et\Vt)} \,\identity \,,
\eeqan
\eseqn
where
\bseq
\beqa
16\pi^2\, \mat{\delta Z_q^{(\Phi f)}} &=& 2\,\bigl(\mat{\lambda_u^\dagger \lambda_u}\cb^2+\mat{\lambda_d^\dagger \lambda_d}\sb^2\bigr)\,\It[q^2]_{\Phi0}^{21}  \,,\\
16\pi^2\, \mat{\delta Z_u^{(\Phi q)}} &=& 4\,\mat{\lambda_u \lambda_u^\dagger}\cb^2\,\It[q^2]_{\Phi0}^{21} \,,\\
16\pi^2\, \mat{\delta Z_d^{(\Phi q)}} &=& 4\,\mat{\lambda_d \lambda_d^\dagger}\sb^2\,\It[q^2]_{\Phi0}^{21} \,,\\
16\pi^2\, \mat{\delta Z_l^{(\Phi e)}} &=& 2\, \mat{\lambda_e^\dagger \lambda_e}\sb^2\,\It[q^2]_{\Phi0}^{21} \,,\\
16\pi^2\, \mat{\delta Z_e^{(\Phi l)}} &=& 4\, \mat{\lambda_e \lambda_e^\dagger}\sb^2\,\It[q^2]_{\Phi0}^{21} \,,
\eeqan
\eseq{dZf-Phif}
\vspace{-20pt}
\bseq
\beqa
16\pi^2\, \mat{\delta Z_q^{(\ft\chit)}} &=& 2\,\bigl(\mat{\lambda_u^\dagger \lambda_u}\,\It[q^2]_{\ut\chit}^{21} +\mat{\lambda_d^\dagger \lambda_d}\,\It[q^2]_{\dt\chit}^{21}\bigr) \,,\\
16\pi^2\, \mat{\delta Z_u^{(\qt\chit)}} &=& 4\,\mat{\lambda_u \lambda_u^\dagger}\,\It[q^2]_{\qt\chit}^{21} \,,\\
16\pi^2\, \mat{\delta Z_d^{(\qt\chit)}} &=& 4\,\mat{\lambda_d \lambda_d^\dagger}\,\It[q^2]_{\qt\chit}^{21} \,,\\
16\pi^2\, \mat{\delta Z_l^{(\et\chit)}} &=& 2\,\mat{\lambda_e^\dagger \lambda_e}\,\It[q^2]_{\et\chit}^{21} \,,\\
16\pi^2\, \mat{\delta Z_e^{(\lt\chit)}} &=& 4\,\mat{\lambda_e \lambda_e^\dagger}\,\It[q^2]_{\lt\chit}^{21} \,,
\eeqan
\eseq{dZf-ftchit}
\vspace{-20pt}
\bseq
\beqa
16\pi^2\, \delta Z_q^{(\qt\Vt)} &=& 4\,\bigl( g_3^2\,\CasC\,\It[q^2]_{\qt\gt}^{21} +g^2\,\CasL\,\It[q^2]_{\qt\Wt}^{21} +g'^2\,Y_q^2\,\It[q^2]_{\qt\Bt}^{21} \bigr) \,,\\
16\pi^2\, \delta Z_u^{(\ut\Vt)} &=& 4\,\bigl( g_3^2\,\CasC\,\It[q^2]_{\ut\gt}^{21} +g'^2\,Y_u^2\,\It[q^2]_{\ut\Bt}^{21} \bigr) \,,\\
16\pi^2\, \delta Z_d^{(\dt\Vt)} &=& 4\,\bigl( g_3^2\,\CasC\,\It[q^2]_{\dt\gt}^{21} +g'^2\,Y_d^2\,\It[q^2]_{\dt\Bt}^{21} \bigr) \,,\\
16\pi^2\, \delta Z_l^{(\lt\Vt)} &=& 4\,\bigl( g_2^2\,\CasL\,\It[q^2]_{\lt\Wt}^{21} +g'^2\,Y_l^2\,\It[q^2]_{\lt\Bt}^{21} \bigr) \,,\\
16\pi^2\, \delta Z_e^{(\et\Vt)} &=& 4\,g'^2\,Y_e^2\,\It[q^2]_{\et\Bt}^{21} \,.
\eeqan
\eseq{dZf-ftVt}
To arrive at Eqs.~\eqref{dZf-ftchit} and~\eqref{dZf-ftVt}, we have used Eqs.~\eqref{MIred-h}, \eqref{MIred-d} and~\eqref{MIred-nc1} to simplify
%
\beqa
&& (2-\epsilon)\,\I[q^2]_{ij}^{12} -M_j^2\,\I_{ij}^{12} 
=(4-\epsilon)\,\I[q^2]_{ij}^{12} -M_j^2\,\I_{ij}^{12} -2\,\I[q^2]_{ij}^{12} 
= \I_{ij}^{11} -2\,\I[q^2]_{ij}^{12}
\CR &&
= \frac{1}{M_i^2-M_j^2} \bigl( \I_i^1 -\,\I_j^1 \bigr) -2\,\I[q^2]_{ij}^{12} 
= \frac{2}{M_i^2-M_j^2} \bigl( \I[q^2]_i^2 -\,\I[q^2]_j^2 \bigr) -2\,\I[q^2]_{ij}^{12} 
\CR && 
= 2\bigl( \I[q^2]_{ij}^{21} +\, \I[q^2]_{ij}^{12} \bigr) -2\,\I[q^2]_{ij}^{12}
= 2\, \I[q^2]_{ij}^{21} \,.
\eeqan
This relation is also valid in the limit $M_j\to 0$,
\beq
(2-\epsilon)\,\I[q^2]_{i0}^{12} = 2\, \I[q^2]_{i0}^{21} \,,
\eeqn
which we have used to obtain Eq.~\eqref{dZf-Phif}.

\paragraph{Gauge boson kinetic terms.}

General results of wavefunction renormalization of gauge fields from integrating out heavy matter fields are well-known, see e.g.~\cite{HLM14}. The covariant diagrams version of the calculation can be found in~\cite{CovDiag}. Specializing to the case of integrating out the MSSM heavy fields, we find
\bseq
\beqa
\delta Z_G &=& g_3^2\bigl( \bar\delta Z_G^{(\ft)} +\bar\delta Z_G^{(\gt)}\bigr) \,,\label{dZG}\\
\delta Z_W &=& g^2\bigl(\bar\delta Z_W^{(\Phi)} +\bar\delta Z_W^{(\ft)} +\bar\delta Z_W^{(\chit)} +\bar\delta Z_W^{(\Wt)} \bigr) \,,\\
\delta Z_B &=& g'^2\bigl(\bar\delta Z_B^{(\Phi)} +\bar\delta Z_B^{(\ft)} +\bar\delta Z_B^{(\chit)} \bigr)\,,
\eeqan
\eseqn
where
\beqa
&& 16\pi^2\,\bar\delta Z_W^{(\Phi)} = \frac{1}{6}\,\It_{\Phi}^2 \,, \qquad
16\pi^2\,\bar\delta Z_B^{(\Phi)} = \frac{2}{3}\,Y_\phi^2\,\It_{\Phi}^2 \,,\\[4pt]
&& 16\pi^2\,\bar\delta Z_G^{(\ft)} = \frac{1}{6}\bigl( 2\,\It_{\qt}^2 +\,\It_{\ut}^2 +\,\It_{\dt}^2 \bigr) \,, \qquad
16\pi^2\,\bar\delta Z_W^{(\ft)} = \frac{1}{6}\bigl( N_c\,\It_{\qt}^2 +\,\It_{\lt}^2 \bigr) \,, \CR
&& 16\pi^2\,\delta Z_B^{(\ft)} = \frac{1}{3}\bigl( 2N_c\,Y_q^2\,\It_{\qt}^2 +N_c\,Y_u^2\,\It_{\ut}^2 +N_c\,Y_d^2\,\It_{\dt}^2 +2\,Y_l^2\,\It_{\lt}^2 +Y_e^2\,\It_{\et}^2 \bigr) \,,\\[4pt]
&& 16\pi^2\,\bar\delta Z_W^{(\chit)} = \frac{2}{3}\,\It_{\chit}^2 \,, \qquad
16\pi^2\,\bar\delta Z_B^{(\chit)} = \frac{8}{3}\,Y_\phi^2\,\It_{\chit}^2 \,,\\[4pt]
&& 16\pi^2\,\bar\delta Z_G^{(\gt)} = 2\,\It_{\gt}^2 \,,\qquad
16\pi^2\,\bar\delta Z_W^{(\Wt)} = \frac{4}{3}\,\It_{\Wt}^2 \,.
\eeqan
We have used Eq.~\eqref{MIred-d} to reduce the bosonic loop integral $\,\I[q^4]_i^4$ to $\frac{1}{24}\,\I_i^2$. The fermionic loops, on the other hand, are proportional to $-M_i^4\,\I_i^4 +8M_i^2\,\I[q^2]_i^4 +(-16+10\epsilon)\,\I[q^4]_i^4$ which, by Eq.~\eqref{MIred-nc-d}, is equal to $8\,\I[q^4]_i^4-\,\I_i^2 =-\frac{2}{3}\,\I_i^2$.

\subsection{One-loop matching: $d=6$ operators $\O_{d\phi,e\phi}$ in the large $\tan\beta$, low $M_\Phi$ limit}
\label{sec:MSSMmatch-dsix}

We can use the same techniques to obtain one-loop-generated $d=6$ operators. There is a large number of them, but not all are equally interesting phenomenologically. In fact, given the loop suppression, together with a possibly high superpartner mass scale $\Lambda$ due to lack of new particle discoveries as well as a SM-like Higgs boson mass of $m_h\simeq125\,\text{GeV}$, a generic $d=6$ operator with $\O(\lfr\frac{1}{\Lambda^2})$ coefficient is likely to have a negligible effect on observables. In this regard, we would like to identify a region of MSSM parameter space where some $d=6$ operators have parametrically enhanced observable effects, and can thus point to realistic experimental targets to be pursued.

To do so, we first note that, as in the case of $\mat{\bar\delta y_{d,e}}$ discussed in the previous subsection, factors of $\tan\beta$ can appear when operator coefficients are written in terms of $\mat{y_{d,e}}$ rather than $\mat{\lambda_{d,e}}$, which can partially overcome the loop suppression if $\tan\beta\gg1$. We are thus led to consider the large $\tan\beta$ limit. At dimension-six level, $\tan\beta$ enhancement occurs for several operators, among which we focus on $\O_{d\phi}$ and $\O_{e\phi}$, motivated by their relevance to precision Higgs physics as they modify $hb\bar b$ and $h\tau^+\tau^-$ couplings; see Eq.~\eqref{dkf}. Note that in contrast, $\O_{u\phi}$, which modifies $ht\bar t$ coupling, does not have a $\tan\beta$ enhanced effect.

To further boost observable effects of the operators $\O_{d\phi}$ and $\O_{e\phi}$, we would like to focus on the scenario where $M_\Phi$, the mass of the heavier Higgs doublet, is somewhat lower than $\Lambda$. In this case, there are contributions to $C_{d\phi,e\phi}$ that are proportional to $\frac{1}{M_\Phi^2}$, which is parametrically larger compared to $\frac{1}{\Lambda^2}$. There are in principle two sources of such contributions --- loops involving $\Phi$ propagators, and operators proportional to $\Phi_\c$. By carefully enumerating covariant diagrams following the procedure of the previous subsection, we are able to show that loops involving $\Phi$ propagators are all free from $\tan\beta$ enhancement, and so will not consider them further.

As for the second option, there are only a few possibilities for writing down $d=6$ operators that are proportional to $\Phi_\c$, since $\Phi_\c^{(1)}$ ($\Phi_\c^{(2)}$) is already dimension three (five). They are, schematically,
\beq
(\Phi_\c^{(1)})^2 \,,\quad \Phi_\c^{(1)}\psi^2 \,,\quad \Phi_\c^{(1)}\phi^3\,,\quad \Phi_\c^{(1)}\phi P^2 \,,\quad \Phi_\c^{(2)}\phi \,.
\eeqn
Among them, $(\Phi_\c^{(1)})^2$ and $\Phi_\c^{(1)}\psi^2$ do not contain $\O_{d\phi,e\phi}$ with $\tan\beta$ enhanced coefficients, while $\Phi_\c^{(2)}\phi$ has already been absorbed into $\L_\text{SMEFT}^\text{tree}$ via the redefinition of $\beta$ discussed before. So we are left with $\Phi_\c^{(1)}\phi^3$ and $\Phi_\c^{(1)}\phi P^2$. To be explicit, we have
\beqa
\L_\text{SMEFT}^\text{1-loop} &\supset& c_{\Phi\phi^3} |\phi|^2 \bigl(\Phi_\c^{(1)*}\phi+\phi^*\Phi_\c^{(1)}\bigr) +c_{\Phi\phi P^2} \bigl[(D^\mu\Phi_\c^{(1)})^*(D_\mu\phi) +(D^\mu\phi)^*(D_\mu\Phi_\c^{(1)})\bigr] \CR[2pt]
&\overset{\text{IBP}}{=}& c_{\Phi\phi^3} |\phi|^2 \bigl(\Phi_\c^{(1)*}\phi+\phi^*\Phi_\c^{(1)}\bigr) -c_{\Phi\phi P^2} \bigl[\Phi_\c^{(1)*}(D^2\phi) +(D^2\phi)^*\Phi_\c^{(1)}\bigr] \CR[2pt]
&\overset{\text{EoM}}{=}& \bigl(c_{\Phi\phi^3}+2\lambda\,c_{\Phi\phi P^2}\bigr)\, |\phi|^2 \bigl(\Phi_\c^{(1)*}\phi+\phi^*\Phi_\c^{(1)}\bigr) +\dots \CR[2pt]
&\supset& \frac{\tan\beta}{M_\Phi^2}\bigl(c_{\Phi\phi^3}+2\lambda\,c_{\Phi\phi P^2}\bigr) \Bigl(\bigl[\mat{y_d^\dagger}\bigr]_{pr}\bigl[\O_{d\phi}\bigr]^{pr} +\bigl[\mat{y_e^\dagger}\bigr]_{pr}\bigl[\O_{e\phi}\bigr]^{pr}\Bigr) \,.
\eeqa{Ldimsix}
Note that there is also a tree-level matching contribution to $\Phi_\c^{(1)}\phi^3$, which we already computed in Section~\ref{sec:MSSMmatch-tree}. Though $\overline{\text{DR}}$ scheme was assumed there, the one-loop difference between $\overline{\text{MS}}$ and $\overline{\text{DR}}$ is not $\tan\beta$ enhanced and negligible.

The operator coefficients $c_{\Phi\phi^3}$ and $c_{\Phi\phi P^2}$ can be computed from the same covariant diagrams that give rise to $\delta\lambda$ and $\delta Z_\phi$, respectively. In fact, we just need to retrieve $\O(\Phi\phi^3)$ and $\O(\Phi\phi)$ pieces from products of $\bf U$ matrix elements, instead of $\O(\phi^4)$ and $\O(\phi^2)$ pieces. From Appendix~\ref{app:Umatrix} we see that, with the exception of diagrams involving $U_{\varphi\varphi}$, this amounts to starting from the latter, and replacing $\sb\phi\to\cb\Phi$, $\cb\phi\to -\sb\Phi$ in all possible ways. In other words, from the form of the $\bf U$ matrix we can infer that
%
\beqa
\L_\text{SMEFT} &\supset& \delta\lambda\, |\phi|^4  
+\frac{1}{2} \Bigl(\frac{\partial}{\partial\beta} \delta\lambda\Bigr) |\phi|^2 \bigl(\Phi_\c^{*}\phi+\phi^*\Phi_\c\bigr) \CR
&& +\delta Z_\phi\, |D_\mu\phi|^2
+\frac{1}{2} \Bigl(\frac{\partial}{\partial\beta} \delta Z_\phi\Bigr)\bigl(D^\mu\Phi_\c^{*}D_\mu\phi+D^\mu\phi^*D_\mu\Phi_\c\bigr) \,,
\eeqa{Ld6Phi}
up to loops involving $\Phi$ propagators. We have verified Eq.~\eqref{Ld6Phi} by explicit calculation.

The simple replacement rule observed above, which connects different operators involving $\phi$ and $\Phi_\c$, can be understood by considering a variation of the EFT matching problem we are dealing with now. Suppose, instead of integrating out all BSM fields of the MSSM, we integrate out only the $R$-parity-odd fields, while keeping both Higgs doublets in the low-energy EFT. The calculation in this case would be more conveniently done in the $(H_u, H_d)$ basis, and the angle $\beta$ does not appear in the effective Lagrangian in the electroweak symmetric phase. Afterward, we can substitute
\beq
H_u\to\sb\phi+\cb\Phi\,,\qquad H_d\to \epsilon\cdot(\cb\phi-\sb\Phi)^* \,,
\eeq{HuHd}
so as to write the effective Lagrangian in terms of $\phi$ and $\Phi$. From Eq.~\eqref{HuHd} it is clear that for each term involving $\sb\phi$ ($\cb\phi$), if we replace $\sb\phi\to\cb\Phi$ ($\cb\phi\to -\sb\Phi$), the result would also be a term in the effective Lagrangian. Further integrating out $\Phi$ to arrive at the SMEFT does not change the conclusion for the terms that already existed, namely those generated by integrating out $R$-parity-odd fields. Meanwhile, additional terms, such as $\delta\lambda_{(\Phi\varphi)}|\phi|^4$ (see Eq.~\eqref{dlambdaPhi}), are generated by loops involving $\Phi$, for which the simple replacement rule above does not apply. However, none of these terms is $\tan\beta$ enhanced, and we will thus neglect them in the $d=6$ part of the EFT Lagrangian.

To sum up, in the limit $\tan\beta\gg1$, $M_\Phi\lesssim\Lambda$, we have
\beq
\mat{C_{d\phi}^\text{1-loop}} \simeq \frac{\tan\beta}{M_\Phi^2}\bigl(c_{\Phi\phi^3}+2\lambda\,c_{\Phi\phi P^2}\bigr) \,\mat{y_d^\dagger} \,,\qquad
\mat{C_{e\phi}^\text{1-loop}} \simeq \frac{\tan\beta}{M_\Phi^2}\bigl(c_{\Phi\phi^3}+2\lambda\,c_{\Phi\phi P^2}\bigr) \,\mat{y_e^\dagger} \,,
\eeqn
where
\beqa
c_{\Phi\phi^3} &\simeq& c_{\Phi\phi^3}^{(\ft\ft)} +c_{\Phi\phi^3}^{(\ft\ft\ft)} +c_{\Phi\phi^3}^{(\ft\ft\ft\ft)} +c_{\Phi\phi^3}^{(\chit\Vt\chit\Vt)} \CR
&=& \frac{1}{2} \frac{\partial}{\partial\beta} \bigl(\delta\lambda^{(\ft\ft)} +\delta\lambda^{(\ft\ft\ft)} +\delta\lambda^{(\ft\ft\ft\ft)} +\delta\lambda^{(\chit\Vt\chit\Vt)} \bigr) \,, \\[4pt]
c_{\Phi\phi P^2} &\simeq& c_{\Phi\phi P^2}^{(\ft\ft)} +c_{\Phi\phi P^2}^{(\chit\Vt)} = \frac{1}{2} \frac{\partial}{\partial\beta} \,\delta Z_\phi \,,
\eeqan
with various contributions to $\delta\lambda$ and $\delta Z_\phi$ computed in the previous subsection.

\section{Bottom-tau Yukawa unification}
\label{sec:YU}

In this section, we study implications of $b$-$\tau$ Yukawa unification on the SUSY spectrum in the EFT framework. To simplify the analyses, we neglect Yukawa couplings of the first two generation fermions, and impose the following relations among MSSM parameters,
\beqa
&&
M_{\qt} = M_{\ut} = M_{\dt} = M_{\lt} = M_{\et} \equiv M_s \,,\\
&&
A_u = A_d = A_e \equiv A_t \,,\\
&&
M_{\gt} = 3M_{\Wt} = 6M_{\Bt} \equiv M_3 \,.
\eeqan
As a result, above the SUSY threshold $\Qth$, we have a theory of 13 parameters:
\bseq
\beqa
&& g'\,,\; g\,,\; g_3\,,\; \lambda_t\,,\; \lambda_b\,,\; \lambda_\tau\,,\; m^2\,,\; A_t\,,\label{gMSSMelim}\\
&& M_\Phi\,,\; M_s\,,\; \mu\,,\; M_3\,,\; \tan\beta\,.
\eeqa{gMSSMin}
\eseq{MSSMparam}
Below the SUSY threshold $\Qth$, they are mapped onto parameters in the SMEFT, as we have worked out in detail in Section~\ref{sec:MSSMmatch}. We shall keep only the renormalizable operators and dimension-six ones that are generated at tree level. The EFT is therefore a theory characterized by 20 parameters:
\bseq
\beqa
&& {g'}^\text{eff}\,,\; g^\text{eff}\,,\; g_3^\text{eff}\,,\; y_t^\text{eff}\,,\; y_b^\text{eff}\,,\; y_\tau^\text{eff}\,,\; \lambda_\text{eff}\,,\; m^2_\text{eff} \,,\label{geff}\\
&& C_\phi\,,\; C_{t\phi}\,,\; C_{b\phi}\,,\; C_{\tau\phi}\,,\; C_{qu}^{(1)}\,,\; C_{qu}^{(8)}\,,\; C_{qd}^{(1)}\,,\; C_{qu}^{(8)}\,,\; C_{le}\,,\; C_{quqd}^{(1)}\,,\; C_{lequ}^{(1)}\,,\; C_{ledq} \,.
\eeqan
\eseq{SMEFTparam}
It is implicit here that all generation indices are set to 3 in the four-fermion operator coefficients.

We numerically evolve the 13 parameters in Eq.~\eqref{MSSMparam} in the regime $Q>\Qth$ according to two-loop RG equations of the MSSM~\cite{MartinVaughnRGE}, and the 20 parameters in Eq.~\eqref{SMEFTparam} in the regime $Q<\Qth$ according to two-loop RG equations of the renormalizable SM~\cite{criticality} and one-loop RG equations of the dimension-six SMEFT~\cite{EFTRGE1,EFTRGE2,EFTRGE3}. At $Q=\Qth$, the two sets of parameters are connected by the matching calculation presented in Section~\ref{sec:MSSMmatch}, together with one-loop scheme conversion between $\overline{\text{MS}}$ (used for RG evolution in the SMEFT) and $\overline{\text{DR}}$ (used for RG evolution in the MSSM)~\cite{MartinVaughnDRbar}.

As boundary conditions for the entire set of RG equations, we set
\beqa
&&
{g'}^\text{eff} = 0.35827\,,\quad
g^\text{eff} = 0.64779\,,\quad
g_3^\text{eff} = 1.1671\,,\CR
&&
y_t^\text{eff}-\frac{v^2}{2}C_{t\phi} = 0.93612\,,\quad
y_b^\text{eff}-\frac{v^2}{2}C_{b\phi} = 0.01539\,,\quad
y_\tau^\text{eff}-\frac{v^2}{2}C_{\tau\phi} = 0.00988\,,\CR
&&
\lambda_\text{eff}-\frac{3v^2}{2}C_\phi = 0.12592\,,\quad
m^2_\text{eff}+\frac{3v^4}{4}C_\phi = -(92.964\,\text{GeV})^2\,,
\eeqa{gSMEFTin}
at $Q=m_t=173.21\,\text{GeV}$, where $v^2=-m^2_\text{eff}/\lambda_\text{eff}$. These linear combinations of SMEFT parameters are what would be actually extracted when mapping the SM Lagrangian to low-energy observables, including $m_W=80.385\,\text{GeV}$, $m_h=125.09\,\text{GeV}$, $\alpha_s(m_Z)=0.1185$, etc. The numbers in Eq.~\eqref{gSMEFTin} are taken from~\cite{criticality}, except for $y_b^\text{eff}-\frac{v^2}{2}C_{b\phi}$, which is taken from~\cite{runningmb}, and $y_\tau^\text{eff}-\frac{v^2}{2}C_{\tau\phi}$, which is fixed by requiring $m_\tau^\text{pole}=1.77686\,\text{GeV}$ is reproduced when the SM is matched onto five-flavor QCD$\times$QED and RG evolved down to the low scale according to~\cite{ACKMPRW}.

The 8 boundary conditions in Eq.~\eqref{gSMEFTin} reduce the number of free parameters from 13 to 5. We choose them to be those in Eq.~\eqref{gMSSMin}. Thus, for any specific values of $M_\Phi$, $M_s$, $\mu$, $M_3$, $\tan\beta$, we can ask whether the entire set of equations admits a solution with all couplings in the perturbative regime, and if it does, whether $\lambda_b$ and $\lambda_\tau$ unify at the grand unification scale $\QGUT$. 

To be precise, we shall set the matching scale $\Qth=M_s$, and determine $\QGUT$ by $(5/3)^{1/2}g'(\QGUT)=g(\QGUT)$. We define ``$b$-$\tau$ Yukawa unification'' by $|\lambda_b(\QGUT)/ \lambda_\tau(\QGUT) -1|<0.02$ here, as it is generally difficult to have a larger GUT threshold correction~\cite{EHPR}. 

We further set $M_\Phi=M_s$ in this section for simplicity, since $M_\Phi$ does not play a significant role in $b$-$\tau$ Yukawa unification. For several choices of $\tan\beta=50,\,10,\,4,\,2$, we scan $M_s$ between $10^3\,\text{GeV}$ and $10^{10}\,\text{GeV}$, and scan $\mu$ and $|M_3|=-M_3$ within a factor of 50 from $M_s$, to search for solutions with $b$-$\tau$ Yukawa unification (no solution exists when $\mu M_3>0$, see below)\footnote{We can fix the signs of $\mu$ and $M_3$, keeping their relative sign, without loss of generality here, because the MSSM Lagrangian is invariant under simultaneous sign change of $\mu$, $M_{3,2,1}$ and $A_{u,d,e}$.}. We refrain from going beyond $M_s=10^{10}\,\text{GeV}$ for the present analysis, because additional GUT-scale input, namely gauge coupling threshold corrections, would be needed to precisely define $\QGUT$. Also, larger mass ratios are disallowed so as not to compromise the validity of our matching calculation, where all BSM fields are assumed to have similar masses and thus integrated out together.

\begin{figure}[tbp]
\centering
\includegraphics[width=\textwidth]{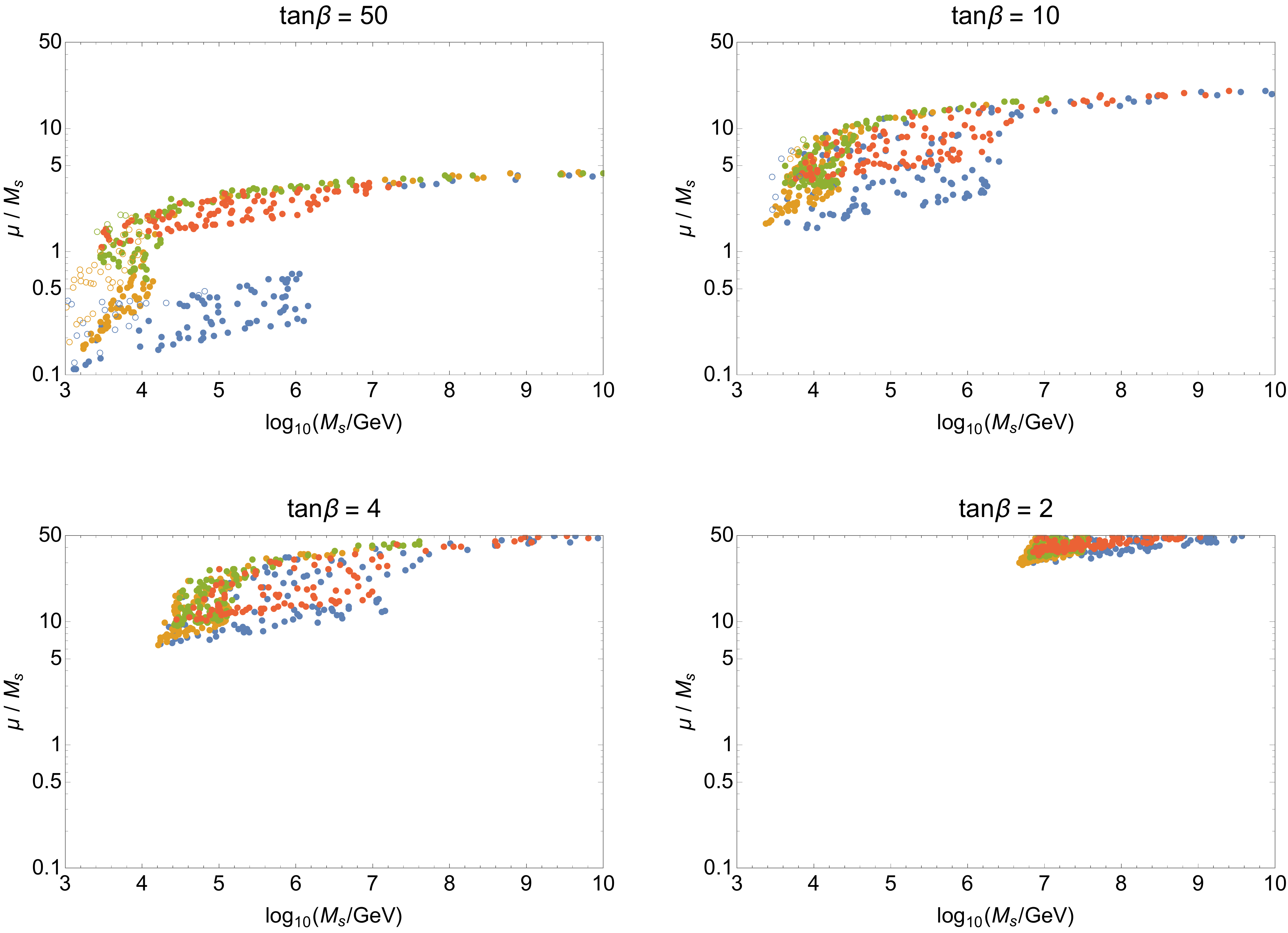}
\caption{\label{fig:HS-rmu}
Points in the MSSM parameter space that allow consistent matching onto the SMEFT and meanwhile realize $b$-$\tau$ Yukawa unification, projected onto $(\log M_s,\,\mu/M_s)$ plane, for several choices of $\tan\beta$. Blue, yellow, green, red points have $x_t\equiv(A_t-\mu\cot\beta)/M_s \,\in\, (-4,\,-\sqrt{6})\,,\; (-\sqrt{6},\,0)\,,\; (0,\,\sqrt{6})\,,\; (\sqrt{6},\,4)$, respectively. Empty circles represent solutions with a gluino lighter than 2\,TeV, potentially already in tension with direct LHC searches, depending on decay kinematics.}
\end{figure}

\begin{figure}[tbp]
\centering
\includegraphics[width=\textwidth]{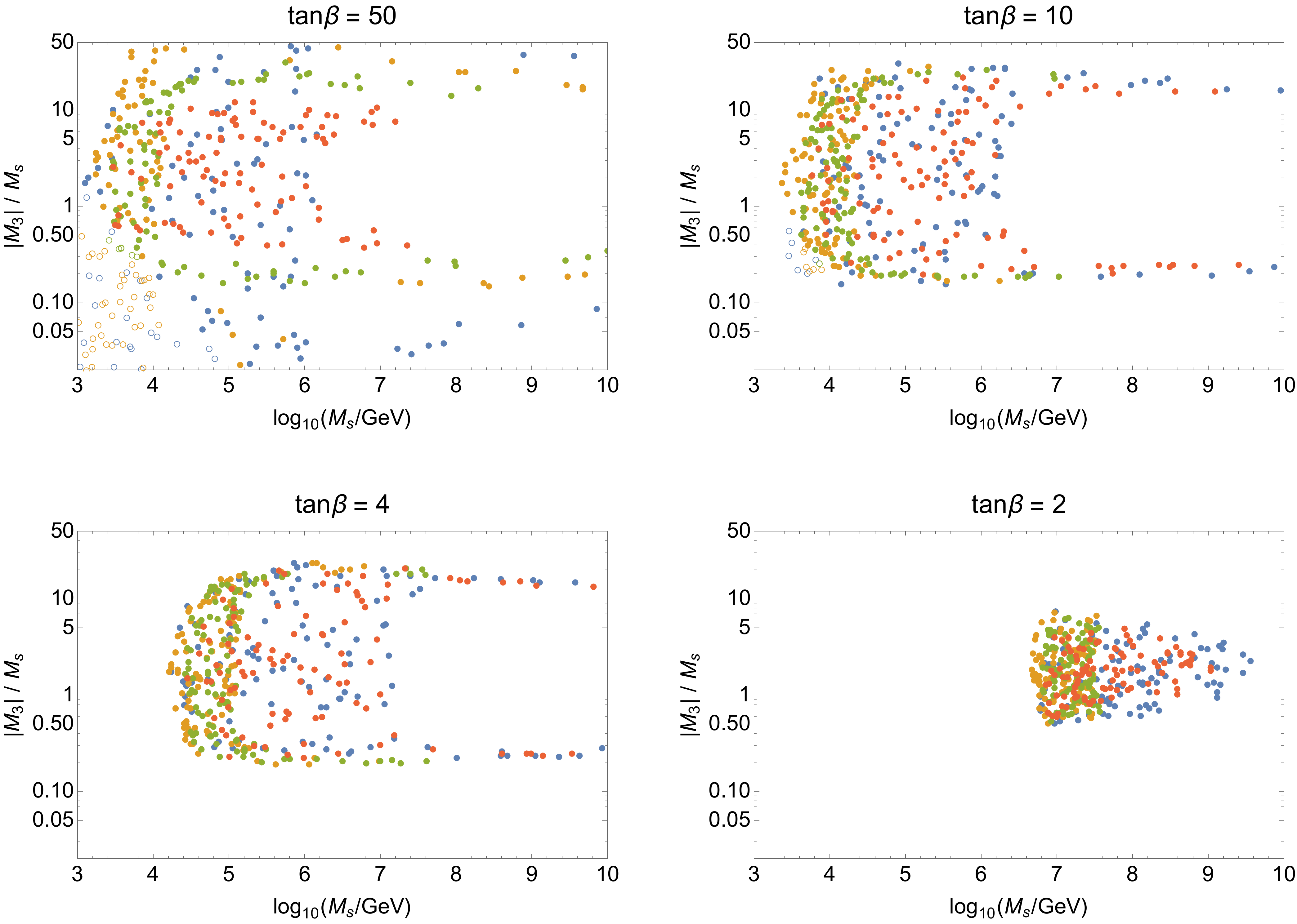}
\caption{\label{fig:HS-rM3}
Same as Figure~\ref{fig:HS-rmu}, now projected onto $(\log M_s,\,|M_3|/M_s)$ plane.}
\end{figure}

Figures~\ref{fig:HS-rmu} and~\ref{fig:HS-rM3} show points in the MSSM parameter space that allow consistent matching of the MSSM onto the SMEFT and meanwhile realize $b$-$\tau$ Yukawa unification, projected onto $(\log M_s,\,\mu/M_s)$ and $(\log M_s,\,|M_3|/M_s)$ planes, respectively. Different colors (blue, yellow, green, red) are used for solutions with $x_t\equiv(A_t-\mu\cot\beta)/M_s$ in different ranges ($-4<x_t<-\sqrt{6}$, $-\sqrt{6}<x_t<0$, $0<x_t<\sqrt{6}$, $\sqrt{6}<x_t<4$, respectively). We have quite conservatively considered a large interval $(-4,4)$ for $x_t$, keeping in mind the caveat that $x_t$ values past maximal mixing $\pm\sqrt{6}$ (blue and red dots) may run afoul of charge and color breaking vacuum constraints~\cite{CCB-CLM,CCB-Hollik}. In addition, points with $|M_3|<2\,\text{TeV}$, potentially already in tension with gluino searches at the LHC (depending on decay kinematics, see e.g.~\cite{SUSYsearch1711}), are represented by empty circles in all plots.

An immediate observation from these figures is that $b$-$\tau$ Yukawa unification is achievable for SUSY scales from TeV all the way up to (at least) $10^{10}\,\text{GeV}$, with suitable choices of mass ratios and $\tan\beta$. It is worth noting, though, that a large Higgsino mass $\mu>M_s$ is always required for $\tan\beta\lesssim10$, which may be less preferable from the point of view of fine-tuned electroweak symmetry breaking.

There are two issues that are key to understanding these results in more detail, which we now discuss in turn.

\begin{figure}[tbp]
\centering
\includegraphics[width=\textwidth]{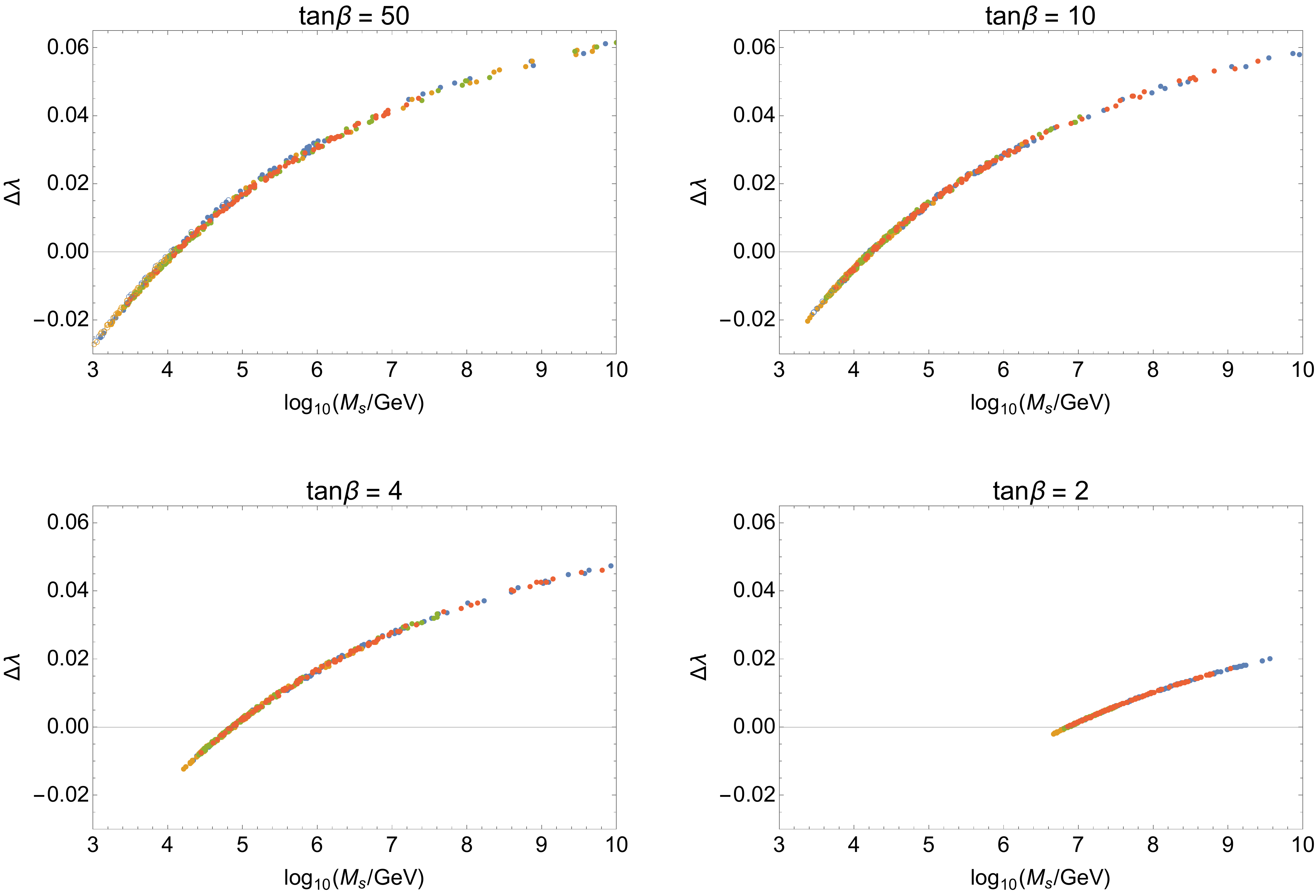}
\caption{\label{fig:HS-dl}
Same as Figure~\ref{fig:HS-rmu}, now showing SUSY threshold correction for the Higgs quartic coupling, defined as $\Delta\lambda \equiv \lambda-\lambda_\text{eff}$ at the matching scale $\Lambda=M_s$.}
\end{figure}

\begin{figure}[tbp]
\centering
\includegraphics[width=\textwidth]{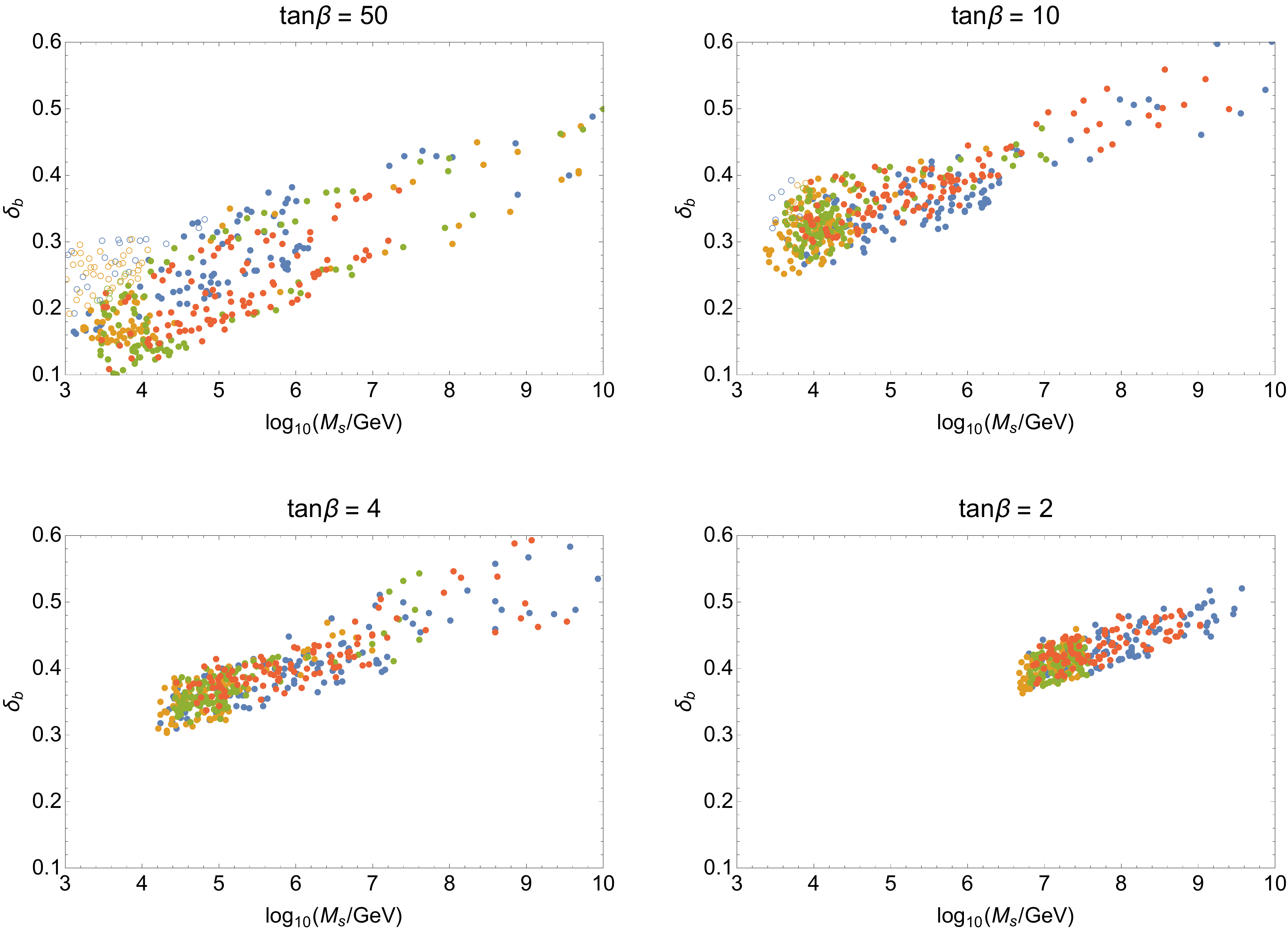}
\caption{\label{fig:HS-db}
Same as Figure~\ref{fig:HS-rmu}, now showing SUSY threshold correction for the bottom Yukawa coupling, defined as $\delta_b \equiv (y_b-y_b^\text{eff})/y_b^\text{eff}$ at the matching scale $\Lambda=M_s$.}
\end{figure}

\subsection{Matching of the Higgs quartic}

First of all, it should noted that it is not always possible to match the MSSM onto the SMEFT while satisfying the boundary conditions of Eq.~\eqref{gSMEFTin}, for arbitrary choices of SUSY parameters. This is largely due to the fact that the Higgs quartic coupling $\lambda$ is a derived quantity in the MSSM, given by $\frac{1}{8}(g^2+{g'}^2)\,\cbb^2$ at tree level. A threshold correction of just the right size is needed for $\lambda_\text{eff}$ to match the low-energy determination, most importantly from $m_h=125\,\text{GeV}$. 

To see this explicitly, we plot in Figure~\ref{fig:HS-dl} the value of
\beq
\Delta\lambda \equiv \lambda-\lambda_\text{eff} \,,
\eeqn
for each point in our sample of $b$-$\tau$ Yukawa unification solutions, evaluated at the matching scale $\Qth=M_s$. For the most part of parameter space, this threshold correction can be approximated by
\beqa
\Delta\lambda
&\simeq& \delta\lambda_{(\ft\ft\ft)} +\delta\lambda_{(\ft\ft\ft\ft)} \CR
&\simeq& \frac{N_c}{16\pi^2} \Bigl\{2\,y_t^4(A_t-\mu\cot\beta)^2\,\It_{\ft}^3 \CR
&&\qquad\quad +\frac{1}{2}\bigl[y_t^4(A_t-\mu\cot\beta)^4 +(y_b\tan\beta)^4(\mu-A_t\cot\beta)^4\bigr]\,\I_{\ft}^4 \Bigr\} \CR
&\simeq& \frac{N_c}{16\pi^2} \frac{1}{12} \Bigl[\Bigl(y_b\tan\beta\frac{\mu}{M_s}\Bigr)^4 +y_t^4\bigl((x_t^2-6)^2 -36\bigr) \Bigr] \,,
\eeqa{dlambda-approx}
see Eqs.~\eqref{dlambda-ft3} and~\eqref{dlambda-ft4}. The dependence on $x_t$ in Eq.~\eqref{dlambda-approx} explains the existence of up to four branches of solutions, separated by $x_t=-\sqrt{6}$, $0$ and $\sqrt{6}$.

Matching of the Higgs quartic essentially selects a range of $\mu/M_s$ for any given $M_s$, for which Eq.~\eqref{dlambda-approx} can possibly be of the right size with suitable choice of $x_t$. Since the required threshold correction increases logarithmically with the SUSY threshold scale, and is meanwhile insensitive to $\tan\beta$ when $\cbb^2\simeq1$, the range of $\mu/M_s$ being selected roughly scales as $\cot\beta\,(\log M_s)^{1/4}$ for $\tan\beta\gg1$. Of course, on each branch of $x_t$, part of this range can be excluded by either lack of $b$-$\tau$ Yukawa unification, or a mass ratio $\mu/M_s$ or $|M_3|/M_s$ outside of the interval $(1/50,\, 50)$. Nevertheless, the general trend of $\mu/M_s\sim\cot\beta\,(\log M_s)^{1/4}$ is still visible in Figure~\ref{fig:HS-rmu}.

Another feature of the figures is that the available parameter space is cut off at low $M_s$. Here the $\Delta\lambda$ needed becomes too small to be achievable by Eq.~\eqref{dlambda-approx}, which is bounded from below, while maintaining a large enough threshold correction for the bottom Yukawa (which is roughly proportional to $(\mu/M_s)\tan\beta$, see below). The issue is more severe at smaller $\tan\beta$ because of a smaller $\lambda\simeq\frac{1}{8}(g^2+{g'}^2)\,\cbb^2$ at any given $\Qth=M_s$. These conclusions are perhaps more familiar when phrased as ``raising the SM Higgs mass to 125\,GeV requires large one-loop corrections from heavy stops.'' Here, instead of computing $m_h$ from the full theory (the MSSM), we have taken an EFT approach, where $m_h$ is computed from the SM to fix $\lambda_\text{eff}$, and the problem becomes matching $\lambda_\text{eff}$ with $\lambda$ with the right amount of threshold correction. See also~\cite{EllisWells1706} for related discussion.

\subsection{Bottom Yukawa threshold correction}

Next, let us take a closer look at the SUSY threshold correction for the bottom Yukawa coupling, which is a key ingredient for $b$-$\tau$ Yukawa unification. Our discussion in what follows in this subsection is consistent with previous studies~\cite{HallRattazziSarid,Hempfling,CarenaOlechowskiPolorskiWagner,KaneKoldaRoszkowskiWells}.

In Figure~\ref{fig:HS-db} we plot
\beq
\delta_b \equiv \frac{y_b-y_b^\text{eff}}{y_b^\text{eff}}\,,
\eeq{dbdef}
evaluated at the matching scale $\Qth=M_s$, for our sample of $b$-$\tau$ Yukawa unification solutions. We see that they correspond to a specific range of $\delta_b$ for any given $M_s$, with numbers ranging from 10\% to 60\%.

At this point, it is worth emphasizing again that threshold corrections, which originate from renormalizable operators generated in EFT matching, do not decouple as the EFT cutoff is raised. In fact, as we see from Figures~\ref{fig:HS-dl} and~\ref{fig:HS-db}, for both the Higgs quartic and the bottom Yukawa, a higher $M_s$ calls for a larger SUSY threshold correction, in order to compensate for a longer period of running in the SMEFT. 

Returning to the issue of bottom Yukawa threshold correction, we note that 
for the most part of parameter space, $\delta_b$ is dominated by contribution from the squark-gluino loop,
\beq
\delta_b \simeq \delta_b^{(\qt\dt\gt)} \simeq \frac{g_3^2}{16\pi^2}\frac{y_b}{y_b^\text{eff}}\cdot 2\,\CasC\,\Bigl(\frac{\mu}{M_s}\tan\beta\Bigr) \Bigl(M_3 M_s\,\It_{\ft\gt}^{21}\Bigr) \,,
\eeq{db-approx}
see Eq.~\eqref{dyb-go}. Since $\,\It_{\ft\gt}^{21}$ is negative-definite, a positive $\delta_b$ is only possible when $\mu M_3<0$, which explains our sign choice. We have checked explicitly that no solutions can be found when the sign of either $\mu$ or $M_3$ is reversed.

The factor $(M_3 M_s\,\It_{\ft\gt}^{21})$ in Eq.~\eqref{db-approx} only depends on the mass ratio. It is approximately $-\frac{M_3}{M_s}$ when $|M_3|/M_s\ll1$, and $-\frac{M_s}{M_3}(\log\frac{M_3^2}{M_s^2}-1)$ when $|M_3|/M_s\gg1$, with a maximum absolute value of about 0.566 at $|M_3|/M_s\simeq2.12$. Thus, for any given value of $(\mu/M_s)\tan\beta$ that is sufficiently large, we expect to have two solutions for $|M_3|/M_s$ -- one on each side of 2.12 -- which lead to the same desired $\delta_b$ (up to higher-order corrections from e.g.\ gluino loop contribution to $g_3$ threshold correction). This degeneracy is clearly visible in Figure~\ref{fig:HS-rM3}, especially in the high $M_s$ regime of the first three plots, where the range of $\mu/M_s$, as determined by the Higgs quartic matching condition, is narrow due to the $x_t$-dependent terms in Eq.~\eqref{dlambda-approx} becoming subdominant. For the $\tan\beta=2$ plot, on the other hand, only a region near $|M_3|/M_s\simeq2.12$ survives because of a much smaller $(\mu/M_s)\tan\beta$ ($\sim100$ as opposed to $\sim200$ for the first three plots, as can be inferred from Figure~\ref{fig:HS-rmu}).

In addition to Eq.~\eqref{db-approx}, there is a subdominant contribution to $\delta_b$ from squark-Higgsino loop, which is responsible for some finer details of the plots. From Eq.~\eqref{dyb-ho} we have
\beq
\delta_b^{(\qt\ut\chit)} = \frac{\lambda_t^2}{16\pi^2}\frac{y_b}{y_b^\text{eff}}\cdot x_t\,\Bigl(\frac{\mu}{M_s}\tan\beta\Bigr) \Bigl(M_s^2\,\It_{\ft\chit}^{21}\Bigr) \,.
\eeq{db-ho}
Comparing Eqs.~\eqref{db-approx} and~\eqref{db-ho}, we see that $\delta_b^{(\qt\dt\gt)}$ and $\delta_b^{(\qt\ut\chit)}$ have opposite (same) signs when $x_t>0$ ($x_t<0$). Thus, higher values of $\mu/M_s$ are required for the $x_t>0$ branches (green and red) to compensate for the cancellation between $\delta_b^{(\qt\dt\gt)}$ and $\delta_b^{(\qt\ut\chit)}$, as we can see from Figure~\ref{fig:HS-rmu}.

\section{Higgs couplings in TeV-scale SUSY}
\label{sec:Higgs}

In the previous section, we have seen that $b$-$\tau$ Yukawa unification alone does not point to a unique scale for the masses of superpartners in the MSSM. However, if in addition, we would like the MSSM to provide a dark matter candidate in the form of the lightest neutralino, that would be further motivation for TeV-scale SUSY. For example, two classic thermal dark matter benchmarks are a $\sim$1\,TeV Higgsino LSP and a $\sim$2.7\,TeV wino LSP~\cite{MDM}. A wider range of masses is allowed if the LSP is a mixture of bino, wino and Higgsino states or if the sfermions do not decouple~\cite{Beneke14,Beneke16}, or if non-thermal production mechanisms are at work. Therefore, we will broadly consider the 1-10\,TeV regime for superpartner masses, while remaining agnostic about the detailed cosmology of dark matter. We will focus on precision Higgs coupling measurements as an indirect probe of TeV-scale SUSY, and discuss how they can be complementary to direct superpartner searches at the LHC.

To compute Higgs coupling modifications, we follow the same numerical procedure as outlined at the beginning of Section~\ref{sec:YU}. Now the 20 SMEFT parameters in Eq.~\eqref{SMEFTparam} should be evolved down to $Q=m_h=125.09\,\text{GeV}$, in order to compute $\delta\kappa_b$ and $\delta\kappa_\tau$ according to Eq.~\eqref{dkf}. As discussed in Section~\ref{sec:MSSMmatch-dsix}, we shall focus on the scenario where $M_\Phi$, the mass of the second Higgs doublet, is relatively low. To be precise, let us first fix $M_\Phi=1\,\text{TeV}$, and allow $M_s$ and $|M_3|$ to vary between $1\,\text{TeV}$ and $10\,\text{TeV}$. The Higgsino mass $\mu$ is determined by requiring exact $b$-$\tau$ Yukawa unification, i.e.\ $\lambda_b(\QGUT)=\lambda_\tau(\QGUT)$. Solutions may exist on multiple branches of $x_t$, in which case we find all of them.

\begin{figure}[tbp]
\centering
\includegraphics[width=\textwidth]{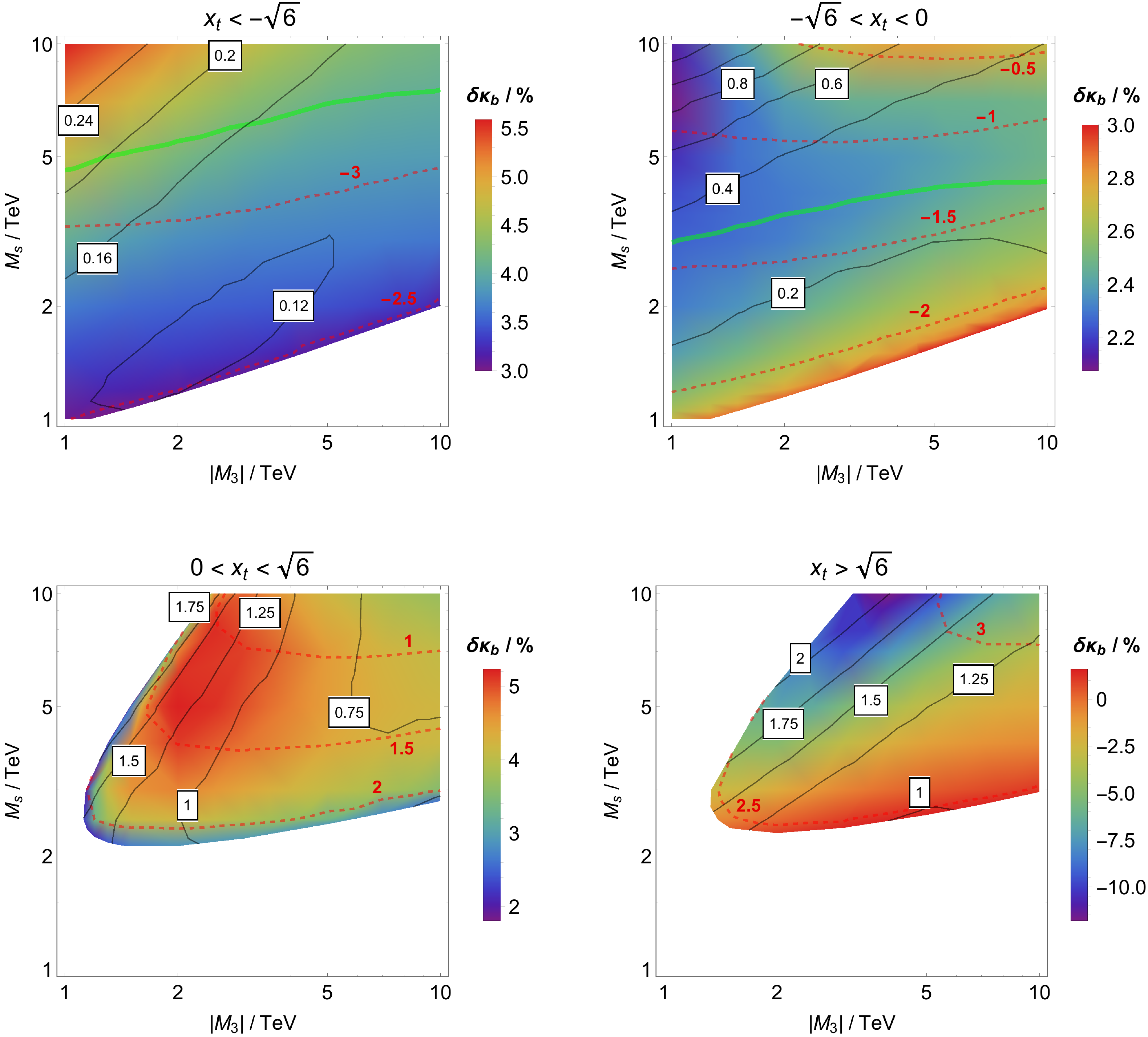}
\caption{\label{fig:TeV-tb50}
Variation of $\delta\kappa_b$ in the region of the $|M_3|$-$M_s$ plane where a solution exists for exact $b$-$\tau$ Yukawa unification, on each $x_t$ branch, with $M_\Phi=1$\,TeV and $\tan\beta=50$. Superimposed are contours of $\mu/M_s$ (black) and $x_t$ (red dashed). Light green curves in the $x_t<0$ plots correspond to the 1\,TeV Higgsino dark matter benchmark. Direct superpartner searches probe lower mass regions of the parameter space (with $|M_3|\lesssim2$\,TeV potentially already excluded at the LHC depending on decay kinematics), while precision Higgs measurements can be more sensitive to higher mass regions where $\delta\kappa_b$ is enhanced by one-loop corrections.}
\end{figure}

\begin{figure}[tbp]
\centering
\includegraphics[width=\textwidth]{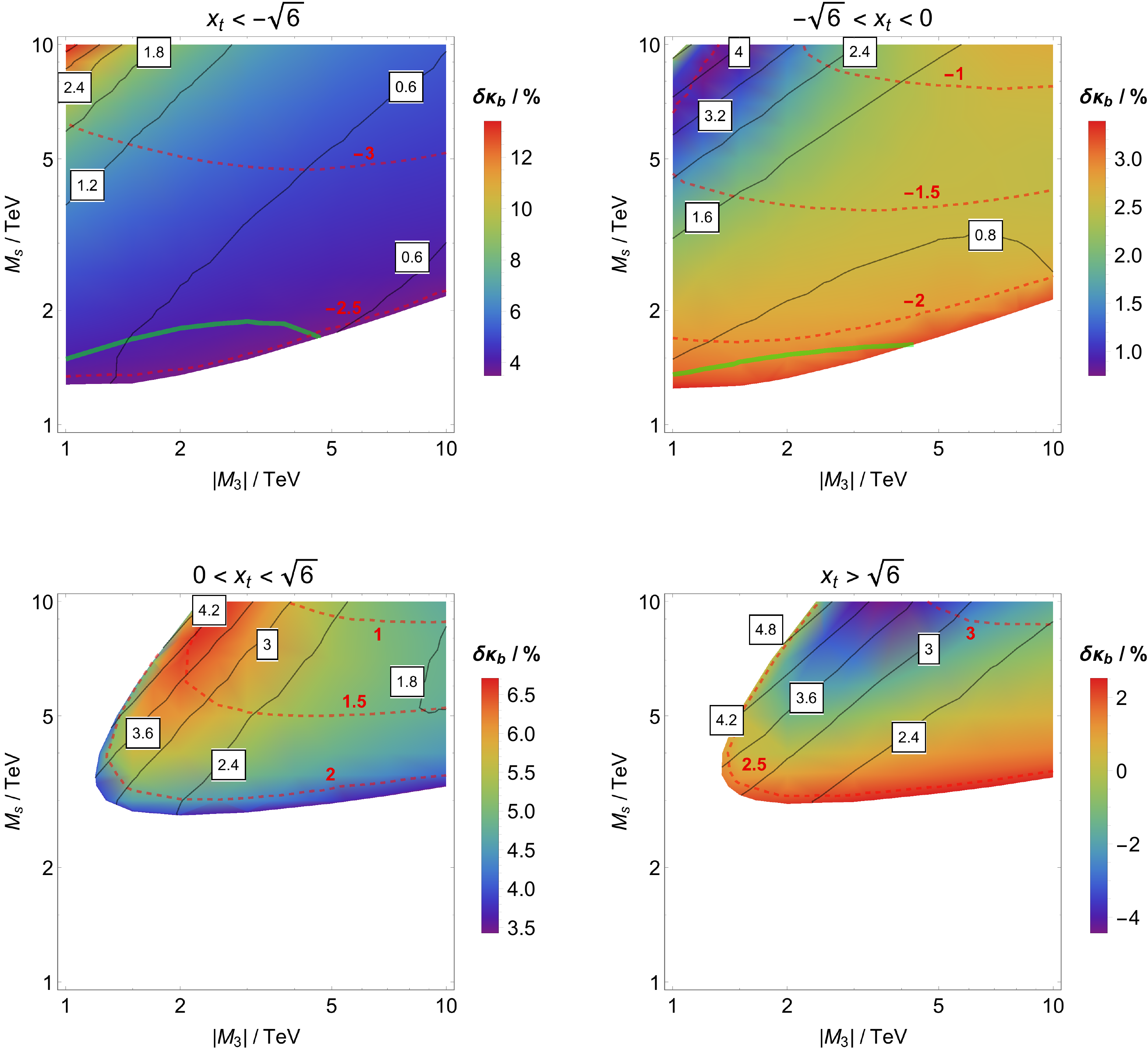}
\caption{\label{fig:TeV-tb20}
Same as Figure~\ref{fig:TeV-tb50}, now with $\tan\beta=20$.
}
\end{figure}

\begin{figure}[tbp]
\centering
\includegraphics[width=\textwidth]{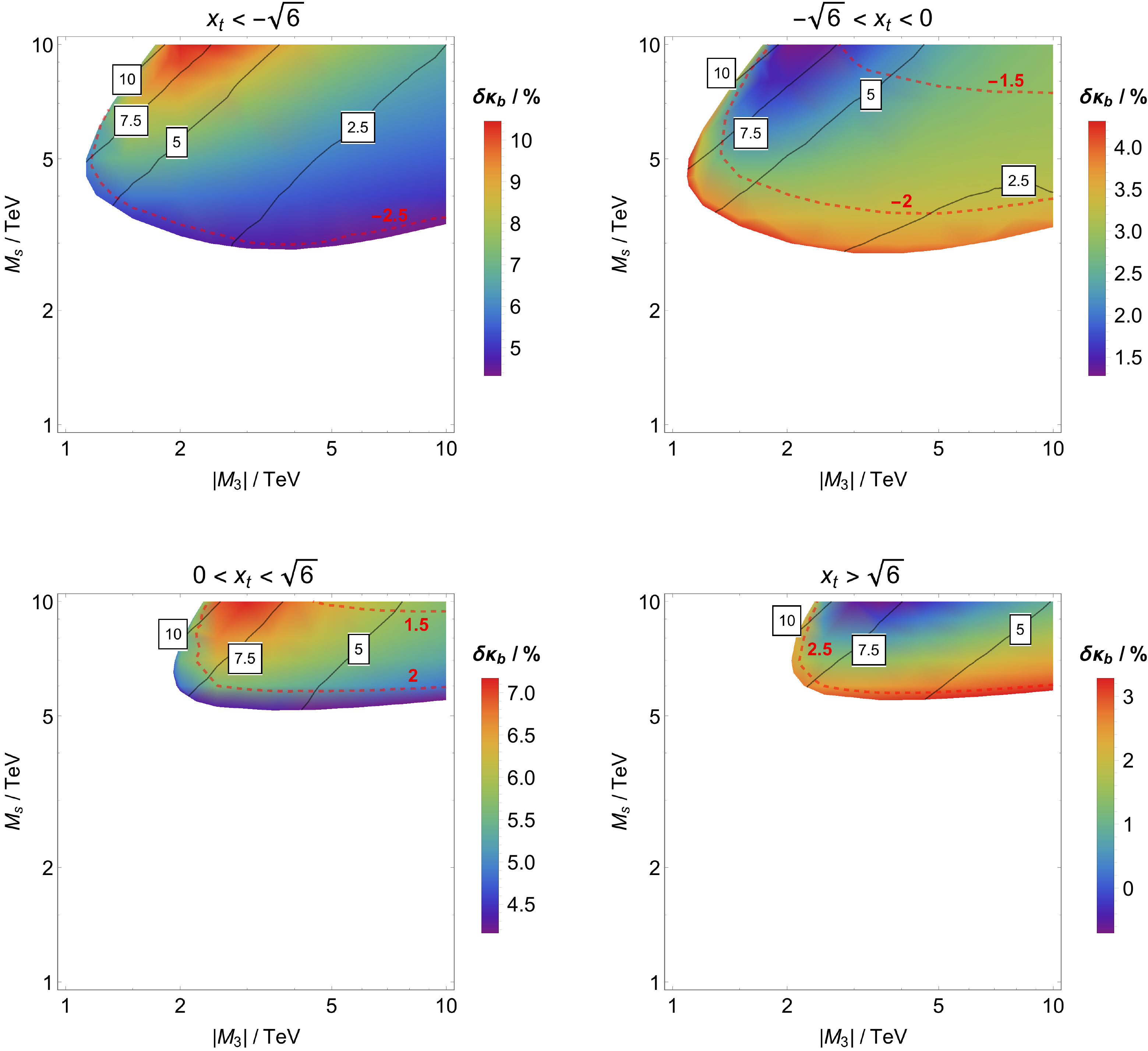}
\caption{\label{fig:TeV-tb8}
Same as Figure~\ref{fig:TeV-tb50}, now with $\tan\beta=8$.
}
\end{figure}

Our results are displayed in Figures~\ref{fig:TeV-tb50}, \ref{fig:TeV-tb20} and~\ref{fig:TeV-tb8}, for $\tan\beta=50,\,20,\,8$, respectively. For each of the four $x_t$ branches, we show variation of $\delta\kappa_b$ in the region of the $|M_3|$-$M_s$ plane where a solution exists. Also shown in the plots are contours of $\mu/M_s$ (black) and $x_t$ (red dashed) which, as we will see shortly, are the key quantities that determine the value of $\delta\kappa_b$. In addition, light green contours represent $\mu=1\,\text{TeV}$, corresponding to the Higgsino thermal dark matter benchmark. Plots of $\delta\kappa_\tau$ (not shown here) exhibit the same patterns of variation in the $|M_3|$-$M_s$ plane, but with smaller overall sizes than $\delta\kappa_b$ as a consequence of $C_{\tau\phi}\propto y_\tau/y_\tau^\text{eff}<y_b/y_b^\text{eff}$.

\begin{figure}[tbp]
\centering
\includegraphics[width=\textwidth]{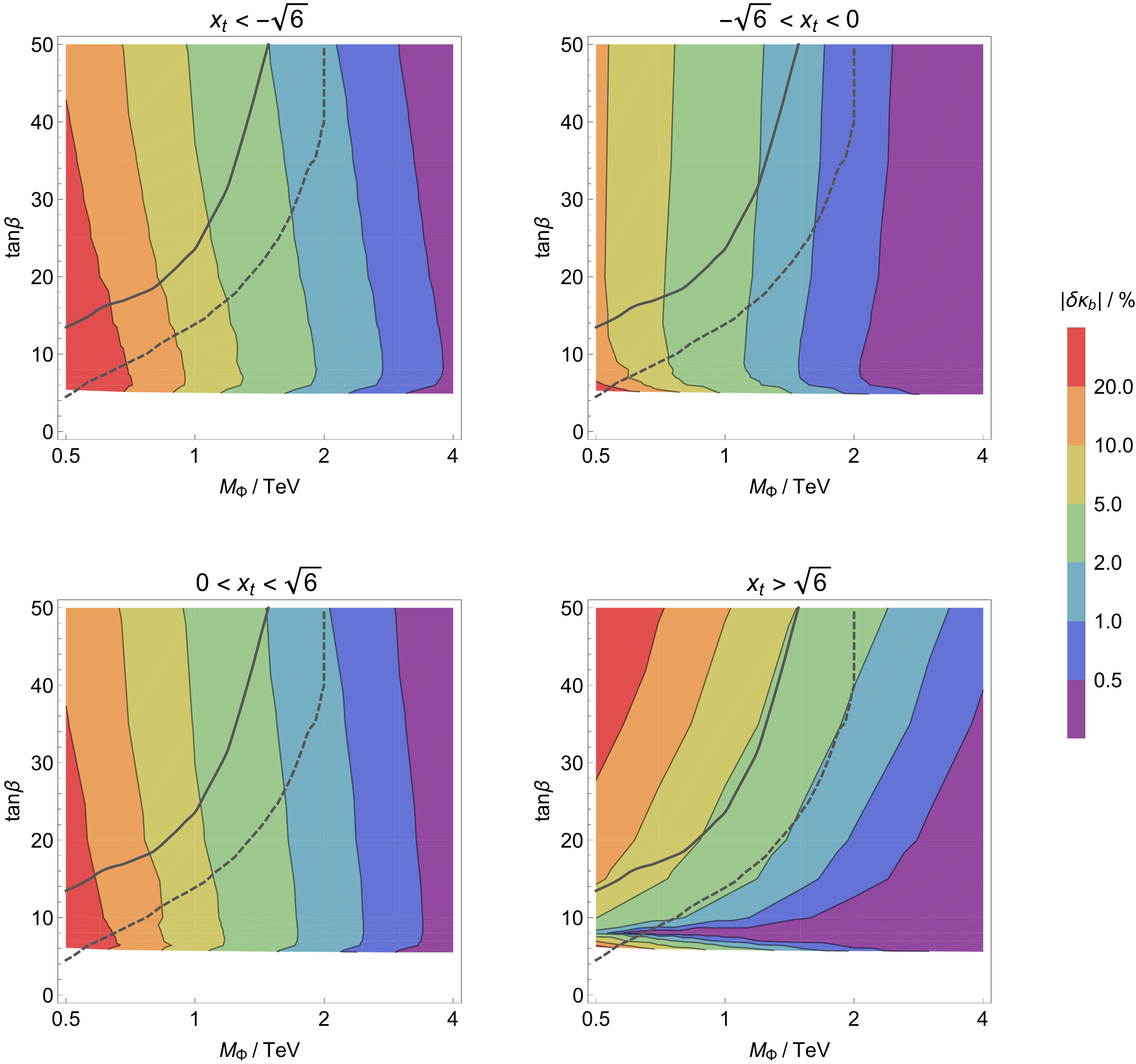}
\caption{\label{fig:HC}
Contours of $|\delta\kappa_b|$ in the $M_\Phi$-$\tan\beta$ plane, for our benchmark scenario $|M_3|=5$\,TeV, $M_s=10$\,TeV, which will evade gluino and stop searches at the LHC. The Higgsino mass is determined by exact $b$-$\tau$ Yukawa unification, for which solutions exist for $\tan\beta\gtrsim5$. Dark solid and dashed curves represent current exclusion limit (95\% CL) and projected high-luminosity reach (95\% CL\ with 3\,ab$^{-1}$ at 14\,TeV) from heavy Higgs searches in the di-tau channel at the LHC, reported assuming the $m_h^\text{mod+}$ benchmark scenario. Future Higgs factories, with 0.5-1\% projected precision for the $hb\bar b$ coupling, will be able to probe much of the parameter space displayed.
}
\end{figure}

From these plots, it is first of all interesting to see how large one-loop effects can be. Indeed, as we have fixed $M_\Phi=1\,\text{TeV}$, a tree-level calculation would yield constant $C_{b\phi}$ (and hence $\delta\kappa_b$) for given $\tan\beta$; see Table~\ref{tab:tlm}. 
The pattern of $\delta\kappa_b$ observed in the figures is a result of interplay between tree- and one-loop-level contributions. For the most part of parameter space (with large $\tan\beta$ and low $M_\Phi$), we can approximate
\beqa
C_{b\phi} &\simeq& C_{b\phi}^\text{tree} +\frac{y_b}{M_\Phi^2}\tan\beta\Bigl(c_{\Phi\phi^3}^{(\ft\ft\ft)} +c_{\Phi\phi^3}^{(\ft\ft\ft\ft)}\Bigr) \CR
&\simeq& -\frac{y_b}{2M_\Phi^2} \Bigl[(g^2+g'^2) -\frac{\tan\beta}{16\pi^2}\,y_t^4\,\Bigl(\frac{\mu}{M_s}\Bigr)\,x_t(x_t^2-6)\Bigr] \,.
\eeqa{Cbphi-approx}
at the matching scale $\Qth=M_s$. We see that tree-level matching always gives a negative contribution to $C_{b\phi}$, and thus a positive contribution to $\delta\kappa_b$. On the other hand, the one-loop piece can have either sign, depending on the value of $x_t$. On two of the four branches, $x_t<-\sqrt{6}$ and $0<x_t<\sqrt{6}$, its contribution to $C_{b\phi}$ is negative, resulting in an enhanced (positive) $\delta\kappa_b$. More specifically, for $x_t<-\sqrt{6}$ (upper-left plot in each figure), $\delta\kappa_b$ is seen to increase monotonically with both $\mu/M_s$ and $|x_t|$, while for $0<x_t<\sqrt{6}$ (lower-left plot in each figure), $\delta\kappa_b$ also increases with $\mu/M_s$, but now exhibits a plateau around $x_t=\sqrt{2}$ where $-x_t(x_t^2-6)$ is maximized, in agreement with Eq.~\eqref{Cbphi-approx}. In contrast, the other two branches feature a negative one-loop contribution to $\delta\kappa_b$: for $-\sqrt{6}<x_t<0$ (upper-right plot in each figure), we have a suppressed but still positive $\delta\kappa_b$, with the suppression being more severe in regions with large $\mu/M_s$ and $x_t$ close to $-\sqrt{2}$; for $x_t>\sqrt{6}$ (lower-right plot in each figure), one-loop correction becomes large enough in part of the parameter space so as to make $\delta\kappa_b$ negative, and, as expected, $\delta\kappa_b$ tends to be smaller (more negative) in regions with larger $\mu/M_s$ and $x_t$.

Precision Higgs measurements -- $h\to b\bar b$ in particular -- are most sensitive to regions of parameter space with the largest $|\delta\kappa_b|$, which in most cases (all $x_t<-\sqrt{6}$ and $0<x_t<\sqrt{6}$ plots, and $x_t>\sqrt{6}$ plots for $\tan\beta=50,\,20$ as well) are those with heavy sfermions and light to intermediate-mass gluino, once $b$-$\tau$ Yukawa unification is stipulated. In these regions, as we have discussed in Section~\ref{sec:YU}, $b$-$\tau$ Yukawa unification calls for relatively large $\mu/M_s$ to boost SUSY threshold correction for $y_b$ (recall $\delta_b\propto |M_3|/M_s$ for $|M_3|/M_s\lesssim2.12$, and larger $\delta_b$ is needed for heavier sfermions), which in turn enhances one-loop contributions to $\delta\kappa_b$ according to Eq.~\eqref{Cbphi-approx}; meanwhile, there is a visible suppression of $|\delta\kappa_b|$ for the largest $\mu/M_s$ (hence smallest $|M_3|/M_s$) due to $|x_t|$ approaching $\sqrt{6}$ in order to match the Higgs quartic (see Eq.~\eqref{dlambda-approx}). 

In comparison, direct searches can most easily access the region of parameter space with light squarks and gluino. Our results show a nice complementarity between direct superpartner searches and precision Higgs measurements, as they probe the SUSY parameter space from different directions.

To further demonstrate this complementarity, let us consider a scenario where the gluino and sfermions are beyond direct LHC reach, even after the high luminosity phase~\cite{HL-ATLAS,HL-CMS}. We choose $|M_3|=5$\,TeV, $M_s=10$\,TeV as a benchmark, and allow $M_\Phi$ and $\tan\beta$ to vary. The Higgsino mass $\mu$ is still determined by exact $b$-$\tau$ Yukawa unification, and is not a free parameter in this analysis.
 
Figure~\ref{fig:HC} shows plots of $|\delta\kappa_b|$ in the $M_\Phi$-$\tan\beta$ plane for this benchmark scenario, on all four $x_t$ branches. The LHC will be able to probe $|\delta\kappa_b|\sim10\%$~\cite{Higgs-ATLAS,Higgs-CMS}, corresponding to part of the sub-TeV regime for $M_\Phi$ (red and orange regions). Meanwhile, direct heavy Higgs searches can put stronger constraints in the high $\tan\beta$ regime. For illustration, we show in Figure~\ref{fig:HC} current exclusion limit from the ATLAS search in the di-tau channel~\cite{heavyHiggs-ATLAS} (the CMS limit~\cite{heavyHiggs-CMS} is slightly weaker) and projected high luminosity LHC reach (up to $M_\Phi=2$\,TeV) in the same channel from the CMS analysis~\cite{heavyHiggs-ECFA} (dark solid and dashed curves, respectively), both of which are reported assuming the ``$m_h^\text{mod+}$ benchmark scenario'' (see~\cite{MSSMbenchmarks}).

On the other hand, a 0.5-1\% level determination of the $hb\bar b$ coupling, as envisioned at possible future Higgs factories (ILC, CLIC, CEPC and FCC-ee --- see e.g.~\cite{leptonicfuture,imporvedHC,ILC250} for recent studies), would extend the sensitivity to $M_\Phi$ potentially up to $\sim$(2-4)\,TeV, even for lower $\tan\beta$, and beyond direct and indirect LHC reach. The existence of well-motivated scenarios, like trans-TeV SUSY with $b$-$\tau$ Yukawa unification studied here, which escape LHC search but nevertheless can manifest themselves as modified Higgs couplings, highlights the opportunity of BSM discoveries through precision Higgs measurements.

To close this section, we finally comment on the availability of a 1\,TeV Higgsino thermal dark matter candidate. From the figures we see that $\mu=1\,\text{TeV}$ (light green curves) can only be achieved on the $x_t<0$ branches for $\tan\beta\gtrsim20$.\footnote{The quantitative discrepancy between our conclusion and that of~\cite{EHPR} is due to differences in the matching calculation for the Higgs quartic. Our results are in good agreement with the more recent calculation in~\cite{SUSYHD}.} The $x_t>0$ branches cannot support such a small Higgsino mass because of cancellation between the squark-gluino and squark-Higgsino loops contributing to $\delta_b$, as discussed below Eq.~\eqref{db-ho}. Meanwhile, when $\tan\beta$ is reduced, a larger $\mu/M_s$ is generally needed to obtain sufficient threshold corrections for both $\lambda$ and $y_b$. The disappearance of the $\mu=1\,\text{TeV}$ curve is further accelerated by a shrinking parameter space where matching of the Higgs quartic is simultaneously possible.

\section{Conclusions}
\label{sec:concl}

As traditional naturalness and weak-scale new physics are under siege, it is worth considering more attentively trans-TeV regimes. Here, effective field theory becomes the tool of choice to accurately connect a vast range of BSM ideas to low-energy observation. In this paper, we have focused on the specific case of the MSSM, and performed a matching calculation onto the SMEFT. In particular, we computed the full set of renormalizable operators of the SMEFT by integrating out heavy superpartners from the path integral up to one-loop level, which allowed us to extract SUSY threshold corrections with ease.

Our calculation highlights the simplicity of recently-developed functional matching and covariant diagrams techniques. In fact, we were able to reproduce one-loop SUSY threshold corrections for all SM parameters from just 30 covariant diagrams (shown in Tables~\ref{tab:cd-U} and~\ref{tab:cd-P}), each of which is straightforward to compute. Essentially, we have taken a more economic route than traditional Feynman diagram calculations, where just the information needed for deriving the low-energy limit of the theory has been extracted from the path integral. In the long run, it is hoped that these novel EFT techniques will aid the program of (automated) precision calculation in trans-TeV supersymmetry, and other BSM scenarios as well.

Taking unification as a key motivation for SUSY, we investigated implications of $b$-$\tau$ Yukawa unification on the MSSM parameter space, while remaining agnostic about further details of the grand unified theory. The EFT approach we have taken allowed us to take advantage of existing precision calculations within the SM, to ensure consistency with low-energy observations, in particular $m_h=125$\,GeV. We found solutions that realize $b$-$\tau$ Yukawa unification for SUSY scales from TeV up to $10^{10}$\,GeV, with suitable choices of superpartner mass ratios and $\tan\beta$ (see Figures~\ref{fig:HS-rmu} and~\ref{fig:HS-rM3}). In this analysis, a key role is played by SUSY threshold corrections to the Higgs quartic and bottom Yukawa couplings, which, when forced to have the correct (finite) sizes (see Figures~\ref{fig:HS-dl} and~\ref{fig:HS-db}), dramatically constrain the predicted SUSY parameter space.

The lower edge of this broad trans-TeV window is further motivated by the possibility of having a dark matter candidate. For superpartners in the (1-10) TeV regime, we showed that one-loop matching contributions can drastically modify tree-level predictions for the $hb\bar b$ (and also $h\tau^+\tau^-$) coupling, rendering some regions of the MSSM parameter space with heavier squarks more accessible to precision Higgs measurements (see Figures~\ref{fig:TeV-tb50}, \ref{fig:TeV-tb20} and~\ref{fig:TeV-tb8}). 

It is interesting to see that, even for superpartner masses out of LHC reach, precision Higgs measurements can offer a powerful indirect probe of TeV-scale SUSY. For example, in a benchmark scenario with a 5\,TeV gluino and 10\,TeV degenerate sfermions that realizes $b$-$\tau$ Yukawa unification, we showed that a 0.5-1\% level determination of the $hb\bar b$ coupling will be able to probe the heavy Higgs mass up to $\sim$(2-4)\,TeV for a wide range of $\tan\beta$ (see Figure~\ref{fig:HC}). This constitutes an unambiguous example of a motivated BSM scenario that may only reveal itself through precision Higgs measurements of the future.

\acknowledgments
We would like to thank S.~A.~R.~Ellis and A.~Pierce for useful discussions. We also thank the DESY Theory Group for hospitality where part of this work was completed. This research was supported in part by the U.S.\ Department of Energy under grant DE-SC0007859. Z.Z.\ also acknowledges support from the Rackham Graduate School of the University of Michigan via a dissertation fellowship.


\appendix

\section{The MSSM U matrix}
\label{app:Umatrix}

In this appendix, we present detailed expressions for the entries of the MSSM $\bf U$ matrix needed in our one-loop matching calculation in Section~\ref{sec:MSSMmatch}. They are obtained from the MSSM Lagrangian by the background field method explained in Section~\ref{sec:CDrev}; see Eqs.~\eqref{bkgfld}, \eqref{LUVexpand}, \eqref{QUV} and~\eqref{Xmat}. 
Keeping in mind that the $\bf U$ matrix is to be used at one-loop level, we do not distinguish between $\beta$ and $\beta'$, and write $\beta$ throughout. Also, tree-level SUSY relations between couplings can be used regardless of scheme choice, e.g.\ gaugino-sfermion-fermion couplings are identified with gauge couplings (which is true beyond tree level in $\overline{\text{DR}}$ but not $\overline{\text{MS}}$ scheme).

In what follows, the heavy Higgs field $\Phi$ is understood as $\Phi_\c$ obtained in Section~\ref{sec:MSSMmatch-tree}. The other heavy fields do not appear because they are set to zero by the classical equations of motion. We carefully keep all \ci{color} and \wi{weak} indices explicit for clarity, using $\ci{i}$ ($\ci{A}$) and $\wi{\alpha}$ ($\wi{I}$) for $\ci{SU(3)_c}$ and $\wi{SU(2)_L}$ fundamental (adjoint) indices on the conjugate fields to appear on the left side of the $\bf U$ matrix, and $\ci{j}$, $\ci{B}$, $\wi{\beta}$, $\wi{J}$ for those on the fields on the right side. We will not explicitly show the entries involving leptons, because they can always be obtained from those involving quarks by the obvious substitutions $q\to l$, $d\to e$, $\mat{\lambda_u}\to0$, $\mat{\lambda_d}\to\mat{\lambda_e}$, $g_3\to 0$.

\subsection{$R$-parity-even block}

\paragraph{Higgs-Higgs entries.} 

From the MSSM Higgs potential, we obtain
\beqa
U_{\Phi\Phi} &=& 
\frac{1}{4}(g^2+g'^2)
\left(\begin{matrix}
-\delta_{\wi{\alpha}}^{\wi{\beta}} \cbb^2 |\phi|^2 +\sbb^2\phi_\wi{\alpha}\phi^{*\wi{\beta}} &
\sbb^2\, \phi_\wi{\alpha}\phi_\wi{\beta} \\
\sbb^2\, \phi^{*\wi{\alpha}}\phi^{*\wi{\beta}} &
-\delta^{\wi{\alpha}}_{\wi{\beta}} \cbb^2 |\phi|^2 +\sbb^2\phi^{*\wi{\alpha}}\phi_\wi{\beta}
\end{matrix}\right)\CR
&& +\frac{1}{2}g^2
\left(\begin{matrix}
\delta_{\wi{\alpha}}^{\wi{\beta}} |\phi|^2 -\phi_\wi{\alpha}\phi^{*\wi{\beta}} & 0 \\
0 & \delta^{\wi{\alpha}}_{\wi{\beta}} |\phi|^2 -\phi^{*\wi{\alpha}}\phi_\wi{\beta}
\end{matrix}\right)\CR
&& +\frac{1}{8}(g^2+g'^2)\, s_{4\beta}
\left(\begin{matrix}
\delta_{\wi{\alpha}}^{\wi{\beta}} && 0 \\
0 && \delta^{\wi{\alpha}}_{\wi{\beta}} 
\end{matrix}\right) \bigl(\phi^*\Phi+\Phi^*\phi\bigr) \CR
&& +\frac{1}{8}(g^2+g'^2)\, s_{4\beta}
\left(\begin{matrix}
\phi_\wi{\alpha}\Phi^{*\wi{\beta}} +\Phi_\wi{\alpha}\phi^{*\wi{\beta}} &&
\phi_\wi{\alpha}\Phi_\wi{\beta} +\Phi_\wi{\alpha}\phi_\wi{\beta} \\
\phi^{*\wi{\alpha}}\Phi^{*\wi{\beta}} +\Phi^{*\wi{\alpha}}\phi^{*\wi{\beta}} &&
\phi^{*\wi{\alpha}}\Phi_\wi{\beta} +\Phi^{*\wi{\alpha}}\phi_\wi{\beta}
\end{matrix}\right) \CR
&& + \frac{1}{4}(g^2+g'^2)\cbb^2 \left(\begin{matrix}
\delta_{\wi{\alpha}}^{\wi{\beta}}|\Phi|^2 +\Phi_\wi{\alpha}\Phi^{*\wi{\beta}} & 
\Phi_\wi{\alpha}\Phi_\wi{\beta} \\
\Phi^{*\wi{\alpha}}\Phi^{*\wi{\beta}}&
\delta^{\wi{\alpha}}_{\wi{\beta}}|\Phi|^2 +\Phi^{*\wi{\alpha}}\Phi_\wi{\beta}
\end{matrix}\right) , \\[6pt]
U_{\Phi\phi} &=& 
-\frac{1}{8}(g^2+g'^2)\, s_{4\beta}
\left(\begin{matrix}
\delta_{\wi{\alpha}}^{\wi{\beta}}|\phi|^2 +\phi_\wi{\alpha}\phi^{*\wi{\beta}} & 
\phi_\wi{\alpha}\phi_\wi{\beta} \\
\phi^{*\wi{\alpha}}\phi^{*\wi{\beta}} &
\delta^{\wi{\alpha}}_{\wi{\beta}}|\phi|^2 +\phi^{*\wi{\alpha}}\phi_\wi{\beta}
\end{matrix}\right) \CR
&& +\frac{1}{4}(g^2+g'^2)\,\sbb^2
\left(\begin{matrix}
\delta_{\wi{\alpha}}^{\wi{\beta}}(\phi^*\Phi+\Phi^*\phi) +\phi_\wi{\alpha}\Phi^{*\wi{\beta}} && 
\phi_\wi{\alpha}\Phi_\wi{\beta}  \\
\phi^{*\wi{\alpha}}\Phi^{*\wi{\beta}} &&
\delta^{\wi{\alpha}}_{\wi{\beta}}(\phi^*\Phi+\Phi^*\phi) +\phi^{*\wi{\alpha}}\Phi_\wi{\beta} 
\end{matrix}\right) \CR
&& -\frac{1}{4}(g^2+g'^2)\,\cbb^2
\left(\begin{matrix}
\Phi_\wi{\alpha}\phi^{*\wi{\beta}} && 
\Phi_\wi{\alpha}\phi_\wi{\beta} \\
\Phi^{*\wi{\alpha}}\phi^{*\wi{\beta}} &&
\Phi^{*\wi{\alpha}}\phi_\wi{\beta}
\end{matrix}\right) \CR
&& +\frac{1}{2} g^2
\left(\begin{matrix}
-\delta_{\wi{\alpha}}^{\wi{\beta}}(\phi^*\Phi) +\Phi_\wi{\alpha}\phi^{*\wi{\beta}} && 
-\phi_\wi{\alpha}\Phi_\wi{\beta} + \Phi_\wi{\alpha}\phi_\wi{\beta} \\
-\phi^{*\wi{\alpha}}\Phi^{*\wi{\beta}} + \Phi^{*\wi{\alpha}}\phi^{*\wi{\beta}} &&
-\delta^{\wi{\alpha}}_{\wi{\beta}}(\Phi^*\phi) + \Phi^{*\wi{\alpha}}\phi_\wi{\beta}
\end{matrix}\right) \CR
&& +\frac{1}{8}(g^2+g'^2)\, s_{4\beta}
\left(\begin{matrix}
\delta_{\wi{\alpha}}^{\wi{\beta}}|\Phi|^2 +\Phi_\wi{\alpha}\Phi^{*\wi{\beta}} & 
\Phi_\wi{\alpha}\Phi_\wi{\beta} \\
\Phi^{*\wi{\alpha}}\Phi^{*\wi{\beta}}&
\delta^{\wi{\alpha}}_{\wi{\beta}}|\Phi|^2 +\Phi^{*\wi{\alpha}}\Phi_\wi{\beta}
\end{matrix}\right) .\label{UPhiphi}
\eeqan
The other two entries $U_{\phi\phi}$ and $U_{\phi\Phi}$ can be obtained from $U_{\Phi\Phi}$ and $U_{\Phi\phi}$ by simply exchanging $\Phi\leftrightarrow\phi$, $\sb\leftrightarrow\cb$.

\paragraph{Higgs-fermion entries.}

From the MSSM Yukawa interactions, we obtain
\bseq
\beqa
&&
U_{\Phi q} = \left(\begin{matrix}
-\sb\,\delta_{\wi{\alpha}}^{\wi{\beta}} \,\bar\psi_d^{\ci{j}} \,\mat{\lambda_d} && 
-\cb\,\epsilon_{\wi{\alpha\beta}} \,\bar\psi_{u\ci{j}}^c \,\mat{\lambda_u^*} \\
-\cb\,\epsilon^{\wi{\alpha\beta}} \,\bar\psi_u^{\ci{j}} \,\mat{\lambda_u} &&
-\sb\,\delta^{\wi{\alpha}}_{\wi{\beta}} \,\bar\psi_{d\ci{j}}^c \,\mat{\lambda_d^*}
\end{matrix}\right) ,\\[6pt]
&&
U_{q\Phi} = \left(\begin{matrix}
-\sb\,\delta_{\wi{\alpha}}^{\wi{\beta}}\,\mat{\lambda_d^\dagger} \,\psi_{d\ci{i}}  &&
-\cb\,\epsilon_{\wi{\beta\alpha}}\,\mat{\lambda_u^\dagger} \,\psi_{u\ci{i}} \\
-\cb\,\epsilon^{\wi{\beta\alpha}} \,\mat{\lambda_u^T} \,\psi_u^{c\ci{i}} && 
-\sb\,\delta^{\wi{\alpha}}_{\wi{\beta}} \,\mat{\lambda_d^T} \,\psi_d^{c\ci{i}}
\end{matrix}\right) ,
\eeqan
\eseqn
\beqa
&&
U_{\Phi u} = \cb\left(\begin{matrix}
\bigl(\bar\psi_q\epsilon\bigr)_{\wi{\alpha}}^{\ci{j}} \,\mat{\lambda_u^\dagger} & 0 \\
0 & \bigl(\bar\psi_q^c\epsilon\bigr)^{\wi{\alpha}}_{\ci{j}} \,\mat{\lambda_u^T}
\end{matrix}\right) ,\quad
U_{u\Phi} = \cb\left(\begin{matrix}
\mat{\lambda_u}\bigl(\psi_q\epsilon\bigr)_{\ci{i}}^{\wi{\beta}} & 0 \\
0 & \mat{\lambda_u^*}\bigl(\psi_q^c\epsilon\bigr)^{\ci{i}}_{\wi{\beta}}
\end{matrix}\right) ,\quad\\[6pt]
&&
U_{\Phi d} = -\sb \left(\begin{matrix}
0 & \bar\psi_{q\ci{j}\wi{\alpha}}^c \,\mat{\lambda_d^T} \\
\bar\psi_q^{\ci{j}\wi{\alpha}}\,\mat{\lambda_d^\dagger} & 0
\end{matrix}\right) ,\quad
U_{d\Phi} = -\sb \left(\begin{matrix}
0 & \mat{\lambda_d}\, \psi_{q\ci{i}\wi{\beta}} \\
\mat{\lambda_d^*}\, \psi_q^{c\ci{i}\wi{\beta}} & 0
\end{matrix}\right) . 
\eeqan
The $\phi f$, $f\phi$ entries (not needed in our calculation) can be obtained from the equations above by simple substitutions $\,\mat{\lambda_u}\cb\to\mat{\lambda_u}\sb$, $\,\mat{\lambda_d}\sb\to-\mat{\lambda_d}\cb$.

\paragraph{Fermion-fermion entries.}

The Yukawa interactions also give rise to
\bseq
\beqa
U_{qu} &=& \sb \left(\begin{matrix}
\delta_{\ci{i}}^{\ci{j}}\,\mat{\lambda_u^\dagger}\bigl(\epsilon\phi^*\bigr)_\wi{\alpha} & 0 \\
0 & \delta_{\ci{j}}^{\ci{i}}\,\mat{\lambda_u^T} \bigl(\epsilon\phi\bigr)^\wi{\alpha}
\end{matrix}\right)
+\cb \left(\begin{matrix}
\delta_{\ci{i}}^{\ci{j}}\,\mat{\lambda_u^\dagger}\bigl(\epsilon\Phi^*\bigr)_\wi{\alpha} & 0 \\
0 & \delta_{\ci{j}}^{\ci{i}}\,\mat{\lambda_u^T} \bigl(\epsilon\Phi\bigr)^\wi{\alpha}
\end{matrix}\right), \\[6pt]
U_{uq} &=& \sb \left(\begin{matrix}
\delta_{\ci{i}}^{\ci{j}}\,\mat{\lambda_u}\bigl(\epsilon\phi\bigr)^\wi{\beta} & 0 \\
0 & \delta_{\ci{j}}^{\ci{i}}\,\mat{\lambda_u^*} \bigl(\epsilon\phi^*\bigr)_\wi{\beta}
\end{matrix}\right)
+\cb \left(\begin{matrix}
\delta_{\ci{i}}^{\ci{j}}\,\mat{\lambda_u}\bigl(\epsilon\Phi\bigr)^\wi{\beta} & 0 \\
0 & \delta_{\ci{j}}^{\ci{i}}\,\mat{\lambda_u^*} \bigl(\epsilon\Phi^*\bigr)_\wi{\beta}
\end{matrix}\right) ,
\eeqan
\eseqn
\vspace{-10pt}
\bseq
\beqa
U_{qd} &=& \cb \left(\begin{matrix}
\delta_{\ci{i}}^{\ci{j}}\,\mat{\lambda_d^\dagger}\,\phi_\wi{\alpha} & 0 \\
0 & \delta_{\ci{j}}^{\ci{i}}\,\mat{\lambda_d^T}\, \phi^{*\wi{\alpha}}
\end{matrix}\right)
-\sb \left(\begin{matrix}
\delta_{\ci{i}}^{\ci{j}}\,\mat{\lambda_d^\dagger}\,\Phi_\wi{\alpha} & 0 \\
0 & \delta_{\ci{j}}^{\ci{i}}\,\mat{\lambda_d^T}\, \Phi^{*\wi{\alpha}}
\end{matrix}\right) ,\\[6pt]
U_{dq} &=& \cb \left(\begin{matrix}
\delta_{\ci{i}}^{\ci{j}}\,\mat{\lambda_d}\,\phi^{*\wi{\beta}} & 0 \\
0 & \delta_{\ci{j}}^{\ci{i}}\,\mat{\lambda_d^*}\, \phi_\wi{\beta}
\end{matrix}\right)
-\sb \left(\begin{matrix}
\delta_{\ci{i}}^{\ci{j}}\,\mat{\lambda_d}\,\Phi^{*\wi{\beta}} & 0 \\
0 & \delta_{\ci{j}}^{\ci{i}}\,\mat{\lambda_d^*}\, \Phi_\wi{\beta}
\end{matrix}\right) .
\eeqan
\eseqn

In addition, there are nonzero entries involving the SM gauge bosons, which are however not needed in our calculation.

\subsection{$R$-parity-odd block}

\paragraph{Sfermion-sfermion entries.}

From the sfermion-sfermion-Higgs interactions, we obtain
\bseq
\beqa
U_{\qt\ut} &=& (A_u\sb-\mu\cb)
\left(\begin{matrix}
\delta_{\ci{i}}^{\ci{j}}\,\mat{\lambda_u^\dagger} \bigl(\epsilon\phi^*\bigr)_{\wi{\alpha}} & 0 \\
0 & \delta_{\ci{j}}^{\ci{i}}\,\mat{\lambda_u^T} \bigl(\epsilon\phi\bigr)^{\wi{\alpha}}
\end{matrix}\right) \CR[2pt]
&&
+(A_u\cb+\mu\sb) 
\left(\begin{matrix}
\delta_{\ci{i}}^{\ci{j}}\,\mat{\lambda_u^\dagger} \bigl(\epsilon\Phi^*\bigr)_{\wi{\alpha}} & 0 \\
0 & \delta_{\ci{j}}^{\ci{i}}\,\mat{\lambda_u^T} \bigl(\epsilon\Phi\bigr)^{\wi{\alpha}}
\end{matrix}\right), \\[6pt]
U_{\ut\qt} &=& (A_u\sb-\mu\cb)
\left(\begin{matrix}
\delta_{\ci{i}}^{\ci{j}}\,\mat{\lambda_u} \bigl(\epsilon\phi\bigr)^{\wi{\beta}} & 0 \\
0 & \delta_{\ci{j}}^{\ci{i}}\,\mat{\lambda_u^*} \bigl(\epsilon\phi^*\bigr)_{\wi{\beta}}
\end{matrix}\right) \CR[2pt]
&&
+(A_u\cb+\mu\sb) 
\left(\begin{matrix}
\delta_{\ci{i}}^{\ci{j}}\,\mat{\lambda_u} \bigl(\epsilon\Phi\bigr)^{\wi{\beta}} & 0 \\
0 & \delta_{\ci{j}}^{\ci{i}}\,\mat{\lambda_u^*} \bigl(\epsilon\Phi^*\bigr)_{\wi{\beta}}
\end{matrix}\right), \label{Uutqt}
\eeqan
\eseqn
\vspace{-4pt}
\bseq
\beqa
U_{\qt\dt} &=& (A_d\cb-\mu\sb)
\left(\begin{matrix}
\delta_{\ci{i}}^{\ci{j}}\,\mat{\lambda_d^\dagger}\, \phi_{\wi{\alpha}} & 0 \\
0 & \delta_{\ci{j}}^{\ci{i}}\,\mat{\lambda_d^T}\, \phi^{*\wi{\alpha}}
\end{matrix}\right)
-(A_d\sb+\mu\cb) 
\left(\begin{matrix}
\delta_{\ci{i}}^{\ci{j}}\,\mat{\lambda_d^\dagger}\, \Phi_{\wi{\alpha}} & 0 \\
0 & \delta_{\ci{j}}^{\ci{i}}\,\mat{\lambda_d^T}\, \Phi^{*\wi{\alpha}}
\end{matrix}\right),\CR \\[6pt]
U_{\dt\qt} &=& (A_d\cb-\mu\sb)
\left(\begin{matrix}
\delta_{\ci{i}}^{\ci{j}}\,\mat{\lambda_d}\, \phi^{*\wi{\beta}} & 0 \\
0 & \delta_{\ci{j}}^{\ci{i}}\,\mat{\lambda_d^*}\, \phi_{\wi{\beta}}
\end{matrix}\right)
-(A_d\sb+\mu\cb) 
\left(\begin{matrix}
\delta_{\ci{i}}^{\ci{j}}\,\mat{\lambda_d}\, \Phi^{*\wi{\beta}} & 0 \\
0 & \delta_{\ci{j}}^{\ci{i}}\,\mat{\lambda_d^*}\, \Phi_{\wi{\beta}}
\end{matrix}\right). \CR
\eeqan
\eseqn
Meanwhile, the scalar quartic interactions give rise to
\beq
U_{\qt\qt} = \left(\begin{matrix}
\delta_{\ci{i}}^{\ci{j}}\, {\mat{U_{\qt}}}_{\wi{\alpha}}^{\wi{\beta}} & 0 \\
0 & \delta_{\ci{j}}^{\ci{i}}\, {\mat{U_{\qt}^T}}_{\wi{\beta}}^{\wi{\alpha}}
\end{matrix}\right) ,\quad
U_{\ut\ut} = \left(\begin{matrix}
\delta_{\ci{i}}^{\ci{j}}\, \mat{U_{\ut}} & 0 \\
0 & \delta_{\ci{j}}^{\ci{i}}\, \mat{U_{\ut}^T}
\end{matrix}\right) ,\quad
U_{\dt\dt} = \left(\begin{matrix}
\delta_{\ci{i}}^{\ci{j}}\, \mat{U_{\dt}} & 0 \\
0 & \delta_{\ci{j}}^{\ci{i}}\, \mat{U_{\dt}^T}
\end{matrix}\right) ,
\eeqn
where
\beqa
{\mat{U_{\qt}}}_{\wi{\alpha}}^{\wi{\beta}} &=& \mat{\lambda_u^\dagger \lambda_u} \Bigl[
\sb^2\bigl(\delta_{\wi{\alpha}}^{\wi{\beta}}|\phi|^2 -\phi_\wi{\alpha} \phi^{*\wi{\beta}}\bigr) 
\CR
&&\qquad\quad 
+\sb\cb \bigl(\delta_{\wi{\alpha}}^{\wi{\beta}}(\phi^*\Phi+\Phi^*\phi) -\phi_\wi{\alpha} \Phi^{*\wi{\beta}} -\Phi_\wi{\alpha} \phi^{*\wi{\beta}}\bigr) 
+\cb^2 \bigl(\delta_{\wi{\alpha}}^{\wi{\beta}}|\Phi|^2 -\Phi_\wi{\alpha} \Phi^{*\wi{\beta}} \bigr)
\Bigr] \CR
&& +\mat{\lambda_d^\dagger \lambda_d} \Bigl[ \cb^2\, \phi_\wi{\alpha} \phi^{*\wi{\beta}} 
-\sb\cb \bigl(\phi_\wi{\alpha} \Phi^{*\wi{\beta}} +\Phi_\wi{\alpha} \phi^{*\wi{\beta}}\bigr) +\sb^2 \,\Phi_\wi{\alpha} \Phi^{*\wi{\beta}}\Bigr] \CR
&& +g^2 \,\frac{1}{4}\, \sigma_{\wi{\alpha}}^{\wi{I\beta}} \Bigl[
(\sb^2-\cb^2) \bigl(\phi^*\sigma^\wi{I}\phi-\Phi^*\sigma^\wi{I}\Phi\bigr)
+2\sb\cb \bigl(\phi^*\sigma^\wi{I}\Phi+\Phi^*\sigma^\wi{I}\phi\bigr)\Bigr] \CR
&& +g'^2\,Y_\phi Y_q\, \delta_{\wi{\alpha}}^{\wi{\beta}}\Bigl[
(\sb^2-\cb^2) \bigl(|\phi|^2-|\Phi|^2\bigr) +2\sb\cb \bigl(\phi^*\Phi+\Phi^*\phi\bigr)\Bigr]
\\[4pt]
\mat{U_{\ut}} &=& \mat{\lambda_u \lambda_u^\dagger} \Bigl[\sb^2\,|\phi|^2 +\sb\cb\bigl(\phi^*\Phi+\Phi^*\phi\bigr) +\cb^2\,|\Phi|^2\Bigr] \CR
&& -g'^2\, Y_\phi Y_u \Bigl[
(\sb^2-\cb^2) \bigl(|\phi|^2-|\Phi|^2\bigr) +2\sb\cb \bigl(\phi^*\Phi+\Phi^*\phi\bigr)\Bigr] ,\\[4pt]
\mat{U_{\dt}} &=& \mat{\lambda_d \lambda_d^\dagger} \Bigl[\cb^2\,|\phi|^2 -\sb\cb\bigl(\phi^*\Phi+\Phi^*\phi\bigr) +\sb^2\,|\Phi|^2\Bigr] \CR
&& -g'^2\, Y_\phi Y_d \Bigl[
(\sb^2-\cb^2) \bigl(|\phi|^2-|\Phi|^2\bigr) +2\sb\cb \bigl(\phi^*\Phi+\Phi^*\phi\bigr)\Bigr] .
\eeqan
There are also off-diagonal entries between $\ut$ and $\dt$,
\bseq
\beqa
U_{\ut\dt} = \left(\begin{matrix}
\delta_{\ci{i}}^{\ci{j}}\,\mat{\lambda_u \lambda_d^\dagger} \bigl(\phi\epsilon\Phi\bigr) & 0 \\
0 & \delta_{\ci{j}}^{\ci{i}}\,\bigl(\mat{\lambda_d \lambda_u^\dagger}\bigr)^T \bigl(\phi^*\epsilon\Phi^*\bigr)
\end{matrix}\right) ,\\[6pt]
U_{\dt\ut} = \left(\begin{matrix}
\delta_{\ci{i}}^{\ci{j}}\,\mat{\lambda_d \lambda_u^\dagger} \bigl(\phi^*\epsilon\Phi^*\bigr) & 0 \\
0 & \delta_{\ci{j}}^{\ci{i}}\,\bigl(\mat{\lambda_u \lambda_d^\dagger}\bigr)^T \bigl(\phi\epsilon\Phi\bigr)
\end{matrix}\right) .
\eeqan
\eseqn
%

\paragraph{Sfermion-Higgsino entries.}

From the sfermion-fermion-Higgsino interactions, we obtain
\beqa
&&
U_{\qt\chit} = \left(\begin{matrix}
-\delta_{\wi{\alpha}}^{\wi{\beta}} \,\bar\psi_{d\ci{i}}^c \,\mat{\lambda_d^*}  && 
\epsilon_{\wi{\alpha\beta}} \,\bar\psi_{u\ci{i}}^c \,\mat{\lambda_u^*} \\
\epsilon^{\wi{\alpha\beta}} \,\bar\psi_u^{\ci{i}} \,\mat{\lambda_u} && 
-\delta^{\wi{\alpha}}_{\wi{\beta}} \,\bar\psi_d^{\ci{i}} \,\mat{\lambda_d}
\end{matrix}\right),\quad
U_{\chit\qt} = \left(\begin{matrix}
-\delta^{\wi{\beta}}_{\wi{\alpha}} \,\mat{\lambda_d^T} \,\psi_d^{c\ci{j}} &&
-\epsilon_{\wi{\alpha\beta}}\,\mat{\lambda_u^\dagger} \,\psi_{u\ci{j}}  \\
-\epsilon^{\wi{\alpha\beta}} \,\mat{\lambda_u^T} \,\psi_u^{c\ci{j}} && 
-\delta_{\wi{\beta}}^{\wi{\alpha}}\,\mat{\lambda_d^\dagger} \,\psi_{d\ci{j}} 
\end{matrix}\right) ,\label{Uqtchit}\\[6pt]
&&
U_{\ut\chit} = \left(\begin{matrix}
\bigl(\bar\psi_q^c\epsilon\bigr)^{\wi{\beta}}_{\ci{i}} \,\mat{\lambda_u^T} & 0 \\
0 & \bigl(\bar\psi_q\epsilon\bigr)_{\wi{\beta}}^{\ci{i}} \,\mat{\lambda_u^\dagger}
\end{matrix}\right) ,\quad
U_{\chit\ut} = \left(\begin{matrix}
\mat{\lambda_u^*}\bigl(\psi_q^c\epsilon\bigr)^{\ci{j}}_{\wi{\alpha}} & 0 \\
0 & \mat{\lambda_u}\bigl(\psi_q\epsilon\bigr)_{\ci{j}}^{\wi{\alpha}}
\end{matrix}\right) ,\label{Uutchit}\\[6pt]
&&
U_{\dt\chit} = \left(\begin{matrix}
0 & -\bar\psi_{q\ci{i}\wi{\beta}}^c \,\mat{\lambda_d^T} \\
-\bar\psi_q^{\ci{i}\wi{\beta}}\,\mat{\lambda_d^\dagger} & 0
\end{matrix}\right) ,\quad
U_{\chit\dt} = \left(\begin{matrix}
0 & -\mat{\lambda_d}\, \psi_{q\ci{j}\wi{\alpha}} \\
-\mat{\lambda_d^*}\, \psi_q^{c\ci{j}\wi{\alpha}} & 0
\end{matrix}\right) .
\eeqan
%

\paragraph{Sfermion-gaugino entries.}

From the sfermion-fermion-gaugino interactions, we obtain
\beqa
&&
U_{\qt\gt} = \sqrt{2}\, g_3\left(\begin{matrix}
\bigl(T^\ci{B}\bar\psi_q^c\bigr)_{\ci{i}\wi{\alpha}} \\
\bigl(\bar\psi_q T^\ci{B}\bigr)^{\ci{i}\wi{\alpha}}
\end{matrix}\right) ,\quad
U_{\gt\qt} = \sqrt{2}\, g_3\left(\begin{matrix}
\bigl(\psi_q^c T^\ci{A}\bigr)^{\ci{j}\wi{\beta}} &&
\bigl(T^\ci{A}\psi_q\bigr)_{\ci{j}\wi{\beta}}
\end{matrix}\right) , \\[6pt]
&&
U_{\qt\Wt} = \sqrt{2}\, g \,\frac{1}{2}\left(\begin{matrix}
\bigl(\sigma^\wi{J}\bar\psi_q^c\bigr)_{\ci{i}\wi{\alpha}} \\
\bigl(\bar\psi_q \sigma^\wi{J}\bigr)^{\ci{i}\wi{\alpha}}
\end{matrix}\right) ,\quad
U_{\Wt\qt} = \sqrt{2}\, g\,\frac{1}{2}\left(\begin{matrix}
\bigl(\psi_q^c \sigma^\wi{I}\bigr)^{\ci{j}\wi{\beta}} &&
\bigl(\sigma^\wi{I}\psi_q\bigr)_{\ci{j}\wi{\beta}}
\end{matrix}\right) , \\[6pt]
&&
U_{\qt\Bt} = \sqrt{2}\, g' \,Y_q \left(\begin{matrix}
\bar\psi_{q\ci{i}\wi{\alpha}}^c \\
\bar\psi_q^{\ci{i}\wi{\alpha}}
\end{matrix}\right) ,\quad
U_{\Bt\qt} = \sqrt{2}\, g'\,Y_q\left(\begin{matrix}
\psi_q^{c\ci{j}\wi{\beta}} &&
\psi_{q\ci{j}\wi{\beta}}
\end{matrix}\right) , \\[6pt]
&&
U_{\ut\gt} = -\sqrt{2}\, g_3\left(\begin{matrix}
\bigl(T^\ci{B}\bar\psi_u^c\bigr)_{\ci{i}} \\
\bigl(\bar\psi_u T^\ci{B}\bigr)^{\ci{i}}
\end{matrix}\right) ,\quad
U_{\gt\ut} = -\sqrt{2}\, g_3\left(\begin{matrix}
\bigl(\psi_u^c T^\ci{A}\bigr)^{\ci{j}} &&
\bigl(T^\ci{A}\psi_u\bigr)_{\ci{j}}
\end{matrix}\right) , \\[6pt]
&&
U_{\ut\Bt} = -\sqrt{2}\, g' \,Y_u \left(\begin{matrix}
\bar\psi_{u\ci{i}}^c \\
\bar\psi_u^{\ci{i}}
\end{matrix}\right) ,\quad
U_{\Bt\ut} = -\sqrt{2}\, g'\,Y_u\left(\begin{matrix}
\psi_u^{c\ci{j}} &&
\psi_{u\ci{j}}
\end{matrix}\right) , \\[6pt]
&& U_{\dt\Vt, \Vt\dt} = U_{\ut\Vt, \Vt\ut} \Bigr|_{u\to d} \,.
\eeqan
%

\paragraph{Higgsino-gaugino entries.}

Finally, from the Higgs-Higgsino-gaugino interactions, we obtain
\beq
U_{\chit\Vt} = U_{\chit\Vt}^{(S)} +U_{\chit\Vt}^{(P)} \gamma^5 \,,\quad
U_{\Vt\chit} = U_{\Vt\chit}^{(S)} +U_{\Vt\chit}^{(P)} \gamma^5 \,,
\eeqn
with
\bseq
\beqa
U_{\chit\Wt}^{(S)} &=& \frac{g}{\sqrt{2}}\,\frac{1}{2}\, (\sb+\cb)\left(\begin{matrix}
\bigl(\sigma^\wi{J}\phi\bigr)_\wi{\alpha} \\
\bigl(\phi^*\sigma^\wi{J}\bigr)^\wi{\alpha}
\end{matrix}\right)
-\frac{g}{\sqrt{2}}\,\frac{1}{2}\,(\sb-\cb)\left(\begin{matrix}
\bigl(\sigma^\wi{J}\Phi\bigr)_\wi{\alpha} \\
\bigl(\Phi^*\sigma^\wi{J}\bigr)^\wi{\alpha}
\end{matrix}\right) , \\[6pt]
U_{\Wt\chit}^{(S)} &=& \frac{g}{\sqrt{2}}\,\frac{1}{2}\, (\sb+\cb)\left(\begin{matrix}
\bigl(\phi^*\sigma^\wi{I}\bigr)^\wi{\beta} &&
\bigl(\sigma^\wi{I}\phi\bigr)_\wi{\beta}
\end{matrix}\right)
-\frac{g}{\sqrt{2}}\,\frac{1}{2}\,(\sb-\cb)\left(\begin{matrix}
\bigl(\Phi^*\sigma^\wi{I}\bigr)^\wi{\beta} &&
\bigl(\sigma^\wi{I}\Phi\bigr)_\wi{\beta}
\end{matrix}\right) ,\CR\\[6pt]
U_{\chit\Wt}^{(P)} &=& \frac{g}{\sqrt{2}}\,\frac{1}{2}\, (\sb-\cb)\left(\begin{matrix}
\bigl(\sigma^\wi{J}\phi\bigr)_\wi{\alpha} \\
-\bigl(\phi^*\sigma^\wi{J}\bigr)^\wi{\alpha}
\end{matrix}\right)
+\frac{g}{\sqrt{2}}\,\frac{1}{2}\,(\sb+\cb)\left(\begin{matrix}
\bigl(\sigma^\wi{J}\Phi\bigr)_\wi{\alpha} \\
-\bigl(\Phi^*\sigma^\wi{J}\bigr)^\wi{\alpha}
\end{matrix}\right) , \\[6pt]
U_{\Wt\chit}^{(P)} &=& -\frac{g}{\sqrt{2}}\,\frac{1}{2}\, (\sb-\cb)\left(\begin{matrix}
\bigl(\phi^*\sigma^\wi{I}\bigr)^\wi{\beta} &&
-\bigl(\sigma^\wi{I}\phi\bigr)_\wi{\beta}
\end{matrix}\right)
-\frac{g}{\sqrt{2}}\,\frac{1}{2}\,(\sb+\cb)\left(\begin{matrix}
\bigl(\Phi^*\sigma^\wi{I}\bigr)^\wi{\beta} &&
-\bigl(\sigma^\wi{I}\Phi\bigr)_\wi{\beta}
\end{matrix}\right) ,\CR
\eeqan
\eseqn
\vspace{-14pt}
\bseq
\beqa
U_{\chit\Bt}^{(S)} &=& \frac{g'}{\sqrt{2}}\,Y_\phi\, (\sb+\cb)\left(\begin{matrix}
\phi_\wi{\alpha} \\
\phi^{*\wi{\alpha}}
\end{matrix}\right)
-\frac{g'}{\sqrt{2}}\,Y_\phi\,(\sb-\cb)\left(\begin{matrix}
\Phi_\wi{\alpha} \\
\Phi^{*\wi{\alpha}}
\end{matrix}\right) , \\[6pt]
U_{\Bt\chit}^{(S)} &=& \frac{g'}{\sqrt{2}}\,Y_\phi\, (\sb+\cb)\left(\begin{matrix}
\phi^{*\wi{\beta}} &&
\phi_\wi{\beta}
\end{matrix}\right)
-\frac{g'}{\sqrt{2}}\,Y_\phi\,(\sb-\cb)\left(\begin{matrix}
\Phi^{*\wi{\beta}} &&
\Phi_\wi{\beta}
\end{matrix}\right) ,\\[6pt]
U_{\chit\Bt}^{(P)} &=& \frac{g'}{\sqrt{2}}\,Y_\phi\, (\sb-\cb)\left(\begin{matrix}
\phi_\wi{\alpha} \\
-\phi^{*\wi{\alpha}}
\end{matrix}\right)
+\frac{g'}{\sqrt{2}}\,Y_\phi\,(\sb+\cb)\left(\begin{matrix}
\Phi_\wi{\alpha} \\
-\Phi^{*\wi{\alpha}}
\end{matrix}\right) , \\[6pt]
U_{\Bt\chit}^{(P)} &=& -\frac{g'}{\sqrt{2}}\,Y_\phi\, (\sb-\cb)\left(\begin{matrix}
\phi^{*\wi{\beta}} &&
-\phi_\wi{\beta}
\end{matrix}\right)
-\frac{g'}{\sqrt{2}}\,Y_\phi\, (\sb+\cb)\left(\begin{matrix}
\Phi^{*\wi{\beta}} &&
-\Phi_\wi{\beta}
\end{matrix}\right) .
\eeqan
\eseqn
%

\section{Master integrals}
\label{app:MI}

Our results for one-loop matching presented in Section~\ref{sec:MSSMmatch} are written in terms of master integrals $\,\It[q^{2n_c}]_{ij\dots 0}^{n_i n_j\dots n_L}\equiv\,\I[q^{2n_c}]_{ij\dots 0}^{n_i n_j\dots n_L}/\frac{i}{16\pi^2}$. They can in general be evaluated via the following decomposition formula,
\beqa
\It[q^{2n_c}]_{ij\dots 0}^{n_i n_j\dots n_L} &=& \sum_{p_i=0}^{n_i-1}\biggl[\frac{1}{p_i!}\biggl(\frac{\partial}{\partial M_i^2}\biggr)^{p_i}\frac{1}{\bigl(M_i^2\bigr)^{n_L}\prod_{a\ne i}\bigl(\dsq{ia}\bigr)^{n_a}}\biggr]\,\I[q^{2n_c}]_{i}^{n_i-p_i} \CR
&& +\sum_{p_j=0}^{n_j-1}\biggl[\frac{1}{p_j!}\biggl(\frac{\partial}{\partial M_j^2}\biggr)^{p_j}\frac{1}{\bigl(M_j^2\bigr)^{n_L}\prod_{a\ne j}\bigl(\dsq{ja}\bigr)^{n_a}}\biggr]\,\I[q^{2n_c}]_{j}^{n_j-p_j} +\dots\qquad
\eeqa{MIdec}
where $\dsq{ij}\equiv M_i^2-M_j^2$. To derive Eq.~\eqref{MIdec}, we first recall the definition,
\beq
\int\frac{d^dq}{(2\pi)^d} \frac{q^{\mu_1}\cdots q^{\mu_{2n_c}}}{(q^2-M_i^2)^{n_i}(q^2-M_j^2)^{n_j}\cdots (q^2)^{n_L}}
\,\equiv\, g^{\mu_1\dots\mu_{2n_c}} \,\I[q^{2n_c}]_{ij\dots 0}^{n_i n_j\dots n_L} ,
\eeq{MIdef-app}
where $g^{\mu_1\dots\mu_{2n_c}}$ is the completely symmetric tensor, e.g.\ $g^{\mu\nu\rho\sigma}=g^{\mu\nu}g^{\rho\sigma} +g^{\mu\rho}g^{\nu\sigma} +g^{\mu\sigma}g^{\nu\rho}$. It is easy to see that
\beqa
\I[q^{2n_c}]_{ij\dots 0}^{n_i n_j\dots n_L} &=& \frac{1}{\dsq{ij}} \bigl( \I[q^{2n_c}]_{ij\dots 0}^{n_i, n_j-1,\dots n_L} -\,\I[q^{2n_c}]_{ij\dots 0}^{n_i-1, n_j\dots n_L} \bigr) \,,\label{MIred-h} \\
\I[q^{2n_c}]_{ij\dots 0}^{n_i n_j\dots n_L} &=& \frac{1}{M_i^2} \bigl( \I[q^{2n_c}]_{ij\dots 0}^{n_i n_j\dots, n_L-1} -\,\I[q^{2n_c}]_{ij\dots 0}^{n_i-1, n_j\dots n_L} \bigr) \,,\label{MIred-l}\\
\frac{\partial}{\partial M_i^2} \I[q^{2n_c}]_{ij\dots0}^{n_i n_j\dots n_L} &=&\, n_i \,\I[q^{2n_c}]_{ij\dots0}^{n_i+1, n_j\dots n_L} \,,\label{MIdiff}
\eeqan
Note that in principle, we can just start from $\I[q^{2n_c}]_{ij\dots 0}^{n_i n_j\dots n_L}$ and use Eqs.~\eqref{MIred-h} and~\eqref{MIred-l} repeatedly to reduce the number of propagators, until arriving at a sum of heavy-only degenerate master integrals of the form $\I[q^{2n_c}]_i^{n_i}$ (recall $\I[q^{2n_c}]_0^{n_L}=0$), which cannot be further reduced. However, the same result can be obtained via a more systematic and often easier path, starting from applying Eq.~\eqref{MIdiff},
\beq
\I[q^{2n_c}]_{ij\dots0}^{n_i n_j\dots n_L} = \frac{1}{(n_i-1)!} \biggl(\frac{\partial}{\partial M_i^2}\biggr)^{n_i-1} \frac{1}{(n_j-1)!} \biggl(\frac{\partial}{\partial M_j^2}\biggr)^{n_j-1} \dots \;\I[q^{2n_c}]_{ij\dots0}^{11\dots n_L} \,.
\eeq{MIred-diff}
The master integrals $\,\I[q^{2n_c}]_{ij\dots0}^{11\dots n_L}$, where each heavy propagator appears only once, are much easier to reduce via Eqs.~\eqref{MIred-h} and~\eqref{MIred-l} compared to the original master integral. In fact, we can show that
\beqa
\I[q^{2n_c}]_{ij\dots0}^{11\dots n_L} &=& \frac{1}{\dsq{ij}\dsq{ik}\dsq{il}\dots} \,\I[q^{2n_c}]_{i0}^{1 n_L} +\frac{1}{\dsq{ji}\dsq{jk}\dsq{jl}\dots} \,\I[q^{2n_c}]_{j0}^{1 n_L} +\dots\CR
&=& \frac{1}{\bigl(M_i^2\bigr)^{n_L}\dsq{ij}\dsq{ik}\dsq{il}\dots} \,\I[q^{2n_c}]_{i}^{1} +\frac{1}{\bigl(M_j^2\bigr)^{n_L}\dsq{ji}\dsq{jk}\dsq{jl}\dots} \,\I[q^{2n_c}]_{j}^{1} +\dots\CR
&=& \frac{1}{\bigl(M_i^2\bigr)^{n_L}\prod_{a\ne i}\dsq{ia}} \,\I[q^{2n_c}]_{i}^{1} +\frac{1}{\bigl(M_j^2\bigr)^{n_L}\prod_{a\ne j}\dsq{ja}} \,\I[q^{2n_c}]_{j}^{1} +\dots
\eeqa{MIred-1}
Plugging Eq.~\eqref{MIred-1} into Eq.~\eqref{MIred-diff} and taking derivatives according to Eq.~\eqref{MIdiff}, we obtain
\beqa
\I[q^{2n_c}]_{ij\dots0}^{n_i n_j\dots n_L} &=& \frac{1}{(n_i-1)!} \biggl(\frac{\partial}{\partial M_i^2}\biggr)^{n_i-1} \frac{1}{\bigl(M_i^2\bigr)^{n_L}\bigl(\dsq{ij}\bigr)^{n_j} \bigl(\dsq{ik}\bigr)^{n_k} \bigl(\dsq{il}\bigr)^{n_l} \cdots} \,\I[q^{2n_c}]_{i}^{1} \CR[2pt]
&& +\frac{1}{(n_j-1)!} \biggl(\frac{\partial}{\partial M_j^2}\biggr)^{n_j-1} \frac{1}{\bigl(M_j^2\bigr)^{n_L}\bigl(\dsq{ji}\bigr)^{n_i} \bigl(\dsq{jk}\bigr)^{n_k} \bigl(\dsq{jl}\bigr)^{n_l} \cdots} \,\I[q^{2n_c}]_{j}^{1} \CR[4pt]
&& +\dots \,,
\eeqan
which can be easily seen to lead to Eq.~\eqref{MIdec}.

Eq.~\eqref{MIdec} allows us to decompose an arbitrary master integral $\I[q^{2n_c}]_{ij\dots0}^{n_i n_j\dots n_L}$ into a sum of {\it degenerate} master integrals of the form $\I[q^{2n_c}]_i^{n_i}$. For example,
\beq
\I[q^2]_{ij0}^{211} = \frac{1}{M_i^2 \dsq{ij}}\,\I[q^2]_i^2 +\frac{\partial}{\partial M_i^2} \biggl(\frac{1}{M_i^2 \dsq{ij}}\biggr)\,\I[q^2]_i^1 +\frac{1}{M_j^2 \bigl(\dsq{ji}\bigr)^2}\,\I[q^2]_j^1 \,.
\eeq{MIred-ex}
The degenerate master integrals $\I[q^{2n_c}]_i^{n_i}$ cannot be decomposed further in this way, but can be worked out explicitly and tabulated; see Table~7 of~\cite{CovDiag}. Here, we note that if $n_i\ge2$ and $n_c\ge1$, $\I[q^{2n_c}]_i^{n_i}$ can in fact be further reduced using
\beq
\I[q^{2n_c}]_i^{n_i} = \frac{1}{2(n_i-1)} \,\I[q^{2(n_c-1)}]_i^{n_i-1} \,,
\eeq{MIred-d}
which follows from the explicit expression
\beq
\I[q^{2n_c}]_i^{n_i} = \lf \bigl(-M_i^2\bigr)^{2+n_c-n_i}
\frac{1}{2^{n_c}(n_i-1)!} \frac{\Gamma(\frac{\epsilon}{2}-2-n_c +n_i)}{\Gamma(\frac{\epsilon}{2})} \Bigl(\frac{2}{\bar\epsilon}-\logm{M_i^2}\Bigr) \,,
\eeqn
where $\frac{2}{\bar\epsilon} \equiv \frac{2}{\epsilon} -\gamma +\log 4\pi$ with $\epsilon = 4-d$, and $Q$ is the renormalization scale. For example, Eq.~\eqref{MIred-ex} can be further reduced to
\beq
\I[q^2]_{ij0}^{211} = \frac{1}{2M_i^2 \dsq{ij}}\,\I_i^1 +\frac{\partial}{\partial M_i^2} \biggl(\frac{1}{M_i^2 \dsq{ij}}\biggr)\,\I[q^2]_i^1 +\frac{1}{M_j^2 \bigl(\dsq{ji}\bigr)^2}\,\I[q^2]_j^1 \,.
\eeqn
We therefore only list irreducible master integrals here. For $n_i = \{ 1, 2, 3, 4, 5, 6 \}$,
\beq
\It_i^{n_i} = \Bigl\{
M_i^2\bigl(1-\log M_i^2 \bigr), 
-\log M_i^2  \,,\;
-\frac{1}{2M_i^2} \,,\;
\frac{1}{6M_i^4} \,,\;
-\frac{1}{12M_i^6} \,,\;
\frac{1}{20M_i^8}\, \Bigr\} \,,
\eeqn
while for $n_c = \{ 1, 2, 3 \}$,
\beq
\It[q^{2n_c}]_i^1 = \Bigl\{
\frac{M_i^4}{4}\Bigl(\frac{3}{2}-\log M_i^2\Bigr) \,,\;
\frac{M_i^6}{24}\Bigl(\frac{11}{6}-\log M_i^2 \Bigr) \,,\;
\frac{M_i^8}{192}\Bigl(\frac{25}{12}-\log M_i^2\Bigr) \,
\Bigr\} \,,
\eeqn
where we have dropped the $\frac{1}{\bar\epsilon}$ poles (as in $\overline{\text{MS}}$ and $\overline{\text{DR}}$ schemes), and abbreviated $\log\frac{M_i^2}{Q^2}$ to $\log M_i^2$. In cases where $\O(\epsilon)$ terms are produced from e.g.\ gamma matrix algebra, the $\frac{1}{\bar\epsilon}$ pieces in the master integrals that have been subtracted off can be recovered by simply replacing $-\log M_i^2\to\frac{2}{\bar\epsilon}-\log M_i^2$.

Using the formulas above, we can compute explicit expressions for the master integrals appearing in our one-loop matching results in Section~\ref{sec:MSSMmatch}. We list them in the following, including also additional master integrals encountered when some of the heavy particles are degenerate in mass (e.g.\ $\It_i^4\,=\lim\limits_{M_j^2\to M_i^2}\It_{ij}^{22}$),
\beqa
\It_i^1 &=& M_i^2 \bigl(1-\log M_i^2\bigr) \,,\label{Ii1}\qquad\qquad
\It_i^2 = -\log M_i^2 \,,\\
\It_i^3 &=& -\frac{1}{2M_i^2} \,,\qquad\qquad
\It_i^4 = \frac{1}{6M_i^4} \,,\qquad\qquad
\It_{i0}^{11} = 1-\log M_i^2 \,,\\
\It[q^2]_i^3 &=& -\frac{1}{4}\log M_i^2 \,,\qquad
\It[q^2]_i^4 = -\frac{1}{12M_i^2} \,,\qquad
\It[q^2]_{i0}^{21} = \frac{1}{8}-\frac{1}{4}\log M_i^2 \,,\qquad\\
\It_{ij}^{11} &=& 1-\frac{1}{\dsq{ij}}\bigl( M_i^2\log M_i^2-M_j^2\log M_j^2\bigr) \,,\label{Iij11}\\
\It_{ij}^{21} &=& -\frac{1}{\dsq{ij}}-\frac{M_j^2}{\bigl(\dsq{ij}\bigr)^2}\log\frac{M_j^2}{M_i^2} \,,\\
\It_{ij}^{22} &=& -\frac{2}{\bigl(\dsq{ij}\bigr)^2} -\frac{M_i^2+M_j^2}{\bigl(\dsq{ij}\bigr)^3}\log\frac{M_j^2}{M_i^2} \,,\\
\It_{ij}^{31} &=& \frac{M_i^2+M_j^2}{2M_i^2\bigl(\dsq{ij}\bigr)^2} +\frac{M_j^2}{\bigl(\dsq{ij}\bigr)^3}\log\frac{M_j^2}{M_i^2} \,,\\
\It[q^2]_{ij}^{21} &=& \frac{M_i^2-3M_j^2}{8\dsq{ij}} -\frac{1}{4\bigl(\dsq{ij}\bigr)^2}\Bigl[M_i^2(M_i^2-2M_j^2)\log M_i^2 +M_j^4\log M_j^2\Bigr] \,,\\
\It[q^2]_{ij}^{22} &=& -\frac{M_i^2+M_j^2}{4\bigl(\dsq{ij}\bigr)^2} -\frac{M_i^2 M_j^2}{2\bigl(\dsq{ij}\bigr)^3}\log\frac{M_j^2}{M_i^2}  \,,\\
\It[q^2]_{ij}^{31} &=& -\frac{M_i^2-3M_j^2}{8\bigl(\dsq{ij}\bigr)^2} +\frac{M_j^4}{4\bigl(\dsq{ij}\bigr)^3}\log\frac{M_j^2}{M_i^2}  \,,\\
\It_{ijk}^{111} &=& \frac{M_j^2}{\dsq{ij}\dsq{jk}}\log\frac{M_j^2}{M_i^2}
+ \frac{M_k^2}{\dsq{jk}\dsq{ki}}\log\frac{M_k^2}{M_i^2} \,,\label{Iijk111}\\
\It_{ijk}^{211} &=& -\frac{1}{\dsq{ij}\dsq{ik}} -\frac{1}{\dsq{jk}} \biggl[\frac{M_j^2}{\bigl(\dsq{ij}\bigr)^2}\log\frac{M_j^2}{M_i^2} -\frac{M_k^2}{\bigl(\dsq{ik}\bigr)^2}\log\frac{M_k^2}{M_i^2} \biggr] \,,\\
\It[q^2]_{ijk}^{211} &=& -\frac{M_i^2}{4\dsq{ij}\dsq{ik}} -\frac{1}{4\dsq{jk}} \biggl[\frac{M_j^4}{\bigl(\dsq{ij}\bigr)^2}\log\frac{M_j^2}{M_i^2} -\frac{M_k^4}{\bigl(\dsq{ik}\bigr)^2}\log\frac{M_k^2}{M_i^2} \biggr]  \,.
\eeqan
In the equations above, we have used the notation $\dsq{ij}\equiv M_i^2-M_j^2$. 

Finally, let us also present some formulas that can be used to decrease $n_c$, because they are often useful for simplifying loops involving fermions. When $n_c=1$, we can contract both sides of Eq.~\eqref{MIdef-app} with $g_{\mu_1\mu_2}$ to obtain
\beqa
(4-\epsilon)\,\I[q^2]_{ij\dots 0}^{n_i n_j \dots n_L} &=& \,\I_{ij\dots 0}^{n_i-1,n_j\dots n_L} +M_i^2\,\I_{ij\dots 0}^{n_i n_j \dots n_L} \label{MIred-nc1}\\
(4-\epsilon)\,\I[q^2]_{ij\dots 0}^{n_i n_j \dots n_L} &=& \,\I_{ij\dots0}^{n_i n_j \dots, n_L-1} \quad (n_L\ge 1) \label{MIred-nc1-0}
\eeqan
Similarly, when $n_c=2$, we can contract both sides of Eq.~\eqref{MIdef-app} with $g_{\mu_1\mu_2}g_{\mu_3\mu_4}$ to obtain
\beqa
&& (24-10\epsilon)\,\I[q^4]_{ij\dots 0}^{n_i n_j \dots n_L} = 
\int\frac{d^dq}{(2\pi)^d} \frac{(q^2-M_i^2)^2+2M_i^2\,q^2-M_i^4}{(q^2-M_i^2)^{n_i}(q^2-M_j^2)^{n_j}\cdots (q^2)^{n_L}} \CR
&&\quad = \,\I_{ij\dots 0}^{n_i-2, n_j\dots n_L} +2(4-\epsilon)M_i^2\,\I[q^2]_{ij\dots 0}^{n_i n_j \dots n_L} -M_i^4 \,\I_{ij\dots 0}^{n_i n_j \dots n_L} \quad (n_i\ge 2)\,.\CR
\eeqa{MIred-nc-d}
Alternatively,
\beqa
&& (24-10\epsilon)\,\I[q^4]_{ij\dots 0}^{n_i n_j \dots n_L} = 
\int\frac{d^dq}{(2\pi)^d} \frac{(q^2-M_i^2)(q^2-M_j^2)+(M_i^2+M_j^2)q^2-M_i^2 M_j^2}{(q^2-M_i^2)^{n_i}(q^2-M_j^2)^{n_j}\cdots (q^2)^{n_L}} \CR
&&\quad = \,\I_{ij\dots 0}^{n_i-1, n_j-1,\dots n_L} +(4-\epsilon)(M_i^2+M_j^2)\,\I[q^2]_{ij\dots 0}^{n_i n_j \dots n_L} -M_i^2 M_j^2 \,\I_{ij\dots 0}^{n_i n_j \dots n_L} \quad (n_i, n_j\ge 1)\,.\CR
\eeqa{MIred-nc}
%

\section{Comparison with Feynman diagram matching}
\label{app:BMPZ}

Our results for one-loop SUSY threshold corrections presented in Section~\ref{sec:MSSMmatch-loop}, which are obtained from computing just 30 covariant diagrams, have been cross-checked against conventional Feynman diagram calculations reported in~\cite{BMPZ}, with full agreement found. In this final appendix, we explain how this comparison is made.

The general procedure is as follows. From~\cite{BMPZ}, we obtain analytical relations between the full theory parameters $g_3$, $g$, $g'$, $y_f$, $m^2$ and $\lambda$ (related to MSSM Lagrangian parameters via Eq.~\eqref{SMparam}) and the standard set of SM input observables (denoted with hats) $\hat\alpha_s(m_Z)$, $\hat m_Z$, $\hat G_F$, $\hat\alpha_\text{em}$, $\hat m_f$ and $\hat m_h$, computed via Feynman diagrams up to one-loop accuracy (we consistently drop higher loop order corrections, some of which are also reported in~\cite{BMPZ}). The same relations, with BSM contributions removed, define the corresponding effective parameters $g_3^\text{eff}$, $g^\text{eff}$, ${g'}^\text{eff}$, $y_f^\text{eff}$, $m^2_\text{eff}$ and $\lambda_\text{eff}$ in the SMEFT, up to power-suppressed corrections from $d\ge4$ operators. One-loop threshold corrections are then obtained by comparing the two, which should agree with what we have found via the more elegant covariant diagrams approach. Note that the tadpole-free scheme for Higgs vevs is adopted in~\cite{BMPZ}, so their results should be compared to ours when $\L_\text{SMEFT}$ is written in terms of $\beta$ (as opposed to $\beta'$), i.e.\ when the one-loop-generated piece $c_{\Phi\phi}(\Phi_\c^*\phi+\phi^*\Phi_\c)$ has been absorbed into $\L_\text{SMEFT}^\text{tree}$.

Let us start with the strong coupling $g_3$, which is simply extracted from $\hat\alpha_s(m_Z)$ via
\beq
g_3^2 = \frac{4\pi\,\hat\alpha_s(m_Z)}{1-\Delta\alpha_s} \,.
\eeqn
Therefore,
\beq
g_3^\text{eff} = g_3 \Bigl[\, 1-\frac{1}{2}\bigl(\Delta\alpha_s\bigr)^\text{SUSY}_{\O(1)} \,\Bigr] \,,
\eeq{g3eff}
where, according to~\cite{BMPZ},
\beqa
\bigl(\Delta\alpha_s\bigr)^\text{SUSY} &=& -\frac{g_3^2}{16\pi^2} \Bigl(\, \frac{1}{6}\sum_{f=u,d}\sum_{i=1}^2\log m_{\ft_i}^2 +2\,\log M_3^2 \,\Bigr) \CR
&=& \frac{g_3^2}{16\pi^2} \Bigl[ \,\frac{1}{6}\bigl( 2\,\It_{\qt}^2 +\,\I_{\ut}^1 +\,\I_{\dt}^2 \bigr) +2\,\I_{\gt}^2  +\O\Bigl(\frac{v^2}{\Lambda^2}\Bigr) \Bigr] \,,
\eeqa{DasSUSY}
with summation over three generations implicit. The $\frac{v^2}{\Lambda^2}$ power-suppressed terms come from electroweak symmetry breaking contributions to squark masses, and are not relevant here. For simplicity, throughout this appendix, we denote non-power-suppressed terms as $\O(1)$ (as in Eq.~\eqref{g3eff}) although they are formally $\O(\frac{1}{16\pi^2})$ when loop counting is also taken into account. It is readily seen that Eq.~\eqref{DasSUSY} is in agreement with our $\delta Z_G$ in Eq.~\eqref{dZG}.

Next, to extract electroweak gauge couplings $g$ and $g'$, we recall the relations
\beq
\alpha = \frac{\hat\alpha_\text{em}}{1-\Delta\alpha} \,,\qquad
\cw^2\sw^2 = \frac{\pi\,\alpha}{\sqrt{2}\,\hat m_Z^2\, \hat G_F\, (1-\Delta r)} \,,
\eeq{acwsw}
where
\beq
\Delta r = \frac{\Pi_{WW}^T(0)}{m_W^2} -\text{Re}\, \frac{\Pi_{ZZ}^{T}(m_Z^2)}{m_Z^2} +\delta_\text{VB} \,.
\eeqn
Here, $\Pi_{WW}^T(p^2)$ and $\Pi_{ZZ}^T(p^2)$ are transverse parts of the $W$ and $Z$ self-energies, which represent ``universal'' contributions to $\mu^-\to e^- \bar\nu_e \nu_\mu$ which determines $\hat G_F$. On the other hand, $\delta_\text{VB}$ contains non-universal contributions from vertex corrections, box diagrams, and wavefunction renormalizations. Only the universal part of $\Delta r$, i.e.
\beq
\Delta r_u \equiv \frac{\Pi_{WW}^T(0)}{m_W^2} -\text{Re}\, \frac{\Pi_{ZZ}^{T}(m_Z^2)}{m_Z^2}
\eeqn
is relevant for $g$, $g'$ threshold corrections, because $\delta_\text{VB}$ has an EFT counterpart in terms of local effective operator contributions to muon decay. Thus, from Eq.~\eqref{acwsw},
\beq
\alpha^\text{eff} = \alpha \Bigl[\, 1-(\Delta\alpha)^\text{SUSY}_{\O(1)}\, \Bigr] \,,\qquad
\bigl(\cw^2\sw^2\bigr)^\text{eff} = \cw^2\sw^2 \,\Bigl[\, 1-(\Delta\alpha)^\text{SUSY}_{\O(1)} -(\Delta r_u)^\text{SUSY}_{\O(1)} \, \Bigr] \,.
\eeqn
The QED coupling and weak mixing angle can be directly translated into $SU(2)_L\times U(1)_Y$ gauge couplings via $4\pi\alpha = g\sw = g'\cw$. We therefore obtain
\beqa
g^\text{eff} &=& g\, \Bigl\{ 1+\frac{1}{2}\frac{1}{\cw^2-\sw^2} \Bigl[ \sw^2 \,(\Delta\alpha)^\text{SUSY}_{\O(1)} +\cw^2 \,(\Delta r_u)^\text{SUSY}_{\O(1)} \Bigr] \Bigr\} \CR
&=& g\, \Bigl\{ 1+\frac{1}{2}\frac{1}{\cw^2-\sw^2} \Bigl[ \sw^2 \,(\Delta\alpha)^\text{SUSY}_{\O(1)} +\cw^2\,\frac{4}{v^2} \Bigl(\,\frac{1}{g^2}\Pi_{WW}^T(0) -\frac{\cw^2}{g^2}\Pi_{ZZ}^T(m_Z^2)\Bigr)^\text{SUSY}_{\O(v^2)} \Bigr] \Bigr\} \,, \qquad \label{g2eff}\\
{g'}^\text{eff} &=& g'\, \Bigl\{ 1-\frac{1}{2}\frac{1}{\cw^2-\sw^2} \Bigl[ \cw^2 \,(\Delta\alpha)^\text{SUSY}_{\O(1)} +\sw^2 \,(\Delta r_u)^\text{SUSY}_{\O(1)} \Bigr] \Bigr\}\CR
&=& g'\, \Bigl\{ 1-\frac{1}{2}\frac{1}{\cw^2-\sw^2} \Bigl[ \cw^2 \,(\Delta\alpha)^\text{SUSY}_{\O(1)} +\sw^2\,\frac{4}{v^2} \Bigl(\,\frac{1}{g^2}\Pi_{WW}^T(0) -\frac{\cw^2}{g^2}\Pi_{ZZ}^T(m_Z^2)\Bigr)^\text{SUSY}_{\O(v^2)} \Bigr] \Bigr\} \,.\label{g1eff}
\eeqan
The SUSY part of the self-energies $\Pi_{WW}^T$ and $\Pi_{ZZ}^T$ are to be expanded in powers of $\frac{v^2}{\Lambda^2}$. 
Analytical expressions of these and other self-energies to appear below can be found in~\cite{BMPZ}. They are rather tedious and will not be displayed here.

Then, moving on to Yukawa couplings $y_f$, we note that
\beq
\hat m_f = \frac{1}{\sqrt{2}} y_f v \Bigl( 1-\text{Re}\,\frac{\Sigma_f(m_f)}{m_f} \Bigr) \,,
\eeqn
where $\Sigma_f(\slashed{p})$ is the fermion self-energy, and the light Higgs vev $v$ is extracted via
\beq
v^2 = 4\,\frac{\hat m_Z^2 +\text{Re}\,\Pi_{ZZ}^T(m_Z^2)}{g^2+g'^2} \,.
\eeqn
In the SMEFT, it is $\hat v$, the vev of the canonically normalized light Higgs field $\hat\phi$, that is extracted via this procedure,
\beq
\hat v^2 = 4\,\frac{\hat m_Z^2 +\text{Re}\,\bigl(\Pi_{ZZ}^T(m_Z^2)\bigr)^\text{SM}}{\bigl(g^\text{eff}\bigr)^2+\bigl({g'}^\text{eff}\bigr)^2} \,.
\eeqn
With Eqs.~\eqref{g2eff} and~\eqref{g1eff}, it is easily seen that
\beq
\hat v^2 = v^2 \biggl[1-\Bigl(\frac{\Pi_{WW}^T(0)}{m_W^2}\Bigr)^\text{SUSY}\biggr] \,.
\eeqn
Therefore,
\beq
y_f^\text{eff} = y_f \,\frac{v}{\hat v} \,\biggl[ 1-\Bigl(\frac{\Sigma_f(m_f)}{m_f}\Bigr)^\text{SUSY}_{\O(1)} \biggr]
= y_f \,\biggl[ 1-\Bigl(\frac{\Sigma_f(m_f)}{m_f}\Bigr)^\text{SUSY}_{\O(1)} +\frac{1}{2}\,\Bigl(\frac{\Pi_{WW}^T(0)}{m_W^2}\Bigr)^\text{SUSY}_{\O(1)} \biggr] \,.
\eeq{yfeff}
When cross-checking with our results, it is worth noting the following correspondence between the terms in Eq.~\eqref{yfeff} and those in Eq.~\eqref{thrcor} (using $f=t$ as an example),
\beqa
&& \Bigl(\frac{\Sigma_t(m_t)}{m_t}\Bigr)^\text{SUSY}_{\O(1),\, B_0\,\text{part}} = \frac{\delta y_t}{y_t} \,,\qquad
\Bigl(\frac{\Sigma_t(m_t)}{m_t}\Bigr)^\text{SUSY}_{\O(1),\, B_1\,\text{part}} = \frac{1}{2} \bigl(\delta Z_q +\delta Z_u\bigr)\,,\CR
&& \Bigl(\frac{\Pi_{WW}^T(0)}{m_W^2}\Bigr)^\text{SUSY}_{\O(1)} = -\delta Z_\phi \,,
\eeqan
where $B_0$ and $B_1$ are different loop integrals that appear in $\Sigma_f$.

Finally, we discuss the Higgs potential parameters $m^2$ and $\lambda$. The minimization condition of the 1PI effective potential,
\beqa
0 &=& \mu^2+\mHusq -b\cot\beta -\frac{1}{8}(g^2+g'^2) \cbb\, v^2 -\frac{t_u}{v_u} \CR
&=& \mu^2+\mHdsq -b\tan\beta +\frac{1}{8}(g^2+g'^2) \cbb\, v^2 -\frac{t_d}{v_d} 
\eeqan
allows us to eliminate $\mu^2+\mHusq$ and $\mu^2+\mHdsq$ in favor of $v$ and $\beta$. From Eq.~\eqref{SMparam} we see that $m^2$ and $\lambda$ are related by
\beq
m^2 = \mu^2 +\mHusq\sb^2 +\mHdsq\cb^2 -b\sbb = -\frac{1}{8}(g^2+g'^2) \cbb^2 v^2 +\frac{t_h}{v} = -\lambda v^2 +\frac{t_h}{v}\,,
\eeq{m2lambdarelation}
where
\beq
\frac{t_h}{v} \equiv \frac{\sb\, t_u+\cb\, t_d}{v} = \sb^2\,\frac{t_u}{v_u} +\cb^2\,\frac{t_d}{v_d} \,.
\eeqn
To extract them from $\hat m_h$, we write the tree-level mass matrix squared in the $(H_u, H_d)$ basis,
\beq
{\bf M}_{H}^2 = \left(
\begin{matrix}
b\cot\beta +2\lambda v^2\cbb^{-2}\sb^2 +\frac{t_u}{v_u} &&& -b-2\lambda v^2\cbb^{-2}\sb\cb \\
-b-2\lambda v^2\cbb^{-2}\sb\cb &&& b\tan\beta +2\lambda v^2\cbb^{-2}\cb^2 +\frac{t_d}{v_d}
\end{matrix}
\right).
\eeqn
Therefore,
\beq
\hat m_h^2 = \text{smaller eigenvalue of}\;\; {\bf M}_H^2 -
\left(\begin{matrix}
\Pi_{uu} & \Pi_{ud} \\
\Pi_{ud} & \Pi_{dd}
\end{matrix}\right)
= 2\lambda v^2 +\frac{t_h}{v} -\Pi_{hh} \,,
\eeq{mh2}
where $\Pi_{uu,ud,dd}$ are one-loop self-energies of the Higgs doublets $H_u$, $H_d$, and
\beq
\Pi_{hh} \equiv \sb^2\,\Pi_{uu} +\cb^2\,\Pi_{dd} +2\sb\cb\,\Pi_{ud} \,.
\eeqn
From Eq.~\eqref{mh2} we obtain
\beq
\lambda_\text{eff} = \lambda\,\frac{v^2}{\hat v^2} +\frac{1}{2v^2} \Bigl(\frac{t_h}{v} -\Pi_{hh}\Bigr)^\text{SUSY}_{\O(v^2)}
= \lambda\,\biggl[ 1+\Bigl(\frac{\Pi_{WW}^T(0)}{m_W^2}\Bigr)^\text{SUSY}_{\O(1)} \biggr] +\frac{1}{2v^2} \Bigl(\frac{t_h}{v} -\Pi_{hh}\Bigr)^\text{SUSY}_{\O(v^2)} \,,
\eeq{lameff}
and then from Eq.~\eqref{m2lambdarelation},
\beq
m^2_\text{eff} = -\lambda_\text{eff}\,\hat v^2 +\Bigl(\frac{t_h}{v} -\Pi_{hh}\Bigr)^\text{SM} 
= m^2 -\frac{1}{2}\Bigl(3\,\frac{t_h}{v} -\Pi_{hh}\Bigr)^\text{SUSY}_{\O(\Lambda^2)} \,.
\eeqn
Note that while both $\frac{t_h}{v}$ and $\Pi_{hh}$ contain $\O(\Lambda^2)$ terms, they cancel in the combination $\frac{t_h}{v}-\Pi_{hh}$ appearing in Eq.~\eqref{lameff}. The subleading $\O(v^2)$ terms needed here come from both expanding the loop integrals involved up to next-to-leading order, and electroweak symmetry breaking contributions to superpartner masses. Also, note the different notation adopted in~\cite{BMPZ}: $t_{1,2}=t_{d,u}$, $\Pi_{s_1s_1,s_1s_2,s_2s_2}=\Pi_{dd,ud,uu}$.



\end{document}